%% file: ms.tex
\documentclass[trackchanges,twocolumn,times]{aastex63}

\expandafter\def\csname editcolor1\endcsname{magenta}
\expandafter\def\csname editcolor2\endcsname{blue}  
\expandafter\def\csname editcolor3\endcsname{violet} 

\usepackage{newtxtext,newtxmath}
\usepackage[T1]{fontenc}
\usepackage{graphicx}	
\usepackage{amsmath}	
\usepackage{float}
\usepackage{bookmark} 
\bookmarksetup{numbered, open,}
\usepackage{enumitem}
\setlist[enumerate]{itemsep=0mm}
\usepackage{hyperref}
\usepackage{subfigure}
\usepackage{xspace}
\usepackage{xcolor} 
\usepackage{soul} 

\newcommand{\ArII}{Ar~{\sc ii}}
\newcommand{\ArIII}{Ar~{\sc iii}}

\newcommand{\SiII}{Si~{\sc ii}}

\newcommand{\SiIV}{Si~{\sc iv}}

\newcommand{\SIII}{S~{\sc iii}}

\newcommand{\TiII}{Ti~{\sc ii}}

\newcommand{\FeI}{Fe~{\sc i}}
\newcommand{\FeII}{Fe~{\sc ii}}
\newcommand{\FeIII}{Fe~{\sc iii}}
\newcommand{\CoI}{Co~{\sc i}}
\newcommand{\CoII}{Co~{\sc ii}}
\newcommand{\CoIII}{Co~{\sc iii}}

\newcommand{\NiI}{Ni~{\sc i}}
\newcommand{\NiII}{Ni~{\sc ii}}

\newcommand{\NiIII}{Ni~{\sc iii}}
\newcommand{\NiIV}{Ni~{\sc iv}}
\newcommand{\Fefs}{$^{56}$Fe}
\newcommand{\Cofs}{$^{56}$Co}
\newcommand{\Nifs}{$^{56}$Ni}

\newcommand{\eg}{e.g.\xspace}
\newcommand{\ie}{i.e.\xspace}

\newcommand{\Msun}{\ensuremath{\mathrm{M}_{\odot}}\xspace}

\newcommand{\kms}{km~s\ensuremath{^{-1}}\xspace}

\newcommand{\gcm}{g cm\ensuremath{^{-3}}\xspace}

\newcommand{\mic}{\ensuremath{\mu}m\xspace}

\graphicspath{{./}{figs/}}

\received{xxx}
\revised{xxx}
\accepted{xxx}

\submitjournal{ApJ}

\shorttitle{A MIRI/MRS spectrum of SN~2021aefx at +415\,$\mathrm{d}$}
\shortauthors{Ashall, Hoeflich, et al.}

\begin{document}
\title{A JWST Medium Resolution MIRI Spectrum and Models of the Type Ia supernova 2021aefx at +415~d}

\correspondingauthor{Chris Ashall}
\email{chris.ashall24@gmail.com}

\input{authors.tex}

\begin{abstract}

We present a JWST MIRI/MRS spectrum (5-27 $\micron$) of the Type Ia supernova (SN Ia), SN 2021aefx at $+415$ days past $B$-band maximum. The spectrum, which was obtained during the iron-dominated nebular phase, has been analyzed in combination with previous \textit{JWST} observations of SN 2021aefx, to provide the first \textit{JWST} time series analysis of an SN Ia. We find the temporal evolution of the [\CoIII] 11.888 $\mic$ feature directly traces the decay of \Cofs. The spectra, line profiles, and their evolution are analyzed with off-center delayed-detonation models. Best fits were obtained with White Dwarf (WD) central densities of $\rho_c=0.9-1.1\times 10^9$\gcm, a WD mass of M$_{\mathrm{WD}}$=1.33--1.35M$_\odot$, a WD magnetic field of $\approx10^6$G, and an off-center deflagration-to-detonation transition at $\approx$ 0.5 $M_\odot$ seen opposite to the line of sight of the observer ($-30^o$). The inner electron capture core is dominated by energy deposition from $\gamma$-rays whereas a broader region is dominated by positron deposition, placing SN 2021aefx at +415 d in the transitional phase of the evolution to the positron-dominated regime. The formerly ``flat-tilted' profile at 9 \mic\ now has significant contribution from [\NiIV], [\FeII], and [\FeIII] and less from [\ArIII], which alters the shape of the feature as positrons excite mostly the low-velocity Ar. Overall, the strength of the stable Ni features in the spectrum is dominated by positron transport rather than the Ni mass. Based on multi-dimensional models, our analysis is consistent with a single-spot, close-to-central ignition with an indication for a pre-existing turbulent velocity field, and excludes a multiple-spot, off-center ignition.
\end{abstract}


\section{Introduction} \label{sec:intro}
Type Ia Supernovae (SNe~Ia) originate from the thermonuclear disruption of a carbon-oxygen (C-O) White Dwarf (WD) in a multiple star system \citep{Hoyle1960}; yet to date, the exact details of their progenitor scenarios or explosion mechanisms are unknown. 
Determining the exact origin of SNe~Ia is essential if we are to understand the nucleosynthesis of  heavy elements and improve the use of SNe~Ia as extragalactic distance indicators (see \citealt{2017hsn..book.1955S} and \citealt{2017hsn..book.2615R} for recent reviews). 

Potential SNe~Ia progenitor scenarios include: (1) the single degenerate (SD) scenario which consists of a C-O WD and a non-degenerate companion star \citep{Whelan1973}, 
(2) the double degenerate (DD) scenario which consists of two WDs \citep{Iben1984,Webbink1984}, 
and (3) a triple/quadruple system consisting of at least two C-O WDs \citep{Thompson2011,2013MNRAS.435..943P}. There is also a complex interplay between the progenitor scenario and the explosion mechanism, where many explosion mechanisms can theoretically occur within each  progenitor scenario. 
Two of the leading explosion mechanisms include the explosion of a near Chandrasekhar mass (M$_{\mathrm{Ch}}$) WD
\citep{Iben1984,Khokhlov91} and the detonation of a sub-M$_{\mathrm{Ch}}$ WD \citep{Livne95}.
Both of which can occur in SD and DD systems.  
In the near-M$_{\mathrm{Ch}}$ explosion the WD  accretes H, He,  or C material from a non-degenerate or degenerate companion star until it approaches the M$_{\mathrm{Ch}}$, during which time densities in the center of the star become high enough for a simmering stage to occur and a thermonuclear disruption begins \citep{Khokhlov91}. The flame can propagate as a deflagration, detonation, or both via a deflagration-to-detonation transition (DDT) \citep{Khokhlov91,Hoeflich_1995,Gamezo03,Poludnenko19}. 
In the sub-M$_{\mathrm{Ch}}$ scenario, a surface He layer detonates and produces an inwards shockwave that disrupts the whole WD. This can occur for a variety of core masses (0.6-1.1 $\Msun$) and He shell (0.01-0.2 $\Msun$) masses. Although, only the higher mass WDs and smaller He shell masses are expected to reproduce the observed properties of SNe~Ia \citep{Livne95,Shen2018,Boos_etal_2021}.

Observations of SNe~Ia during the nebular phase reveal their inner layers. 
Spectra at these epochs can be used to measure the  high-density burning regions  in the ejecta \citep[\eg][]{Axelrod_1980,2007Sci...315..825M,2016MNRAS.463.1891A,2020MNRAS.494.2809M,2020Sci...367..415J,Maguire18,Hoeflich_2021_20qxp,Kumar23}.
A spectral region of particular interest is at mid-infrared (MIR) wavelengths ($\sim$5-27~\mic), as it contains lines from critical ions that 
do not have suitable transitions in the optical or near-infrared (NIR). 
These ions can be used to distinguish between the leading progenitor and explosion scenarios. Prior to the launch of the  {\it James Webb Space Telescope} ({\it JWST})
there were four SNe Ia which had MIR ($\lambda>$5~\mic) spectral observations, and there were seven published MIR spectra of SNe Ia in total (see the introduction of \citet{DerKacy_etal_2023_21aefx} for more details).
These data provided new insights into SNe~Ia explosions but the interpretation was hampered by low S/N and low spectral resolution.

\begin{figure*}
    \centering
    \includegraphics[width=0.99\textwidth]{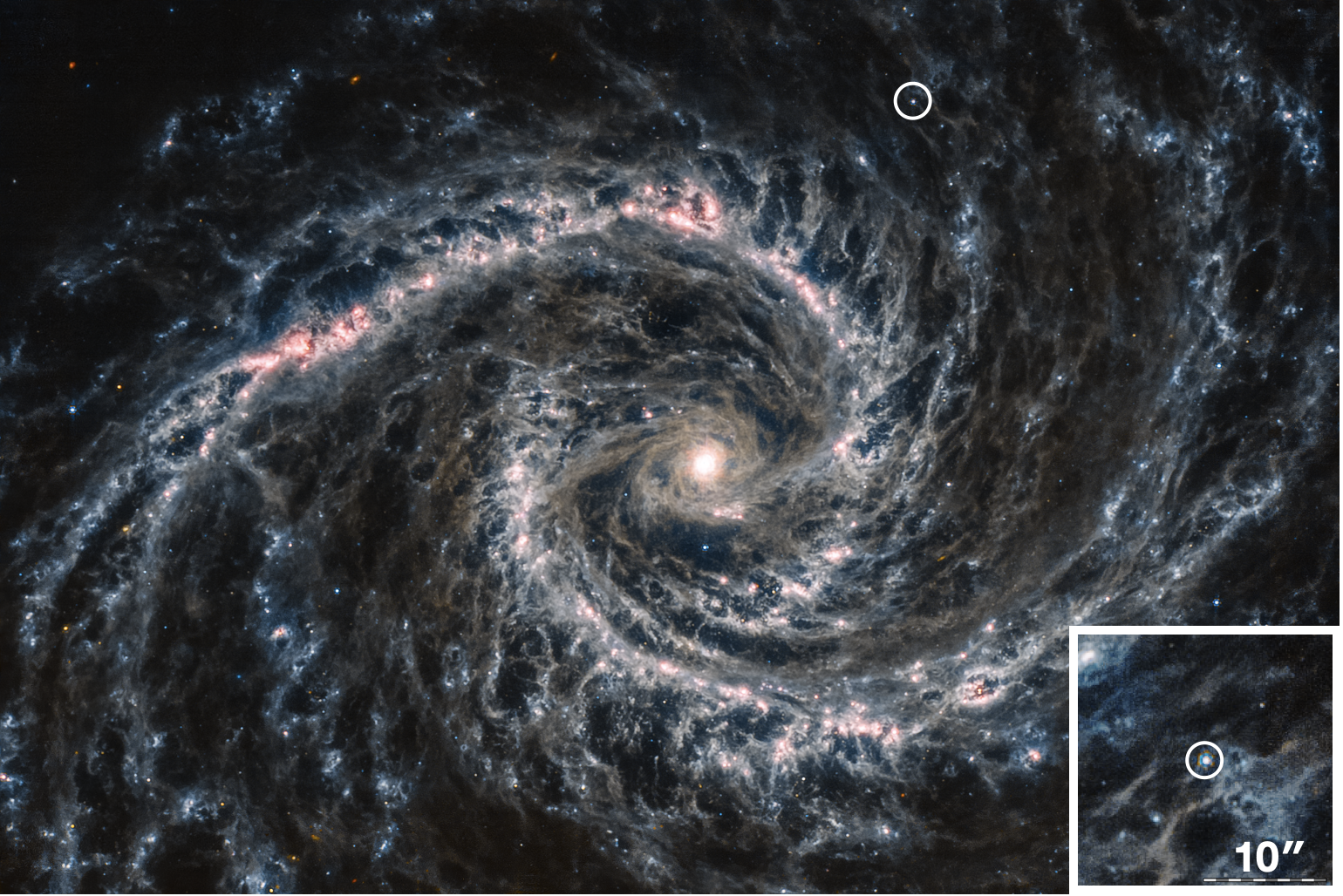}
    \caption{ A stacked composite image of NGC~1566 with photometric data obtained with {\it JWST}’s MIRI instruments, individual bands are published in \citep{Chen23}. Here, SN~2021aefx is at a phases of +357~d past $B$-band maximum and is highlighted in the white circle. North is 1.5° counter-clockwise from left. An inset around the SN is shown in the bottom right corner.
For the original image   see  \href{https://www.flickr.com/photos/geckzilla/52523099436/}{here}.}
    \label{fig:image}
\end{figure*}

The launch of {\it JWST} has opened up a new era of transient astronomy. To date, {\it JWST} spectra of three nearby SNe~Ia have been published. These are:  SN~2022xkq \citep{DerKacy23xkq}, SN~2022pul \citep{Siebert23, Kwok23}, and  SN~2021aefx \citep{Kwok2022,DerKacy_etal_2023_21aefx}.
Of these, SN~2021aefx is the best observed SN~Ia. Broad-band NIR+MIR {\it JWST} imaging was obtained at +255~d after $B$-band maximum, which enabled the late time NIR+MIR decline rates and MIR flux contribution in the explosion to be determined  \citep{Mayker23}.  Spectra of SN~2021aefx were  acquired  at +255~d \citep{Kwok2022}, and +323~d \citep{DerKacy_etal_2023_21aefx}  past $B$-band maximum. 
These spectra revealed many unique features including multiple stable Ni lines, which are indicative of high-density burning, a ``flat-tilted'' [\ArIII] 8.991~\micron\ profile, and a strong [\CoIII] 11.888~$\micron$\ resonance feature \citep{Kwok2022,DerKacy_etal_2023_21aefx}. Spectral modeling of the {\it JWST} data indicated that SN~2021aefx is consistent with a delayed detonation near-M$_{\mathrm{Ch}}$ explosion of a C-O WD, that had an off-center DDT, and produced 5.9 $\times$ 10$^{-2}~\Msun$ of $^{58}$Ni (\citealt{DerKacy_etal_2023_21aefx}; although see \citealt{Blondin23}).  However, all of the previous spectra of SN~2021aefx  were obtained with the low-resolution spectral mode and at wavelengths less than 14~\mic. 
The medium resolution spectrometer (MRS)  on the Mid-Infrared Instrument (MIRI) has a resolving power of $\sim$2700,  a wavelength coverage from $\sim$5-27~$\micron$ and is therefore ideal for obtaining precision velocity measurements as well as observations at long wavelengths.

Here we present the first observed {\it JWST} MIRI/MRS spectrum of a SN~Ia\footnote{The first published {\it JWST} MIRI/MRS spectra of a SN~Ia was SN~2022xkq \citep{DerKacy23xkq}.}.
In \autoref{sec:Data} we discuss the observations and data reduction.  
In \autoref{sec:IDs} we present line identification, followed by spectral comparisons in  \autoref{sec:speccomp}. In \autoref{sec:specevolution} we discuss the time series spectral evolution. In \autoref{sec:models} we use self-consistent multi-dimensional models to produce synthetic spectra, where we demonstrate how the MIRI/MRS data can be used in conjunction  with synthetic spectra and line profiles to understand the  underlying explosion physics, and the conditions at the thermonuclear runaway.
A summary of our results and our conclusions are presented  in \autoref{sec:conclusions}. 

\section{Observations \& Data Reduction} \label{sec:Data}

\begin{deluxetable}{cccc}
  \tablecaption{Log of {\it JWST} observations. The spectrum was taken at MJD=59964.34 which is rest-frame +415~d  relative to $B$-band maximum \label{tab:obs}} 
  \tablehead{\colhead{Parameter} & \colhead{Value}& \colhead{Value}& \colhead{Value}}
  \startdata
    \multicolumn{4}{c}{MIRI/MRS Spectra}  \\
        \hline
    Sub-band& Short &Medium & Long\\
    Groups per Integration& 35&35&36 \\
    Integrations per Exp. &3&3&3\\
    Exposures per Dither &1&1&1\\
    Total Dithers & 12 & 12& 12\\
    Exp Time [s] & 10224.88& 10224.88 & 10511.565 \\
        Readout Pattern&SLOWR1&SLOWR1&SLOWR1 \\
  \enddata
\end{deluxetable}

SN~2021aefx was discovered on 2021 Nov 11.3 (MJD=59529.5) by the Distance Less Than 40 Mpc Survey \citep{DLT40}. It was located at 
$\alpha = 04^{h}19^{m}53^{s}.40$, $\delta=-54\arcdeg56\arcmin53\arcsec.09$, south-west of the center of its host galaxy NCG~1566 (z=0.0050).  The location of SN~2021aefx in its host galaxy is shown in Fig. \ref{fig:image}. 
Early time observations and analysis of SN~2021aefx 
indicated it was discovered within hours of the explosion, had an early excess emission in the $u$-band, a quickly evolving color curve, and extremely high spectral velocities \citep{Ashall22,Hosseinzadeh22,Ni23}.  By maximum light it resembled a normal luminosity SNe~Ia, demonstrating that SN~2021aefx is only unusual in the outermost layers  \citep{Ashall22,Hosseinzadeh22,Ni23}

 {\it JWST}  MIRI/MRS spectral observations of SN~2021aefx were triggered through {\it JWST}-GO-2114  \citep{2021jwst.prop.2114A}. Observations began on 20th January 2023 at 03:34:38 (MJD=59964.15), and ended on 
 20th January 2023 at  12:47:50 (MJD=59964.53). We take our time of observation as the mid-point, MJD = 59964.34. Throughout this work, we use a time of $B$-band maximum of MJD = 59547.25 \citep{DerKacy_etal_2023_21aefx} implying the spectrum was obtained at rest-frame +415~d past $B$-band maximum light.

\begin{figure*}
    \centering
    \includegraphics[width=\textwidth]{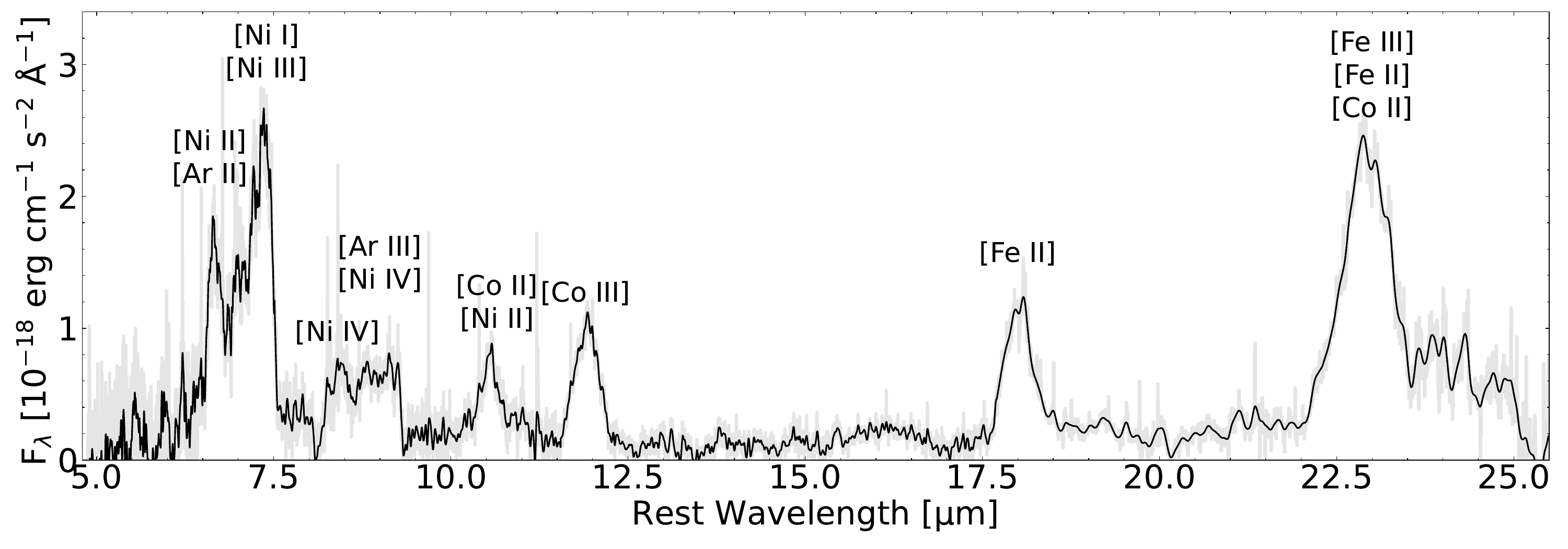}
    \caption{The MIRI/MRS spectrum of SN~2021aefx at +415~d past maximum light. The raw data is plotted in light gray behind. Due to the fact that the spectrum is heavily over sampled, it has been smoothed to the instrumental resolution (black).  The dominant ions that contribute to each feature are labeled.   \label{fig:fullspec}}
\end{figure*}

Spectral data were acquired using the MRS on MIRI with the short, medium, and long gratings and with each channel to produce a continuous spectrum from $\sim$5-27~$\micron$. The details of the instrument setup can be found in Table \ref{tab:obs}. The total exposure time was 8.6 hours. 

The data were reduced using a custom-built pipeline designed to extract observations of faint point sources that have complex backgrounds in MIRI/MRS data cubes (see \citealt{Shahbandeh24}). The details of the data reduction can be found in Appendix \ref{sec:Datared}. This data reduction technique dramatically reduced the background flux level by two orders of magnitude across all channels. This both increases the S/N in the extraction of the SN, and ensures that the continuum is dominated by SN and not instrumental flux.

The  final spectrum
has been smoothed with Spextractor \citep{Burrow20}
channel-by-channel to properly account for the differences in
resolution across the full MIRI/MRS wavelength coverage. Throughout this work, the spectra were corrected to rest frame using the recessional velocity of 1500~\kms, and a rotational galactic velocity of $65 \pm 60$~\kms at the location of the SN \citep{Elagali2019}. 

Comparing the MIRI/low-resolution spectrometer (LRS) and MIRI/MRS spectra  of  SN~2021aefx is useful for checking both the flux and wavelength calibration. It is known from the comparison between \textit{JWST} MIRI photometry and \textit{JWST} MIRI/LRS spectra that the absolute flux calibration of the MIRI/LRS data of SN~2021aefx is within 2\% \citep{Kwok2022,Chen23}. 
Generally, for MIRI/MRS data it is thought that the flux calibration is accurate to 5.6$\pm0.7$\% \citep{Argyriou23}. 
This was also confirmed using MIRI/MRS data of SN~2022acko where the spectral flux in Channels 1, 2 and 3 were found to be consistent with  simultaneous MIRI broad-band photometry \citep{Shahbandeh24}. 
However, it is not yet known how well the pipeline extracts the  SN flux in Channel 4, and how successful it is at accurately removing the background from both the instrument and the underlying host galaxy. 
Overall, the flux of SN~2021aefx MIRI/MRS spectrum at +415~d is lower than that at the previous epoch (see \autoref{sec:specevolution}). It is also apparent that the decrease in peak flux is more rapid at earlier epochs compared to later ones. This is consistent with the light curve flattening with time.
Although there is no simultaneous photometry at +415~d, this time series behavior suggests that the flux calibration of the MIRI/MRS spectrum is accurate. 
However, for future literature comparison we  provide synthetic photometry using the \textit{JWST} passbands, see Table \ref{tab:synphot}.
Finally, we  note that, the wavelength calibration of MIRI/MRS spectra is accurate to within 9~\kms at 5~$\micron$ and to 27~\kms at 28~$\micron$ \citep{Argyriou23}.

\begin{deluxetable}{cc}
  \tablecaption{Synthetic photometry produced from the +415~d spectrum of SN~2021aefx  \label{tab:synphot}} 
  \tablehead{\colhead{Filter} & \colhead{Flux}\\ \colhead{}& \colhead{mJy}}
  \startdata
        \hline
    F770W& 0.1724\\
    F1000W& 0.3931\\
    F1130W& 0.3412\\
    F1280W& 0.1330\\
    F1500W&  0.1165
  \enddata
\end{deluxetable}

\begin{figure*}
    \centering
    \includegraphics[width=.9\textwidth]{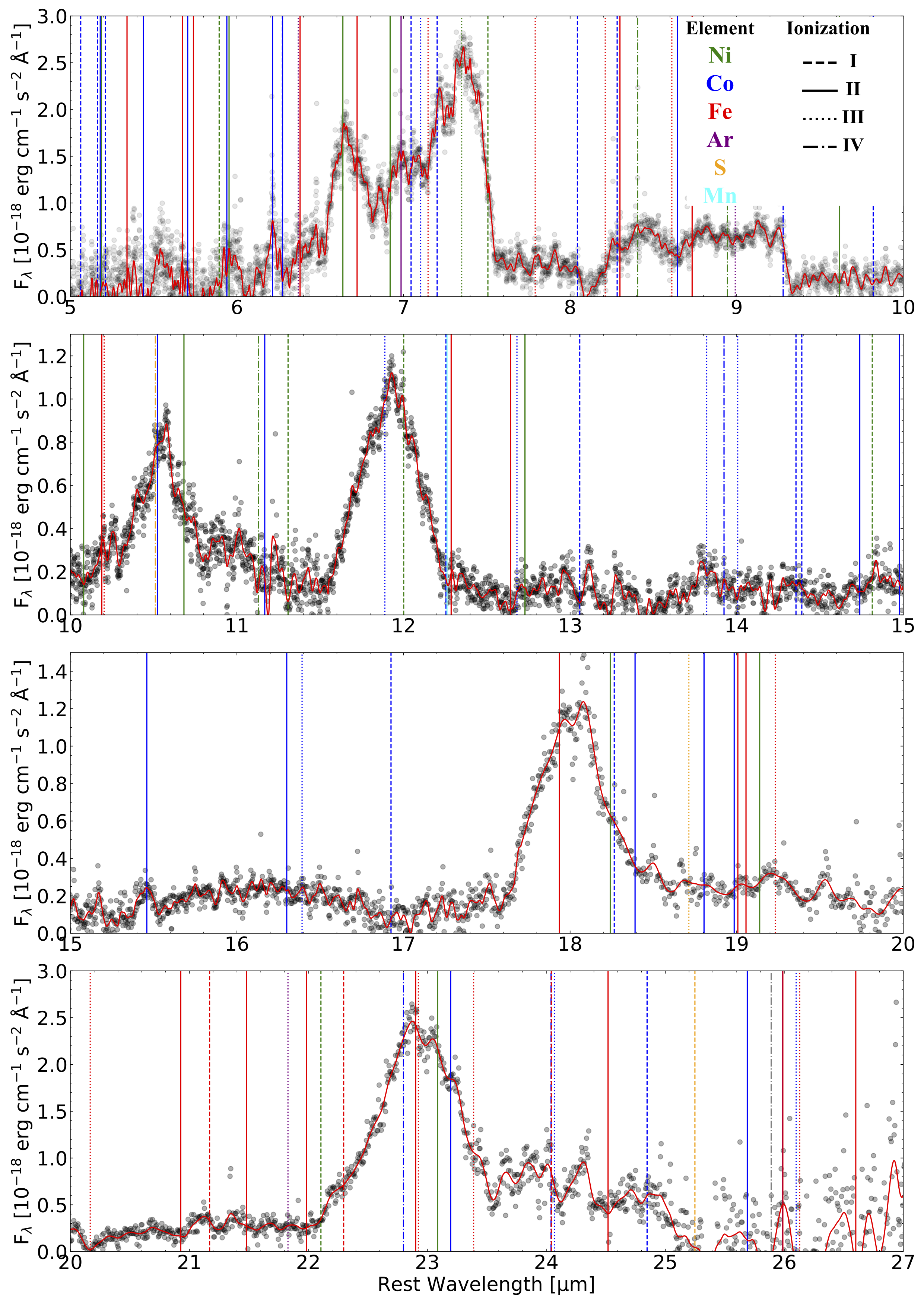}
    \caption{The +415~d spectrum of SN~2021aefx with the main contributing ions labeled. }
    \label{fig:LineID}
\end{figure*}

\section{Line Identifications} \label{sec:IDs}

The fully stitched 4-channel MIRI/MRS spectrum is presented in Fig. \ref{fig:fullspec}.
The strongest lines that contribute to the formation of the spectral features have been labeled. 
These line identifications are based on radiation hydrodynamical models discussed in section \autoref{sec:models}.
At short wavelengths ($<14~\micron$), the line IDs are consistent with those presented in \citet{Kwok2022} and \citet{DerKacy_etal_2023_21aefx}.
However, the  individual line strengths and profiles have evolved over the intervening
92~d between observations --- demonstrating that the physical processes driving the spectral formation have profoundly changed (see \autoref{sec:models}).

We identify four dominant regions of line formation, which are presented in Fig. \ref{fig:LineID}. The strongest lines contributing to each spectral region are described below. For the line identification, we use the models presented in \autoref{sec:models}. The full list of line IDs, including weaker lines, can be found in Table \ref{tab:ir_lines}.

Between 5--7.5$~\micron$ the spectrum is dominated by stable Ni lines including [\NiI] 5.893 $~\micron$; [\NiII] 6.636 and 6.920 $~\micron$; as well as [\FeII] 5.674$~\micron$, and  [\ArII]  6.985$~\micron$. Between 7.5--10.0$~\micron$ lines of [\NiIV] 8.405, 8.945$~\micron$, and [\ArIII] 8.991$~\micron$ are the strongest.

In the wavelength range of 10.0--15.0$~\micron$ the strongest features are, 
[\FeII] 10.189, 12.286, 12.642$~\micron$;
[\NiII] 10.682, 12.729$~\micron$;
[\NiIV] 11.13$~\micron$;
[\NiI] 12.001$~\micron$;
[\CoIII] 11.888$~\micron$; and
[\CoI] 12.255 $~\micron$.

Between 15--20$~\micron$ the strongest spectral lines contributing to the formation of the features are, [\FeII] 17.936, 19.056$~\micron$, and [\SIII] 18.713$~\micron$. 
Beyond 20$~\micron$  the spectrum is dominated by
[\FeIII] 22.925$~\micron$;
[\FeI] 24.052$~\micron$;
[\CoIII] 24.070$~\micron$; and
[\FeII] 24.519, 25.988$~\micron$.

It should be noted that the line IDs beyond 20~$\micron$
 are tentative, as the background subtraction is uncertain at these wavelengths.  Furthermore, the flux calibration at these wavelengths is highly uncertain and the spectrum at wavelengths  $>25$~$\micron$ should not be trusted until a more accurate reduction in Channel 4 is available.
  Furthermore, it is not possible to confidently identify all spectral features, as data for many atomic line transitions in these MIR regions are missing.
However, through spectral modeling we can identify some of the unknown line strengths as well as estimates of the flux contribution from the weaker lines (see \autoref{subsect:Atomic} and \autoref{sec:opt_lines}).

\begin{figure*}
    \centering
    \includegraphics[width=.99\textwidth]{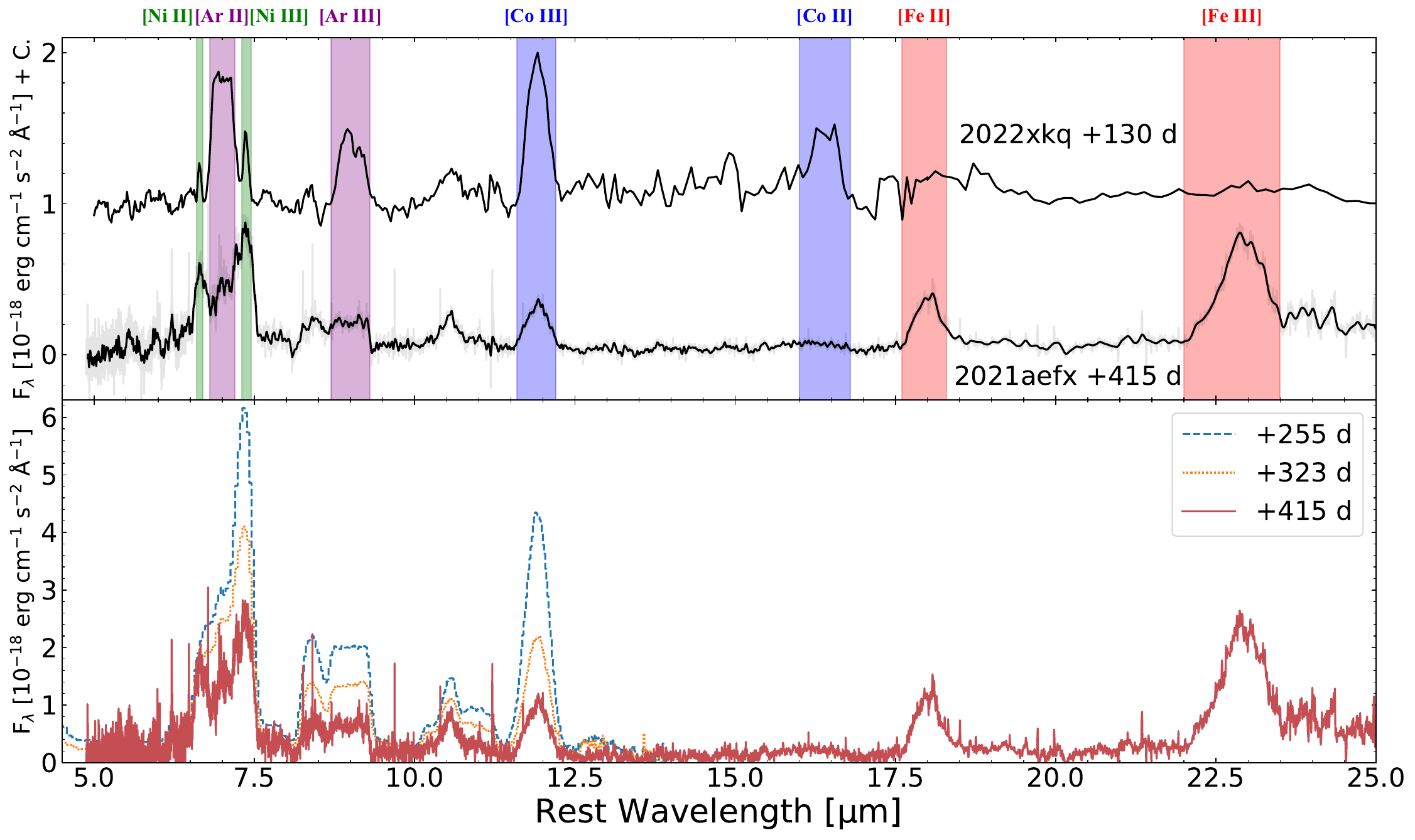}
    \caption{(\textit{Top:}) A spectral comparison between the MIRI/MRS data of SN~2022xkq and SN~2021aefx. The dominant ions contributing to various spectral regions are highlighted with vertical colored bars. We note that SN~2022xkq is a low luminosity SN~Ia. (\textit{Bottom:}) Time series comparison of all published MIR spectra of SN~2021aefx. }
    \label{fig:aefxcomp}
\end{figure*}

\section{Spectral Comparison} \label{sec:speccomp}
\begin{figure}
    \centering
    \includegraphics[width=.45\textwidth]{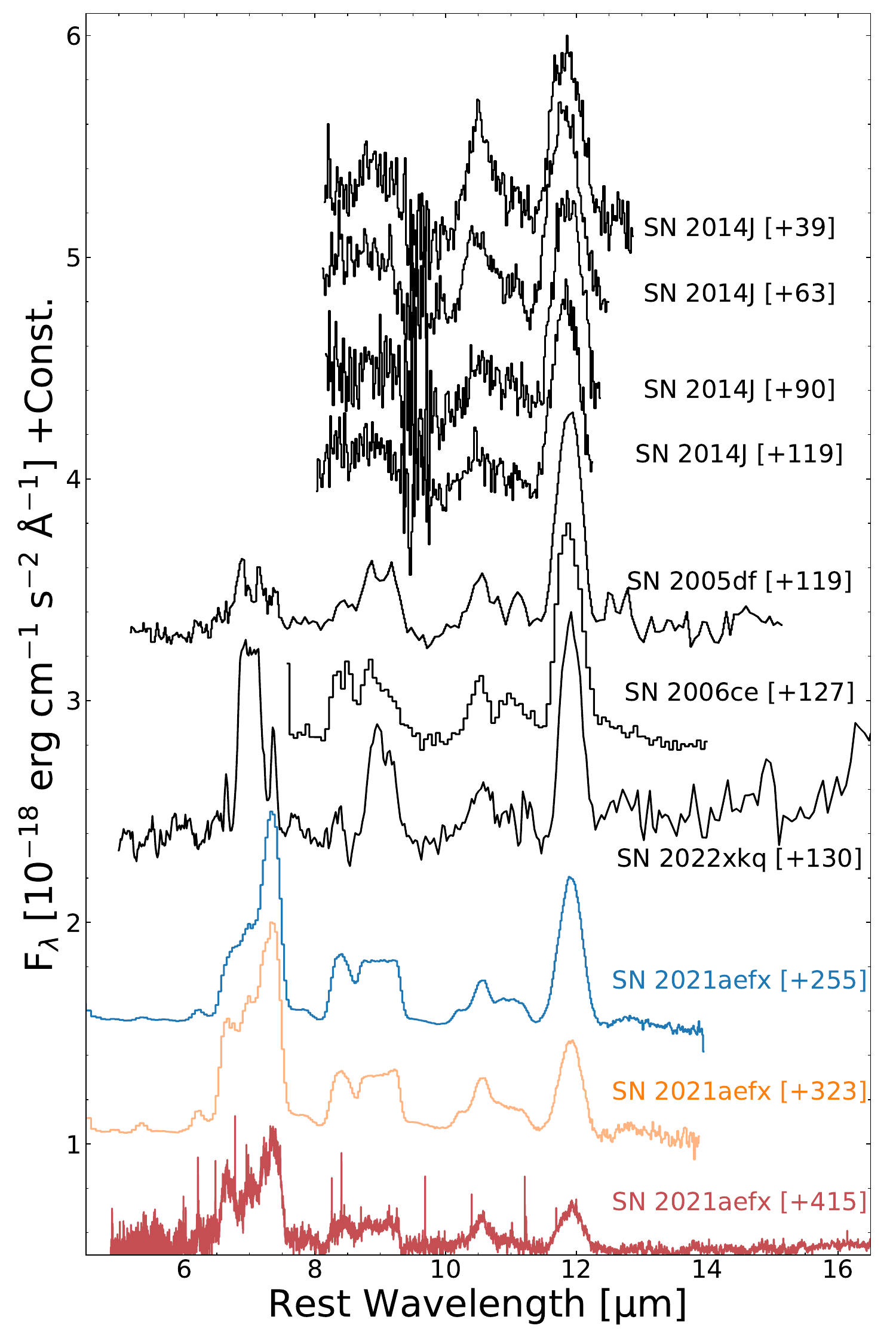}
    \caption{A selection of MIR SNe~Ia spectra. Times relative to $B$-band maximum are provided next to each observation. \label{fig:fullcomp} All SNe~Ia in this plot are normal SNe~Ia except for SN~2022xkq which is sub-luminous.  }
\end{figure}

Figure \ref{fig:fullcomp} shows a spectral comparison plot between SN~2021aefx and a sample of published MIR SNe~Ia observations. 
The previous sample covers the evolution of SNe~Ia from +39 to +323~d past $B$-band maximum. All previous MIRI/LRS spectral observations of SN~2021aefx were re-reduced using the method outlined in \citep{DerKacy_etal_2023_21aefx}, but using pipeline version 1.13.4 and \textsc{crds} version 1223.pmap. Overall the line profiles and strengths are unaffected by the new reduction, but this version of the pipeline and calibration files fixes previously known issues with the MIRI/LRS wavelength calibration.

\begin{figure}
    \centering
    \includegraphics[width=.45\textwidth]{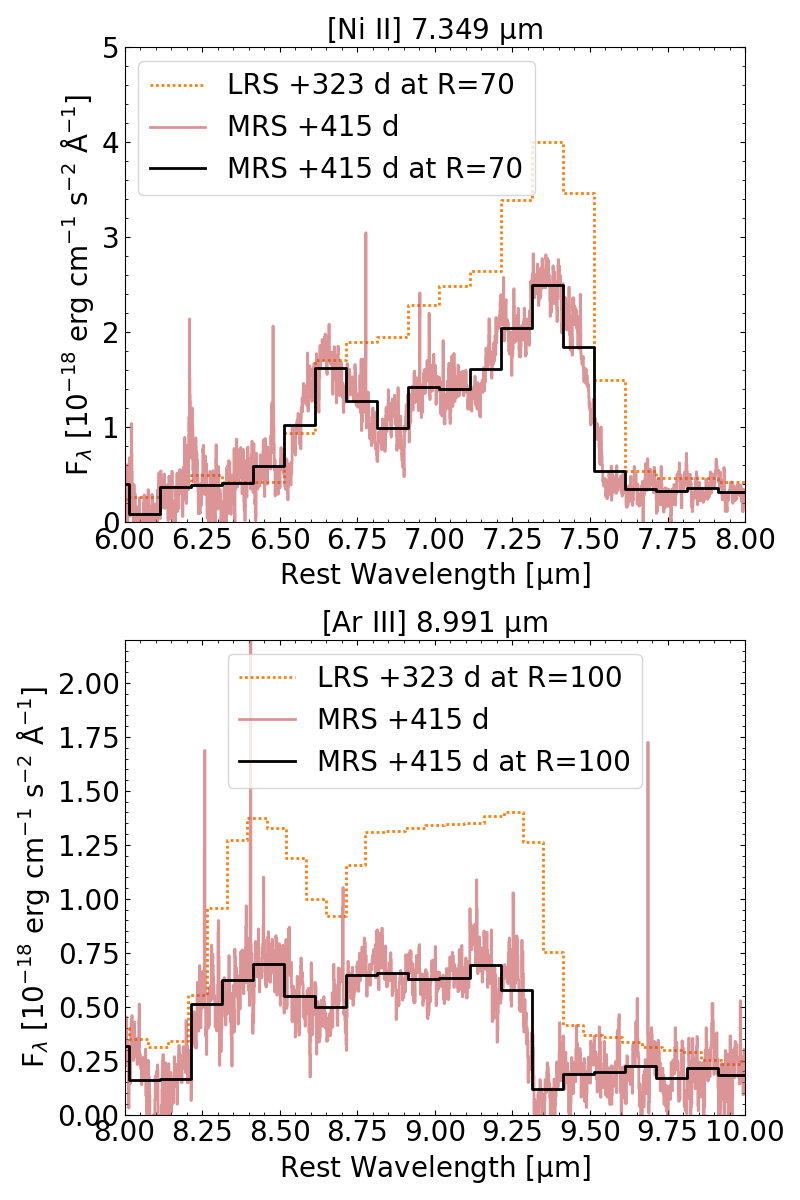}
    \caption{ The MIRI/MRS data has been resampled to the spectral resolution of the MIRI/LRS data. For the [\NiII] 7.349~$\micron$ profile (top panel) many of the isolated features cannot be resolved in the MIRI/LRS resolution, but are visible in the MIRI/MRS spectrum. 
    For the [\ArIII] 8.991~$\micron$ feature (bototm panel), the shape of the profile goes from  ``flat-tilted'' to ``flat-topped'' between +323~d and +415~d, demonstrating that the shape change is not caused by a resolution effect and is consistent with the physical interpretation provided in \autoref{sec:models}. We also note here that all MIRI/LRS data is heavily oversampled and should be resampled to the spectral resolution, as we have done in these panels, before direct comparison to the MIRI/MRS data.} 
    \label{fig:resolution}
\end{figure}

All spectra share similarities and consist of broad emission features that are composed of forbidden line transitions from Ar, Co, Fe, and Ni. 
The strongest of these is the  [\CoIII] 11.888$~\micron$ resonance feature. 
However, the capabilities and power of \textit{JWST} become apparent with the first two spectra of SN~2021aefx.
These spectra were observed using MIRI/LRS (R$\sim$100) and  revealed many spectral features that had not been observed before.
In particular,  a ``flat-tilted'' [\ArIII] 8.991$~\micron$ profile and 
multiple ionization states of Ni were observed \citep{Kwok2022,DerKacy_etal_2023_21aefx}.
The MIRI/MRS spectrum presented in this work is dominated by similar spectral features as earlier epochs, although with the higher resolution (R$\sim$2700) we can more accurately determine the exact region of the ejecta in which these ions are located as well as examine the data for resolved features. Overall, the MIRI/MRS data allow for detailed line IDs and analysis of ejecta structure at a level not possible with MIRI/LRS data, see Fig \ref{fig:resolution}. 

To date, the spectrum of SN~2022xkq at 130~d past maximum light is the only published  MIRI/MRS spectrum of a SN~Ia \citep{DerKacy23xkq}.  The top panel of Fig. \ref{fig:aefxcomp} shows a comparison between the low-luminosity SN~2022xkq and the normal-luminosity SN~2021aefx. 
Although they are taken 285~d apart, many of the same ions form the spectrum, but with different strengths and ratios; reflecting the variations in ionization balances due to phase and sub-type differences. In  SN~2021aefx the stable Ni features are stronger than those in SN~2022xkq relative to both the Ar and Co features.
There is also a significant contribution from Fe in the spectral formation of SN~2021aefx, which is prevalent in these epochs due to the decay of \Cofs.
The spectra of SN~2021aefx and SN~2022xkq also vary in the ionization state of Ar where for the low-luminosity  SN~2022xkq [\ArII] is the dominant ionization state  but for the normal-luminosity SN~2021aefx [\ArIII] is the dominant ionization state of Ar. For SN~2021aefx the Ar features have significantly more line blending from Fe-group elements than in SN~2022xkq.

\section{Temporal Evolution of SN~2021\lowercase{aefx}} \label{sec:specevolution}

The time evolution of the \textit{JWST} MIR spectra of SN~2021aefx from +215 to +415~d relative to rest-frame $B$ band maximum light in absolute flux is presented in the bottom panel of Fig. \ref{fig:aefxcomp}. As the data of SN~2021aefx are the first MIR time series of an SN from \textit{JWST}, they allows us
to follow the evolution of features as a function of time. We start by analyzing the [\CoIII] 11.888~$\micron$ feature, as it is the most prominent feature in the MIR spectra, and then proceed to the  evolution of other dominant lines.

\subsection{[Co III] 11.888~$\micron$} \label{sec:Cofsdecay}
\begin{figure}
    \centering
    \includegraphics[width=0.45\textwidth]{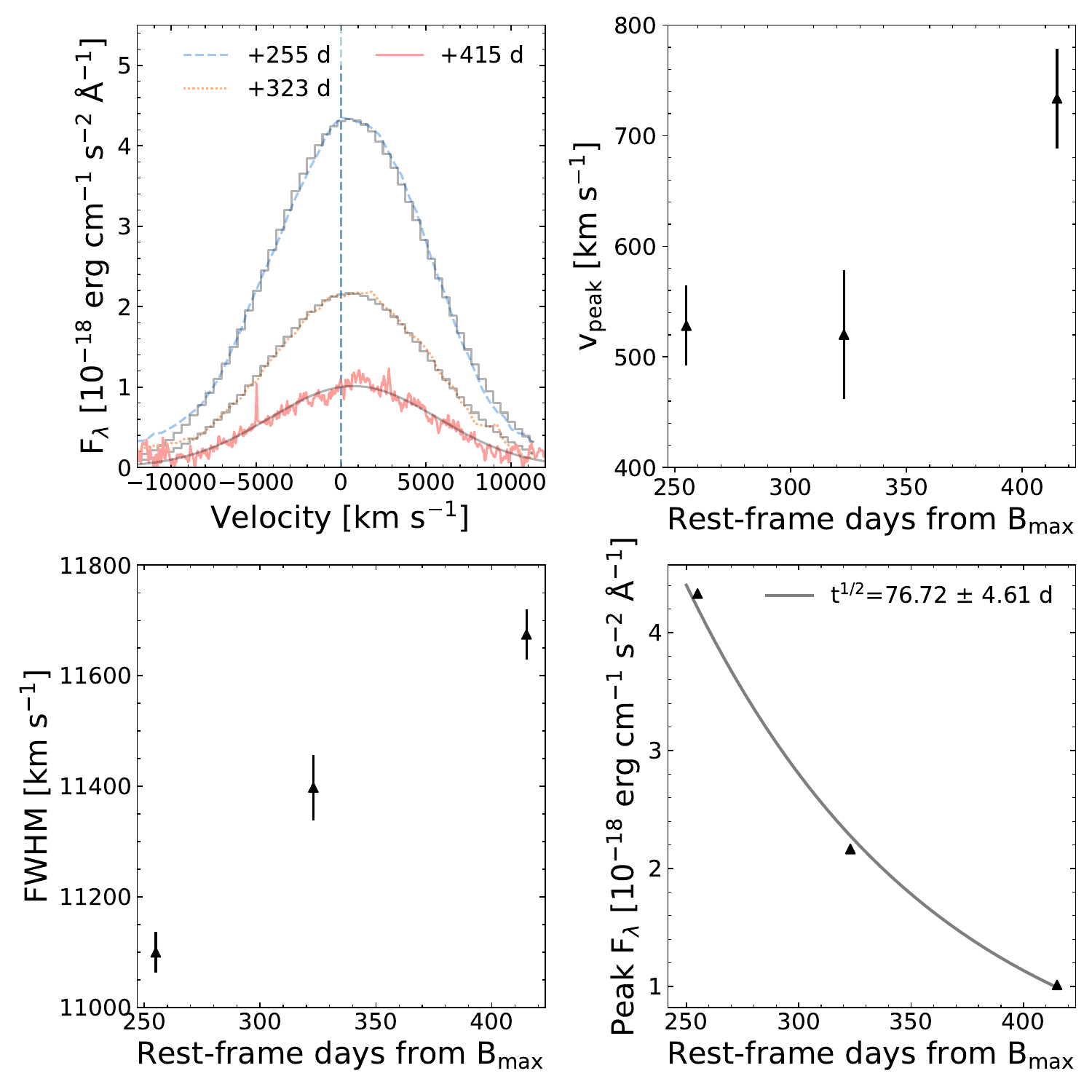}
    \caption{(\textit{Top Left:}) Evolution of the  [\CoIII] 11.888~$\micron$ feature as a function of time, for simplicity it is fit with a single Gaussian profile. (\textit{Top Right:}) The velocity evolution of the peak of the emission profile.  (\textit{Bottom Left:}) The evolution of the feature's FWHM. (\textit{Bottom Right:}) The peak flux of the feature  as a function of time, along with a line of best-fit with a decaying exponential function.    For all panels, if the error bars are not visible they are smaller than the markers. Furthermore, we have only plotted the fit error and not resolution error.}
    \label{fig:Covstime.pdf}
\end{figure}

The  [Co III] 11.888~$\micron$ feature gets weaker as a function of time 
(see the top left panel of Fig. \ref{fig:Covstime.pdf}).
We quantitatively examine the evolution of this feature. 
For simplicity, we assume the feature can be modeled by a single Gaussian function, although there is likely some  blending with the [\NiI] 12.001~$\micron$ line. 

The peak velocity (v$_{\mathrm{peak}}$) of the [\CoIII] feature is  $\sim$500~\kms at +255~d and +323~d, and increases to $\sim$750~\kms by +415~d. The values in $v_{\mathrm{peak}}$ are consistent between the MIRI/LRS and MIRI/MRS data, once the error due to the low spectral resolution of the LRS data is considered.

The measured FWHM of the [\CoIII] 11.888~$\micron$ feature increases from $\sim$11,100~\kms\ at +255~d to $\sim$11,700~\kms\ at +415~d. 
The cause of this is likely driven by the strength of the [\CoIII] 11.888~$\micron$ line decreasing as a function of time and the [\NiI] 12.001~$\micron$ line getting stronger. 

The peak flux of the [\CoIII] feature decreases from $\sim4.2 \times 10^{-18}$\,erg\,cm$^{-1}$\,s$^{-2}$\,\AA$^{-1}$ at +255~d to 
$\sim1 \times 10^{-18}$\,erg\,cm$^{-1}$\,s$^{-2}$\,\AA$^{-1}$ at +415~d. 
The evolution of the feature is well described by an exponential decay, with a half-life of 76.72$\pm$4.61~d. This is consistent with the half-life of \Cofs\ which is 77.27~d, demonstrating that the time evolution of this feature can be used to directly trace the energy deposition of the \Cofs\ in the ejecta. 
This trend is not found in the  optical [\CoIII] feature at 6200~\AA\ which has been found to decrease faster than the \Cofs\ cooling rate  \citep{2013ApJ...767..119M}. This highlights the uniqueness of the  [\CoIII] 11.888~$\micron$ feature. It is a resonance line and all of the recombination passes though it. 
It also demonstrates that the [\CoIII] line dominates this feature throughout, as well as confirming that the flux calibration of spectra is consistent between the MIRI/LRS and MIRI/MRS data.
Continued observations of this feature in Cycle 2 and 3 (\citealt{2023jwst.prop.3726D,2024jwst.prop.6582D}) up to +1150 days past maximum light will probe the region where radioactive $^{57}$Co may dominate the heating.  If there is little mixing in the inner part of the ejecta this $^{57}$Co will be located in the central $\sim$1,000~\kms \citep{Hoeflich_2021_20qxp}. 
This may cause the [\CoIII] feature to become stronger and narrower over time.  Following the evolution of this feature with future spectral observations will allow for the distribution and strength of the $^{57}$Co to be characterized.

\subsection{The evolution of other important features}
\begin{figure}
    \centering
    \includegraphics[width=.45\textwidth]{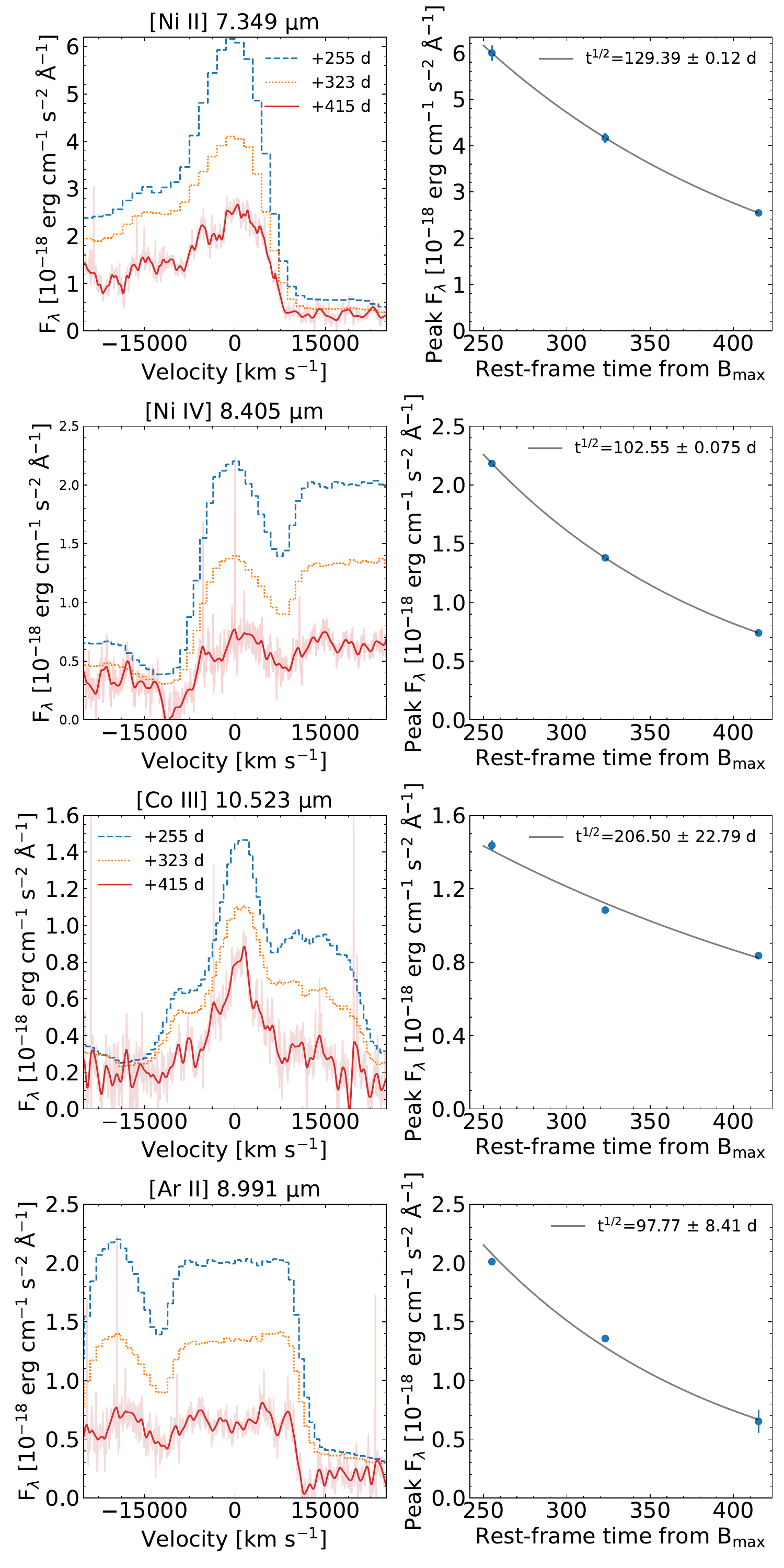}
    \caption{(\textit{Left:}) The velocity profile evolution of four important wavelength ranges within MIR spectra of SN~2021aefx.  For most regions the MIRI/MRS data can resolve structure within features which appear blended in the MIRI/LRS mode, see also Fig. \ref{fig:resolution}. 
    (\textit{Right:}) The peak flux of the profiles as a function of time fit with exponential decay curves.  }
    \label{fig:aefxvelevolutionprof}
\end{figure}

We now turn our attention to the temporal evolution of four other features. These are the [\NiII] 7.349~$\micron$, the [\NiIV] 8.405~$\micron$, the [\CoIII] 10.523~$\micron$, and the [\ArII] 8.991~$\micron$ regions. 
Due to the more complex structures and blends in these features, we choose to concentrate on the evolution of only the peak flux. Figure \ref{fig:aefxvelevolutionprof} shows the evolution of these features in velocity space, as well as their peak flux as a function of time.  

In the earlier phases, the [\NiII]~7.349~$\micron$ feature appears to be dominated by one transition, but by day +323 a blue wing appears in the emission profile. 
In the MIRI/LRS data, this blue wing is blended with the main [\NiII]~7.349~$\micron$ feature, however, it is resolved in the MIRI/MRS data at +415~d, and shows two distinct components with many smaller overlapping emission features in the bluest wavelengths. We identify this blue feature as [\CoI]~7.202$\micron$.
It is clear through this comparison that MIRI/MRS data is required to fully resolve profiles, but even with the higher resolution, the emission profiles are a complex blend of many lines, some of which are unknown. The peak flux of this wavelength region as a function of time does not follow the radioactive decay of \Cofs. 

The [\NiIV]~8.405~~$\micron$ feature appears to be dominated by one spectral ion throughout the MIRI/LRS time series of data. However, the MIRI/MRS data reveal that there are at least two resolved profiles contributing to this feature, one closer to the rest wavelength of  [\NiIV]~8.405~$\micron$ and one at $\sim$9000~\kms. We identify the blue feature to be a blend of [\CoI]~8.283~$\micron$ and [\FeII]~8.299~$\micron$.
The peak flux of this feature does decrease as a function of time, but unlike the [\CoIII]~11.888~$\micron$ feature, it does not follow the half-life of \Cofs.

The [\CoIII]~10.523~$\micron$ feature is the second most prominent [\CoIII] region in the MIR spectrum, the shape of this profile changes dramatically between the MIRI/LRS and MIRI/MRS data. It is much narrower and peaked within the MIRI/MRS data. The evolution of the peak flux of this feature follows a half-life of 206.50$\pm$22.79~d, which is more than twice that of what would be expected of \Cofs\ decay. This demonstrates the uniqueness of the [\CoIII]~11.888~$\micron$ feature, as it is both isolated and a resonance line, so all recombination passes through the transition. It also highlights that an apparently isolated feature, such as the  [\CoIII]~10.523~$\micron$, is still heavily blended, and not dominated by one single transition, meaning any velocity extracted from such features will be highly uncertain.

The [\ArII]~8.991~$\micron$ feature is unique as it shows a flat top profile. This flat-top profile (at earlier epochs this was referred to as a ``flat-tilted'' profile)  is seen throughout all of the spectra in the time series, and has been interpreted as being caused by a shell of Ar in the ejecta \citep{Kwok2022}, which may come from an off-center deflagration to detonation transition \citep{DerKacy_etal_2023_21aefx}. 
The blue side of this feature is clearly contaminated by emission from the  [\NiIV]~8.405~$\micron$ transition. We also see that the peak of this feature does not follow the half life of \Cofs.  Overall, we stress that all of the line profiles shown in this section are highly blended, and only with improved atomic data can the true components that contribute to the  MIRI/MRS spectrum of SN~2021aefx be extracted.

\subsection{[\NiII]~/~[\CoIII] ratio as a function of time}

\begin{figure}
    \centering
    \includegraphics[width=.49\textwidth]{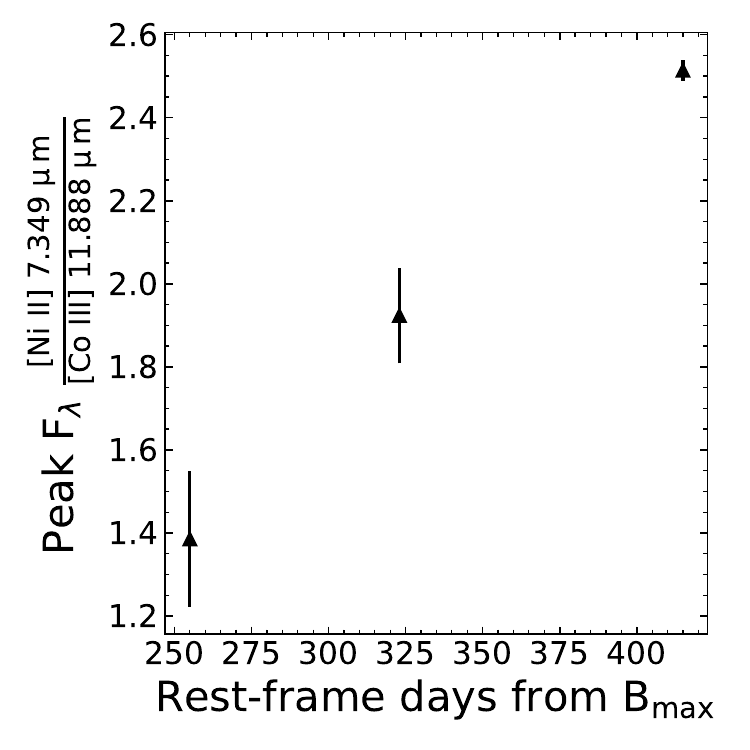}
    \caption{The ratio of the peak flux of the [\NiII]~7.349~$\micron$ to [\CoIII]~11.888~$\micron$ features as a function of time demonstrates that emission from stable Ni increasingly dominates the spectral formation. We chose to examine the peak flux here as it follows the bulk of the emission, and the features are too blended to obtain accurate values of HWHM. }
    \label{fig:NitoCo}
\end{figure}

Examining the peak flux ratio  [\NiII]~7.349~$\micron$ / [\CoIII]~11.888~$\micron$ as a function of time is another useful measurable.
It allows us to determine how the temporal variation in the energy deposition, heating, and mixing  between the  \Nifs\ region and the  electron capture element region evolves.  As can be seen in Fig. \ref{fig:NitoCo}, this ratio gradually increases throughout the time series. At later epochs if the decay of $^{57}$Co begins to dominate it could be expected that this trend changes direction. However, this may heavily depend on the location of the $^{57}$Co with relation to the stable Ni in the ejecta.

\subsection{Half Width at Half Maximum vs v$_{\mathrm{peak}}$}

There are many overlapping spectral lines that contribute to the formation of the nebular phase spectrum of SN~2021aefx. MIRI/MRS observations allow for more of these lines to be resolved. However, many of the atomic line transitions and strengths are not known.  Regardless of this, we attempt to fit the emission profiles of 
spectral features in order to get a broad understanding of where in the ejecta the emitting regions of certain ions are located. 
The full analysis is given in Appendix \ref{sec:vel}. We stress  that the lack of full line IDs and the assumption that all emitting line profiles can be explained by simple functions (where we inherently assume they are symmetrical), and the priors given in the fitting procedure make the results highly uncertain. 

\begin{figure}
    \centering
    \includegraphics[width=.49\textwidth]{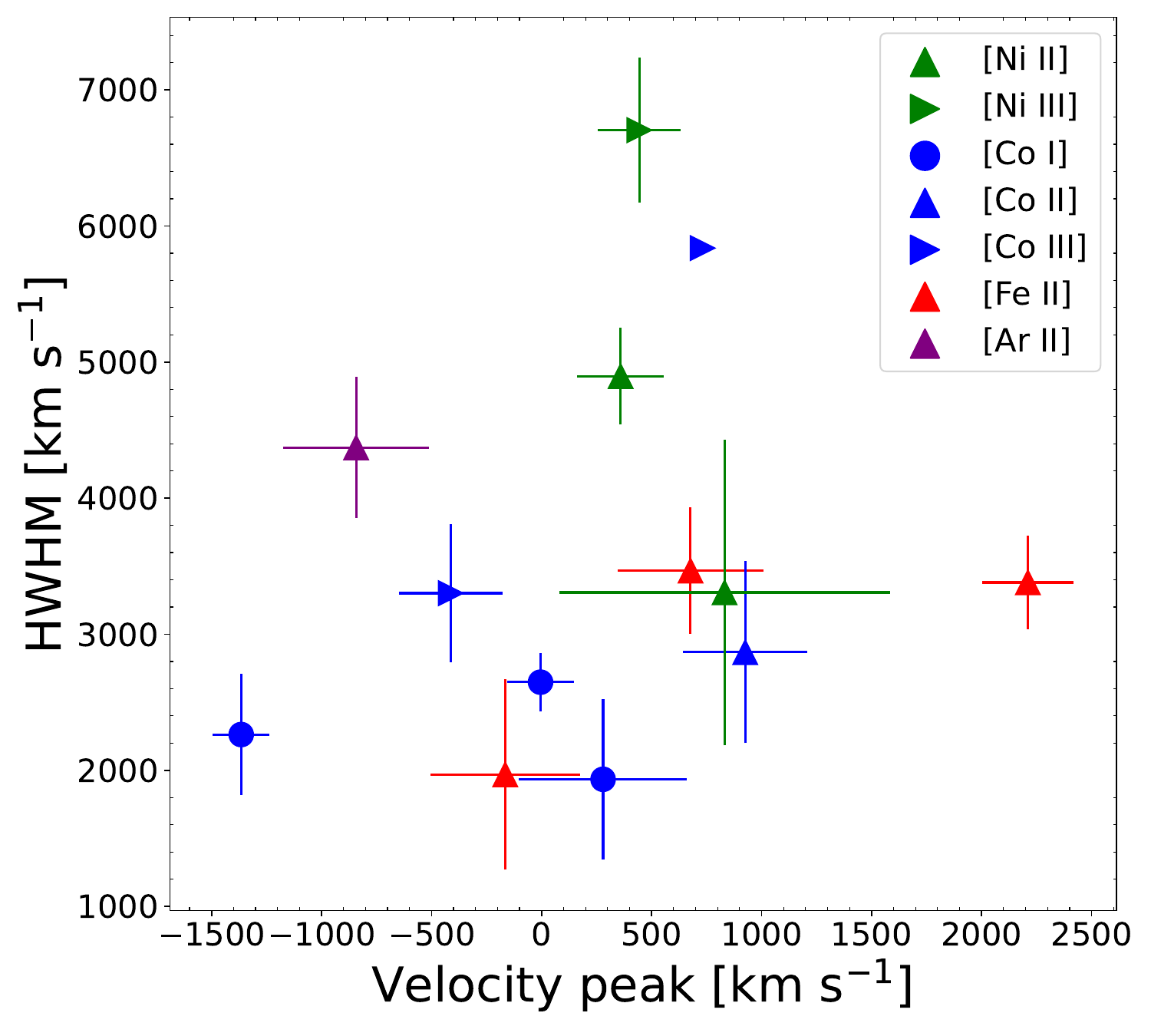}
    \caption{ The peak velocity vs HWHM of selected spectral features. As discussed in the text, these values are highly uncertain, and the detailed spectral modeling presented in \autoref{sec:models} is needed to extract further information about the explosion.  }
    \label{fig:peakvsFWHM}
\end{figure}

Figure \ref{fig:peakvsFWHM} shows the Half Width at Half Maximum (HWHM) vs the peak velocity of ions that have been fit in Appendix \ref{sec:vel}. Overall, we see no general trend with any of the ions. This is likely to be caused by the number of Gaussians required to reproduce the observed spectra, along with unknown spectral lines, and their corresponding strengths. Therefore, we emphasize that fitting spectra with multiple Gaussians has very little physical meaning, except in the case of isolated strong features such as the [\CoIII]~11.888~$\micron$ line. 
Hence, we turn our attention to the models below, to further understand the physics and formation of the spectra. We also emphasize that both the high spectral resolution and the extended wavelength coverage of the MRS mode on MIRI are critical for the physical interpretation described below.

\section{Models} \label{sec:models}
In the following section, we compare the data to models. 
The simulations utilized in this work are computed
using the HYDrodynamical RAdiation code (HYDRA; e.g. \citealt{2003ASPC..288..371H})
that consists of physics-based modules which provide
solutions for: the rate equations that determine the nuclear reactions, the statistical equations needed to determine the atomic level populations, the equation-of-state, the matter opacities, the hydrodynamic evolution,
and the radiation-transport equations (RTE). Here, the RTE
is treated using Variable Eddington Tensor
methods, with a Monte-Carlo (MC) scheme providing
the necessary closure relation to the momentum equations needed to solve the generalized scattering and non-LTE problem \citep{2002astro.ph..7103H,Hoeflich2003,Penney_etal_2014,Diamond_etal_2015,Hoeflich_2017,Hristov_etal_2021}. The relevant physics and current limitations of the  simulations in the nebular phase are discussed in detail in 
 \citet{Hoeflich_2021_20qxp}. The growth of the WD mass towards  $M_{Ch}$ has been simulated following the approach by \citet{1979wdvd.coll..280S} and \citet{1982ApJ...253..798N}. The accretion rate and composition of the accreting material during the final stages has been tuned so that the ignition is triggered at a central density $\rho_c$ \citep{2002astro.ph..7103H,Diamond_etal_2015}. To trigger  a thermonuclear runaway at a $\rho_c$ below $\approx 0.8-1.0  \times 10^9 $ \gcm requires accretion rates in excess of $2 \times 10^{-6} ~\mathrm{M}_\odot/\mathrm{yr}$. To avoid over-Eddington luminosity of the accreting WD,
 He or C-rich matter is required, at least during the final stages.
 This limit depends on the details of the physical processes which occur during the accretion, such as the Urca-cooling by neutrinos (see \citealt{1970mwla.book.....G,1995SSRv...74..427H} and, Sect. 5.7 of \citealt{Diamond_etal_2015}). 
For a review on a wide variety of progenitor evolutions see  \citet{2017hsn..book.....A} and references therein.

In this work, the explosion model is based on simulations of off-center
delayed-detonation models following \citet{1999ApJ...527L..97L}. This class of models has been successfully used to reproduce the photospheric and nebular phase flux and polarization spectra of SNe~Ia. It has also been used to reproduce the abundance distributions in SN-remnants and to study the role of SNe~Ia as producers of the positrons observed in our Galaxy \citep{2006NewAR..50..470H,2015ApJ...804..140F,Telesco_etal_2015,Hristov_etal_2021,Hoeflich_2021_20qxp,2022ApJ...930..107M,Penney_etal_2014}.

Here, the simulations are used with parameters that have been shown to reproduce the evolution of the polarization and flux spectra of the normal-luminosity  SN~2019np \citep{Hoeflich_etal_2023_19np}. 
These model parameters have been successfully applied 
to SN~2021aefx  at +323~d if seen from $\Theta \approx -30^o$ \citep{DerKacy_etal_2023_21aefx} where $\Theta$ is the angle between the equatorial plane defined by the orthogonal vector between the kinematic center and the location of the deflagration-to-detonation transition. 
Note that a variety of  $\Theta$ values have been found for different SNe using both line profiles and spectropolarimetry, e.g. SN~2003hv \citep{2006NewAR..50..470H,2006ApJ...652L.101M}, SN~2003du \citep{Hoeflich_etal_2004},  SN~2012ke \citep{2012A&A...545A...7P}, SN~2019np \citep{Hoeflich_etal_2023_19np}, and SN~2020qxp \citep{Hoeflich_2021_20qxp}. For nebular profiles of different SNe, the distribution of $\Theta $ is
consistent with a random orientation relative to the observer. Line-polarization during the photospheric phase favors angles close to the northern pole (\ie ~ $\Theta $ larger than $30 ^o$ \citealt{Hoeflich_etal_2023_19np}). This can be understood by selective
line absorption by a large-scale abundance asymmetry, 
as produced in an off-center DDT. Both effects combined support
the notion of a loop-sided asymmetry. 
For objects with multi-epoch observations, consistent values of $\Theta$ have been found, suggesting a large-scale abundance asymmetry. We note that  the point of ignition and the deflagration-to-denotation transition are distinct and are related to very different physical processes. 

In this work, the basic model parameters (the initial magnetic field of the WD, B; and the central density $\rho_c$) have been extended to cover a larger range in values, to take advantage of the increase in spectral resolution of the MIRI/MRS data. 

\begin{figure*}[ht]
  \includegraphics[scale=0.7]{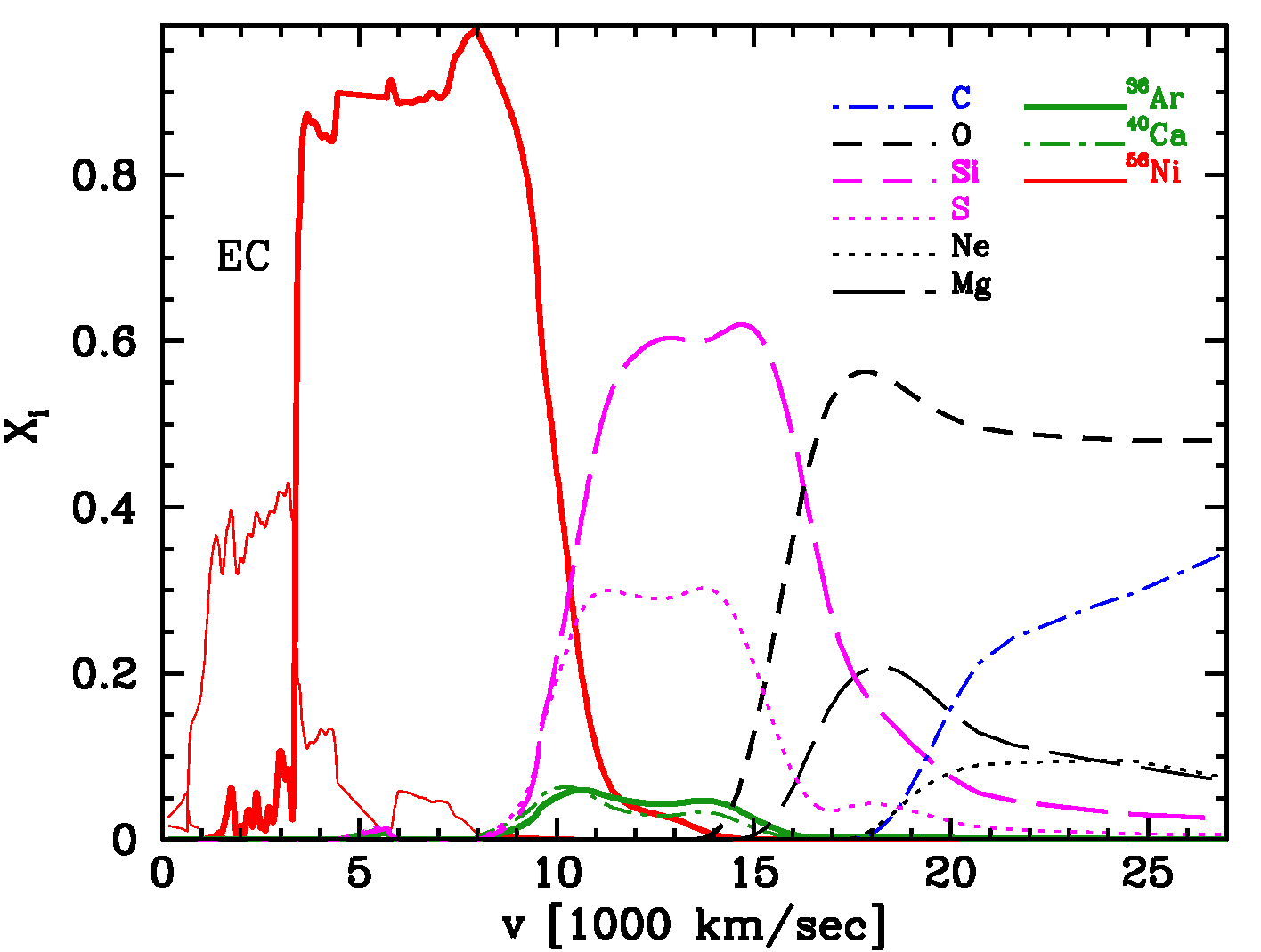}
  \includegraphics[scale=0.7]{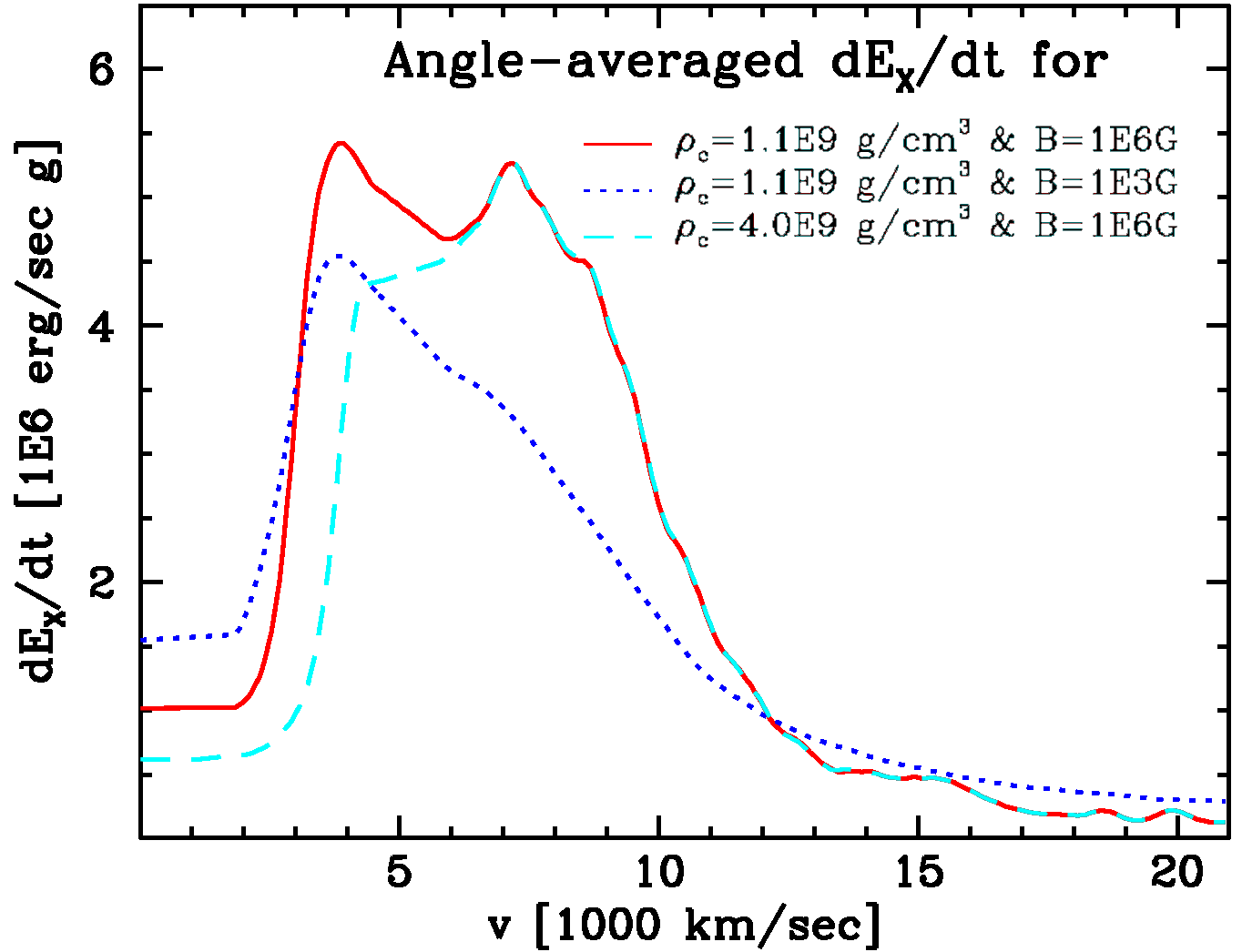}
  \caption
  {
  ({\it Left}:) The angle-averaged composition of our best fit model, Model~25
    from \citet{Hoeflich_etal_2017} and \citet{Hoeflich_etal_2023_19np}.
    The abundances are stratified, with asymmetries limited
    to the outer edge of the  quasi-statistical equilibrium and nuclear statistical equilibrium layers in velocity. EC marks the region of electron capture elements with the thin, red line being $^{58}$Ni. 
    ({\it Right:}) Angle-averaged energy deposition by $\gamma$-rays and positrons, $E_X$, for our reference model with ($\rho_c, B)= (1.1 \times 10^9$ \gcm, $10^6$ G ; red). Also shown are explosion models with a large $\rho_c$, $(4 \times 10^9$\gcm and $B = 10^6$ G ; cyan), and a lower WD initial magnetic field $(1.1 \times 10^9$ \gcm, $10^3$ G ; blue). Note that the shift of the sharp drop of $E_X$ corresponds to the inner edge of the \Nifs\ distribution as a function of $\rho_c$ \citep{Diamond_etal_2015}.
    Positron-transport effects become important for small B. (E.g., on top of  $E_\gamma$), $E_X$ in the central electron capture-region depends sensitively on the magnetic B-field. As a result, the emission from electron capture elements will change by a factor of 2.
    Though large $\rho_c$ increases the $^{58}${Ni} production by a factor of 2, the specific energy
    input is halved, leading to similar electron capture line strengths in the corresponding features.
    The degeneracy
    can be broken by line profiles with a resolution of better than 1000 \kms,  or a time series of data and models that cover the 
    $\gamma$-ray to the positron-dominated regime.
     The excitation in the Ar-region depends sensitively
    on the magnetic field of the WD, which has consequences for the spectra and a profound impact on the Ar line profiles (see Sect. 6.3). 
    }
  \label{fig:models}
\end{figure*}

The model parameters are given in Table \ref{tab:model_params}. The base model, Model 25, serves as a reference to a series of simulations with various $\rho_c$, see Fig. \ref{fig:models}.
To first order, the central density of the WD at the time of the runaway depends on the accretion rate,
and  WD magnetic field. Note that the accretion on the WD is governed by the configuration of the progenitor,
and the range of possible accretion depends on the composition of  the accreted material  \citep{1979wdvd.coll..280S,1982ApJ...253..798N}. For  comprehensive overviews, see the textbook by \citet{2017hsn..book.....A}.
The WD masses of models considered here are close to $M_{Ch}$.

\begin{deluxetable}{l|ccccc}
  \tablecaption{Model 25 Parameters ($^*$: base model 25 is used as reference).}
  \label{tab:model_params}
  \tablehead{\colhead{Parameter}  &\colhead{Mod.25a}& \colhead{Mod.25b}& \colhead{Mod.25}& \colhead{Mod.25c}& \colhead{Mod.25d}}
  \startdata
    M$_{\rm ej}$ [\Msun] & $1.30$& $1.33$ & $1.35^*$   & $1.35$& $1.38$ \\
    $\rho_\mathrm{c}$ [$10^9$\gcm] &  $0.5$ &  $0.9$ &  $1.1^*$ & $1.1$ &  $4.0$  \\
    M$_{\rm tr}$ [\Msun]    & $0.24$ & $0.24$ & $0.24 ^*$ & $0.24$ & $0.24$ \\
    M$_{\rm DDT}$ [\Msun]   &  $0.5$ & $0.5$ & $0.5^*$& $0.5$ & $0.5$ \\
    B(WD)  [$10^3$ G]  &  $1000$ & $1000$ & $1000^*$ &  $1$ & $1000$ \\
\enddata
\end{deluxetable}
\begin{figure}
    \centering
    \includegraphics[width=.43\textwidth]{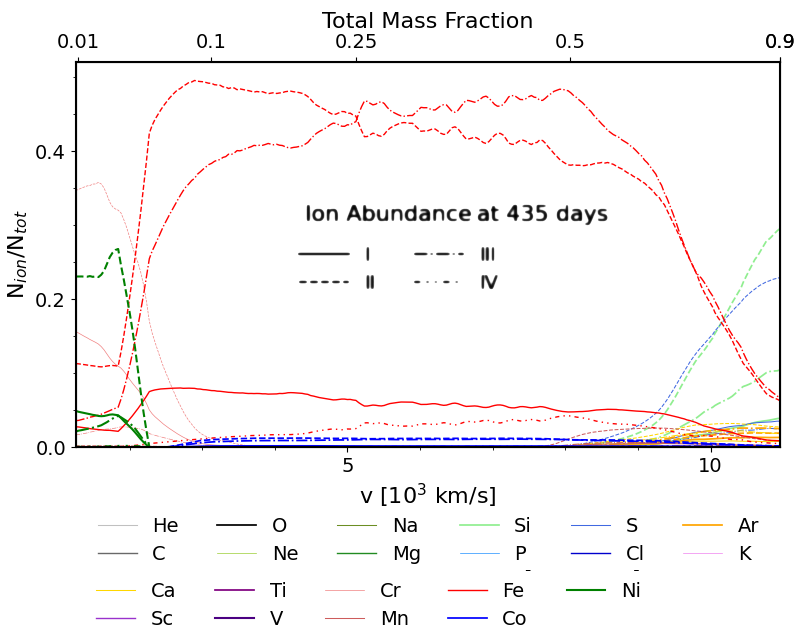} 
    \caption{Same as the left panel from \autoref{fig:models} but with the angle-averaged ionization levels of I-IV per particle zoomed in on the spectra-forming region. At day +415, most of the  \Cofs\ (blue) has decayed to    \Fefs\ with  II-III being the dominant ionization stages where a small amount of I/IV are present. Various isotopes of electron capture elements (Fe/Co/Ni) are found in stages  I-III and are seen at low velocities.}
    \label{fig:ion}
\end{figure}   

\subsection{Resolution of Simulations}
Medium-resolution spectra are crucial for accurately interpreting the physical characteristics of individual observed and synthetic features.
For our discussion below, where we compare observations with theory, the spectral resolution sets the limit to resolve small features and their profiles.
The spectral resolution of the data  varies with wavelength but $R \sim 2700$ corresponds to a velocity resolution of about $\sim$ 111 \kms. 
As in previous works ( e.g. \citealt{Hoeflich_etal_2023_19np}), the model domain is spatially resolved by a  equidistant Cartesian grid of 300 points per dimension which covers an expansion velocity of $\approx 26,000$~\kms. Taking into account the second order  discretization in spatial coordinates of the simulations, 
this translates into a spectral resolving power $R\approx 600$. This corresponds to a velocity of $\approx 500$~\kms for underlying physical features in the full-domain simulation. This resolution is employed for solving for the temperature structure and rate equations because optical depth effects in the UV are important 
for the ionization balance due to the incomplete Rosseland cycle \citep{Hoeflich_2021_20qxp}\footnote{Even at 1000+ days, the UV optical depth remains large in the iron-rich region.}.

 For the emitted synthetic spectra and line profiles, more than 99.9\% of  the flux in the optical to MIR originates within 15,000 \kms because the energy is produced from the radioactive decay of $^{56}$Ni/$^{56}$Co (see Fig. \ref{fig:models}). 
 Therefore, the computed domain has been reduced to boost the resolving power to $R\approx 1200$, which corresponds to a velocity of $\approx 300$ \kms. This allows for a direct comparison with the MIRI/MRS spectra without
artificial convolution of the observed data. The model has approximately the same resolving power as the Channel 4 data. 

The typical scale over which the density and abundances vary in our models  (see Fig. \ref{fig:models}), is $\approx 1,000 $ \kms. 
Physically, smaller scales
down to  100~\kms can be expected as a result of downwards cascading in Rayleigh-Taylor instabilities in the outer layers.
However, due to the large number of small plumes, a detection of density and chemical inhomogeneities  requires a S/N better than 300 in the polarization spectra \citep{Hoeflich_etal_2023_19np} or in the peak flux of strong emission lines in flux spectra.

\subsection{Energy Deposition and Ionization}

The angle-averaged ionization structures at day +415  are shown in \autoref{fig:ion}.
\Fefs ~dominates  the nuclear statistical equilibrium region through singly and doubly ionized species,
as can be expected for normal-luminosity SNe~Ia and under-luminous SNe~Ia at earlier times \citep{Wilk2018,Shingles2020,Hoeflich_2021_20qxp}. Because the density increases inwards, and the recombination rate depends on the square of the density, the ionization decreases towards  the center.
In the models, the ionization rate shifts towards more neutral ions compared to nebular calculation at $\approx$ 200 -- 300\,d 
\citep{Wilk2018,Shingles2020,DerKacy_etal_2023_21aefx}. At day +415, we see  $\sim$5 to 10\%  of neutral iron group elements. One of the reasons for the lower ionization is that temperatures decrease by $\sim$1000-1500 K at day +415 from $\sim 5000$ K  at day 200 -- 300 \citep{Kozma_Fransson_1992,2015ApJ...814L...2F}. The temperature structure of SN~2021aefx at +415~d resembles an under-luminous SNe~Ia at about 190 days \citep{Hoeflich_2021_20qxp}. 

The significant difference between the new spectrum of SN~2021aefx and the previous ones is that +415~d marks the transition from the energy input being dominated by hard $\gamma $-rays
to being dominated by positrons. This will be fully reached at $\sim$+500~d \citep{Penney_etal_2014}. Note that this time also corresponds to a shift in optical emission properties in SN~2011fe \citep{2022ApJ...926L..25T}. Hence the magnetic field of the WD is a critical variable in the nebular simulations \citep{Penney_etal_2014}.
We also note that in our models between 400–500 days $\gamma $-rays are just as important as positrons to the energy position in the most central regions and both have a significant effect on the spectral formation.

\begin{deluxetable*}{rcl|rcl|rcl|rcl|rcl}
\label{tab:lines_model}
  \tablecaption{Line contributions to the \textit{JWST} spectrum  at Day +415 from the reference model with $\rho_c=1.1\times 10^9 $ \gcm and $B=10^6$G. }\label{tab:ir_lines}
  \tabletypesize{\scriptsize}
  \tablehead{\colhead{\bf S} & \colhead{$\lambda$~[\mic]} & \colhead{Ion} 
    & \colhead{\bf S} & \colhead{$\lambda$~[\mic]} & \colhead{Ion}
    & \colhead{\bf S} & \colhead{$\lambda$~[\mic]} & \colhead{Ion}
    & \colhead{\bf S} & \colhead{$\lambda$~[\mic]} & \colhead{Ion}
    & \colhead{\bf S} & \colhead{$\lambda$~[\mic]} & \colhead{Ion}}
    \startdata
 	&	4.8603	&	 [\ion{Fe}{3}] 	&
 {$\ast\ \ast\ $} 	&	4.8891	&	 [\ion{Fe}{2}] 	&
 	&	5.0623	&	 [\ion{Co}{1}] 	&
 	&	5.1635	&	 [\ion{Co}{1}] 	& 
 	&	5.1796	&	 [\ion{Co}{2}] 	\\
 {$\ast\ \ast\ $} 	&	5.1865	&	 [\ion{Ni}{2}] 	&
 	&	5.2112	&	 [\ion{Co}{1}] 	&
{$\ast\ \ast\ $}  	&	5.3402	&	 [\ion{Fe}{2}] 	&
 	&	5.4394	&	 [\ion{Co}{2}] 	&
 	&	 5.4652$^{\dagger}$ 	&	 [\ion{V}{1}] 	\\
 {$\ast\ \ast\ $} 	&	5.6739	&	 [\ion{Fe}{2}] 	&
 {$\ast\ $} 	&	 5.6870$^{\dagger}$ 	&	 [\ion{V}{1}] 	&
 	&	5.7044	&	 [\ion{Co}{2}] 	&
 	&	5.7391	&	 [\ion{Fe}{2}] 	&
{$\ast\ \ast\ $} 	&	5.8933	&	 [\ion{Ni}{1}] 	\\
 	&	5.9395	&	 [\ion{Co}{2}] 	&
{$\ast\  $}  	&	5.9527	&	 [\ion{Ni}{2}] 	&
 {$\ast\ $} 	&	6.2135	&	 [\ion{Co}{2}] 	&
 	&	6.273	&	 [ \ion{Co}{1}] 	&
 	&	6.2738	&	 [\ion{Co}{2}] 	\\
 	&	6.3683	&	 [\ion{Ar}{3}] 	&
 	&	6.379	&	 [\ion{Fe}{2}] 	&
 {$\ast\ \ast\ \ast\ $} 	&	6.636	&	 [\ion{Ni}{2}] 	&
 	&	6.7213	&	 [\ion{Fe}{2}] 	&
 {$\ast\ \ast\ $} 	&	6.9196	&	 [\ion{Ni}{2}] 	\\
 {$\ast\ \ast\  $} 	&	6.9853	&	 [\ion{Ar}{2}] 	&
 	&	7.0454	&	 [\ion{Co}{1}] 	&
  	&	7.103	&	 [\ion{Co}{3}] 	&
 {$\ast\ $} 	&	7.1473	&	 [\ion{Fe}{3}] 	&
 	&	7.2019	&	 [\ion{Co}{1}] 	\\
 {$\ast\ $} 	&	7.3492	&	 [\ion{Ni}{3}] 	&
 {$\ast\ \ast\ $} 	&	7.5066	&	 [\ion{Ni}{1}] 	&
 {$\ast\ $} 	&	7.7906	&	 [\ion{Fe}{3}] 	&
 	&	8.044	&	 [\ion{Co}{1}] 	&
 {$\ast\ $} 	&	8.211	&	 [\ion{Fe}{3}] 	\\
 	&	8.2825	&	 [\ion{Co}{1}] 	&
 {$\ast\ $} 	&	8.2993	&	 [\ion{Fe}{2}] 	&
 {$\ast\ \ast\ \ast\ $} 	&	8.405	&	 [\ion{Ni}{4}] 	&
 {$\ast\ $} 	&	8.6107	&	 [\ion{Fe}{3}] 	&
 {$\ast\ $} 	&	8.6438	&	 [\ion{Co}{2}] 	\\
 {$\ast\ $} 	&	8.7325	&	 [\ion{Fe}{2}] 	&
 {$\ast\ $} 	&	 8.9147$^{\dagger}$ 	&	 [\ion{Ti}{2}] 	&
 {$\ast\ \ast\ $} 	&	8.945	&	 [\ion{Ni}{4}] 	&
 {$\ast\ \ast\ $} 	&	8.9914	&	 [\ion{Ar}{3}] 	&
 {$\ast\ $} 	&	 9.1969$^{\dagger}$ 	&	 [\ion{Ti}{2}] 	\\
 	&	 9.279$^{\dagger}$ 	&	 [\ion{Co}{4}] 	&
 {$\ast\ $} 	&	9.618	&	 [\ion{Ni}{2}] 	&
 	&	9.8195	&	 [\ion{Co}{1}] 	&
 {$\ast\ $} 	&	10.08	&	 [\ion{Ni}{2}] 	&
 {$\ast\ \ast\ $} 	&	 10.1637$^{\dagger}$ 	&	 [\ion{Ti}{2}] 	\\
 {$\ast\ \ast\ $} 	&	10.189	&	 [\ion{Fe}{2}] 	&
 	&	10.203	&	 [\ion{Fe}{3}] 	&
 {$\ast\ $} 	&	10.5105	&	 [\ion{Ti}{2}] 	&
 {$\ast\ $} 	&	10.5105	&	 [\ion{S}{4}] 	&
 {$\ast\ $} 	&	10.523	&	 [\ion{Co}{2}] 	\\
 {$\ast\ \ast\ $} 	&	10.682	&	 [\ion{Ni}{2}] 	&
 {$\ast\ \ast\ $} 	&	11.13	&	 [\ion{Ni}{4}] 	&
 	&	11.167	&	 [\ion{Co}{2}] 	&
 {$\ast\ $} 	&	 11.238$^{\dagger}$ 	&	 [\ion{Ti}{2}] 	&
 	&	11.307	&	 [\ion{Ni}{1}] 	\\
 {$\ast\ \ast\ \ast\ $} 	&	11.888	&	 [\ion{Co}{3}] 	&
 {$\ast\ \ast\ $} 	&	12.001	&	 [\ion{Ni}{1}] 	&
 {$\ast\ $} 	&	 12.1592$^{\dagger}$ 	&	 [\ion{Ti}{2}] 	&
 {$\ast\ \ast\ $} 	&	12.255	&	 [\ion{Co}{1}] 	&
 {$\ast\ $} 	&	12.261	&	 [\ion{Mn}{2}] 	\\
 {$\ast\ \ast\ $} 	&	 12.286$^{\dagger}$ 	&	 [\ion{Fe}{2}] 	&
 {$\ast\ \ast\ $} 	&	12.642	&	 [\ion{Fe}{2}] 	&
   	&	12.681	&	 [\ion{Co}{3}] 	&
 {$\ast\ \ast\ $} 	&	12.729	&	 [\ion{Ni}{2}] 	&
 	&	12.736	&	 [\ion{Ni}{4}] 	\\
  	&	12.811	&	 [\ion{Ne}{2}] 	&
 	&	13.058	&	 [\ion{Co}{1}] 	&
 	&	13.82	&	 [\ion{Co}{3}] 	&
 	&	 13.924$^{\dagger}$ 	&	 [\ion{Co}{4}] 	&
 	&	14.006	&	 [\ion{Co}{3}] 	\\
 	&	14.356	&	 [\ion{Co}{1}] 	&
 	&	14.391	&	 [\ion{Co}{1}] 	&
 {$\ast\  $} 	&	14.739	&	 [\ion{Co}{2}] 	&
 {$\ast\ $} 	&	14.814	&	 [\ion{Ni}{1}] 	&
 {$\ast\  $} 	&	14.977	&	 [\ion{Co}{2}] 	\\
 {$\ast\  $} 	&	15.459	&	 [\ion{Co}{2}] 	&
 	&	16.299	&	 [\ion{Co}{2}] 	&
 {$\ast\  $} 	&	16.391	&	 [\ion{Co}{3}] 	&
 {$\ast\ $} 	&	16.925	&	 [\ion{Co}{1}] 	&
 {$\ast\ \ast\ \ast\ $} 	&	17.936	&	 [\ion{Fe}{2}] 	\\
 {$\ast\ $} 	&	18.241	&	 [\ion{Ni}{2}] 	&
 	&	18.265	&	 [\ion{Co}{1}] 	&
 {$\ast\ $} 	&	18.39	&	 [\ion{Co}{2}] 	&
 {$\ast\ \ast\  $} 	&	18.713	&	 [\ion{S}{3}] 	&
 {$  $} 	&	18.804	&	 [\ion{Co}{2}] 	\\
 	&	18.985	&	 [\ion{Co}{2}] 	&
 {$\ast\ $} 	&	19.007	&	 [\ion{Fe}{2}] 	&
 {$\ast\ \ast\ $} 	&	19.056	&	 [\ion{Fe}{2}] 	&
 	&	19.138	&	 [\ion{Ni}{2}] 	&
 	&	19.232	&	 [\ion{Fe}{3}] 	\\
 	&	20.167	&	 [\ion{Fe}{3}] 	&
 	&	 20.928$^{\dagger}$ 	&	 [\ion{Fe}{2}] 	&
  	&	 21.17$^{\dagger}$ 	&	 [\ion{Fe}{1}] 	&
 	&	 21.986$^{\dagger}$ 	&	 [\ion{Fe}{2}] 	&
 	&	 22.106$^{\dagger}$ 	&	 [\ion{Ni}{1}] 	\\
 {$\ast\ $} 	&	21.481	&	 [\ion{Fe}{2}] 	&
 	&	20.167	&	 [\ion{Fe}{3}] 	&
 	&	 20.928$^{\dagger}$ 	&	 [\ion{Fe}{2}] 	&
  	&	 21.17$^{\dagger}$ 	&	 [\ion{Fe}{1}] 	&
 	&	 21.986$^{\dagger}$ 	&	 [\ion{Fe}{2}] 	\\
 	&	 22.106$^{\dagger}$ 	&	 [\ion{Ni}{1}] 	&
 {$\ast\ $} 	&	21.481	&	 [\ion{Fe}{2}] 	&
 {$\ast\ $} 	&	21.829	&	 [\ion{Ar}{3}] 	&
 {$\ast\ $} 	&	22.297	&	 [\ion{Fe}{1}] 	&
 	&	 22.80$^{\dagger}$ 	&	 [\ion{Co}{4}] 	\\
 {$\ast\ $} 	&	22.902	&	 [\ion{Fe}{2}] 	&
 {$\ast\ \ast\ \ast\ $} 	&	22.925	&	 [\ion{Fe}{3}] 	&
 	&	23.086	&	 [\ion{Ni}{2}] 	&
 {$\ast\ $} 	&	23.196	&	 [\ion{Co}{2}] 	&
 	&	23.389	&	 [\ion{Fe}{3}] 	\\
 	&	 24.04$^{\dagger}$ 	&	 [\ion{Co}{4}] 	&
 {$\ast\ \ast\ \ast $} 	&	24.042	&	 [\ion{Fe}{1}] 	&
 {$\ast\ \ast\ $} 	&	24.07	&	 [\ion{Co}{3}] 	&
 {$\ast\ \ast\ \ast\ $} 	&	24.519	&	 [\ion{Fe}{2}] 	&
 	&	24.847	&	 [\ion{Co}{1}] 	\\
 {$\ast\ \ast\  $} 	&	25.249	&	 [\ion{S}{1}] 	&
 {$\ast\ $} 	&	25.689	&	 [\ion{Co}{2}] 	&
 {$\ast\ \ast\ $} 	&	25.89	&	 [\ion{O}{4}] &
 	&	25.986	&	 [\ion{Co}{2}] 	&
 {$\ast\ \ast\ \ast\ $} 	&	25.988	&	 [\ion{Fe}{2}] 	\\
 	&	26.1	&	 [\ion{Co}{3}] 	&
 	&	26.13	&	 [\ion{Fe}{3}] 	&
 	&	26.601	&	 [\ion{Fe}{2}] 	&
 	&	27.53	&	 [\ion{Co}{2}] 	&
 	&	27.55	&	 [\ion{Co}{1}] 	\\
  {$\ast\  $}	&	28.466	&	 [\ion{Fe}{1}] 	&
 	&	29.675	&	 [\ion{Mn}{2}] 	&
 {$\ast\ \ast\ \ast\ $} 	&	33.038	&	 [\ion{Fe}{3}] 	&
 	&	33.481	&	 [\ion{S}{3}] 	&
 	&	34.66	&	 [\ion{Fe}{2}] 	\\
 {$\ast\ $} 	&	34.713	&	 [\ion{Fe}{1}] 	&
 {$\ast\ \ast\  $} 	&	34.815	&	 [\ion{Si}{2}] 	&
 {$\ast\ \ast\ \ast\ $} 	&	35.349	&	 [\ion{Fe}{2}] 	&
 {$\ast\ $} 	&	35.777	&	 [\ion{Fe}{2}] 	&
 {$\ast\ $} 	&	38.801	&	 [\ion{Fe}{1}] 	\\
 	&	39.272	&	 [\ion{Co}{2}] 	&
 {$\ast\ \ast\ \ast\ $} 	&	51.301	&	 [\ion{Fe}{2}] 	&
 {$\ast\ \ast\ \ast\ $} 	&	51.77	&	 [\ion{Fe}{3}] 	&
 {$\ast\  $} 	&	54.311	&	 [\ion{Fe}{1}] 	&
 {$\ast\ $} 	&	56.311	&	 [\ion{S}{1}] 	\\
 	&	60.128	&	 [\ion{Fe}{2}] 	&
    \enddata
    \tablecomments{The relative strengths are indicated by the number of $*$. For transitions without known 
    lifetimes (marked by $^{\dagger}$), $A_{i,j}$ are assumed from the equivalent iron levels.}
\end{deluxetable*}

\subsection{Flux Spectra and Profiles}\label{subsec:Flux_Profiles}

\begin{figure*}[ht]
  \includegraphics[width=\textwidth]{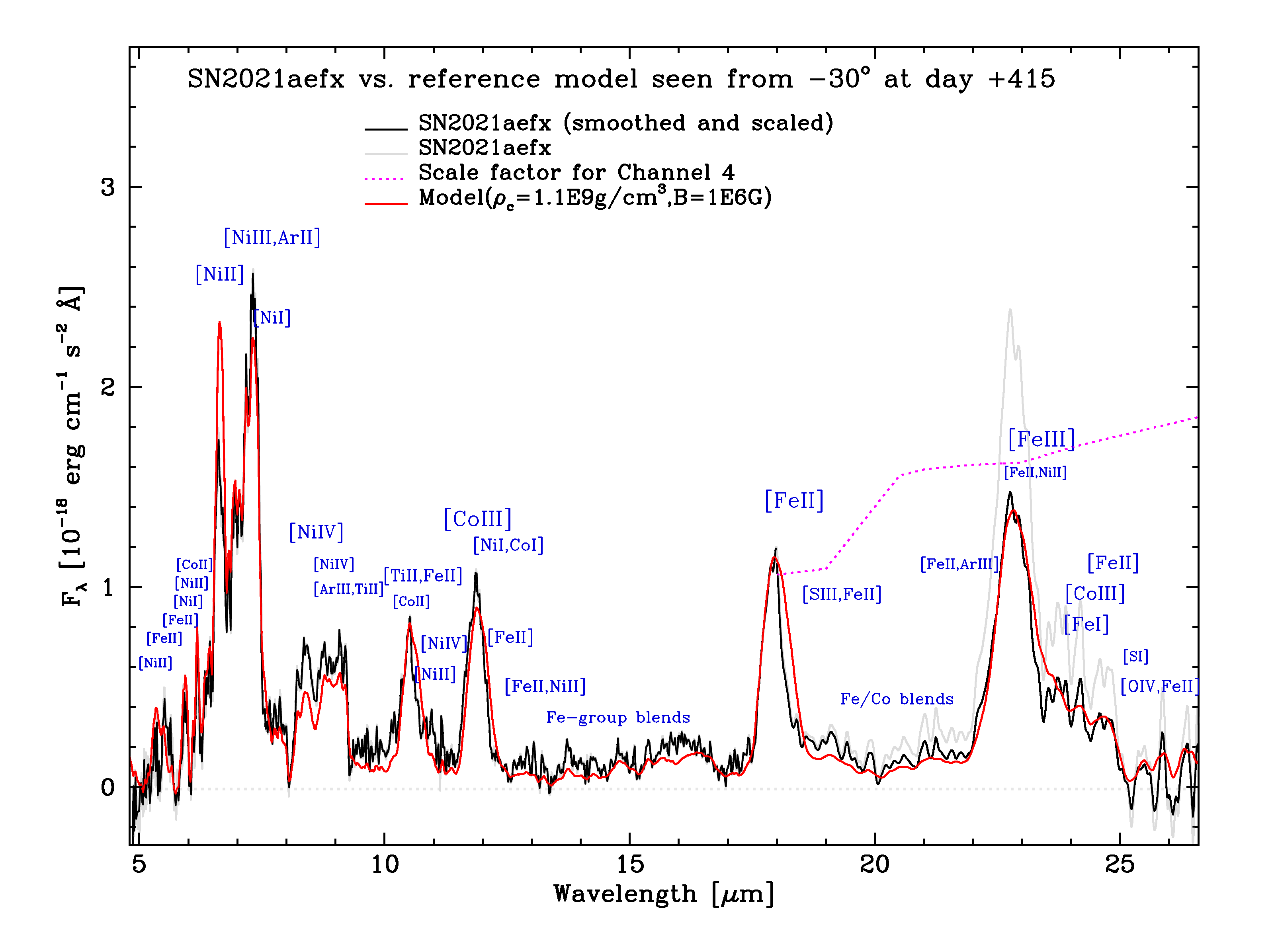}
  \caption{Comparison of the overall synthetic MIR spectra of our  off-center reference
    Model 25 seen from $-30^{\circ}$ 
    and the \textit{JWST} MIRI/MRS  smoothed (black) and raw (gray) spectra of SN~2021aefx at $+415$ days relative to $B$-band maximum. 
    The synthetic flux has been calibrated to the observed flux by adjusting the distance and using $M-m$ to $31.64$~mag.  In addition, the scaling factor (magenta) is shown as applied to Channel 4. This is appropriate because the background reduction in Channel 4 is highly uncertain. The size of the labels for the contributions corresponds roughly to the line strength.  Most of the MIR features show a complex structure and peaked morphology. This comes from a combination of blending and a contribution of electron capture elements. The observations agree reasonably well with the synthetic spectrum, though some shortcomings are apparent (see Sect. 6.4).} 
  \label{fig:specmod}
\end{figure*}
\begin{figure*}[ht]
  \includegraphics[width=\textwidth]{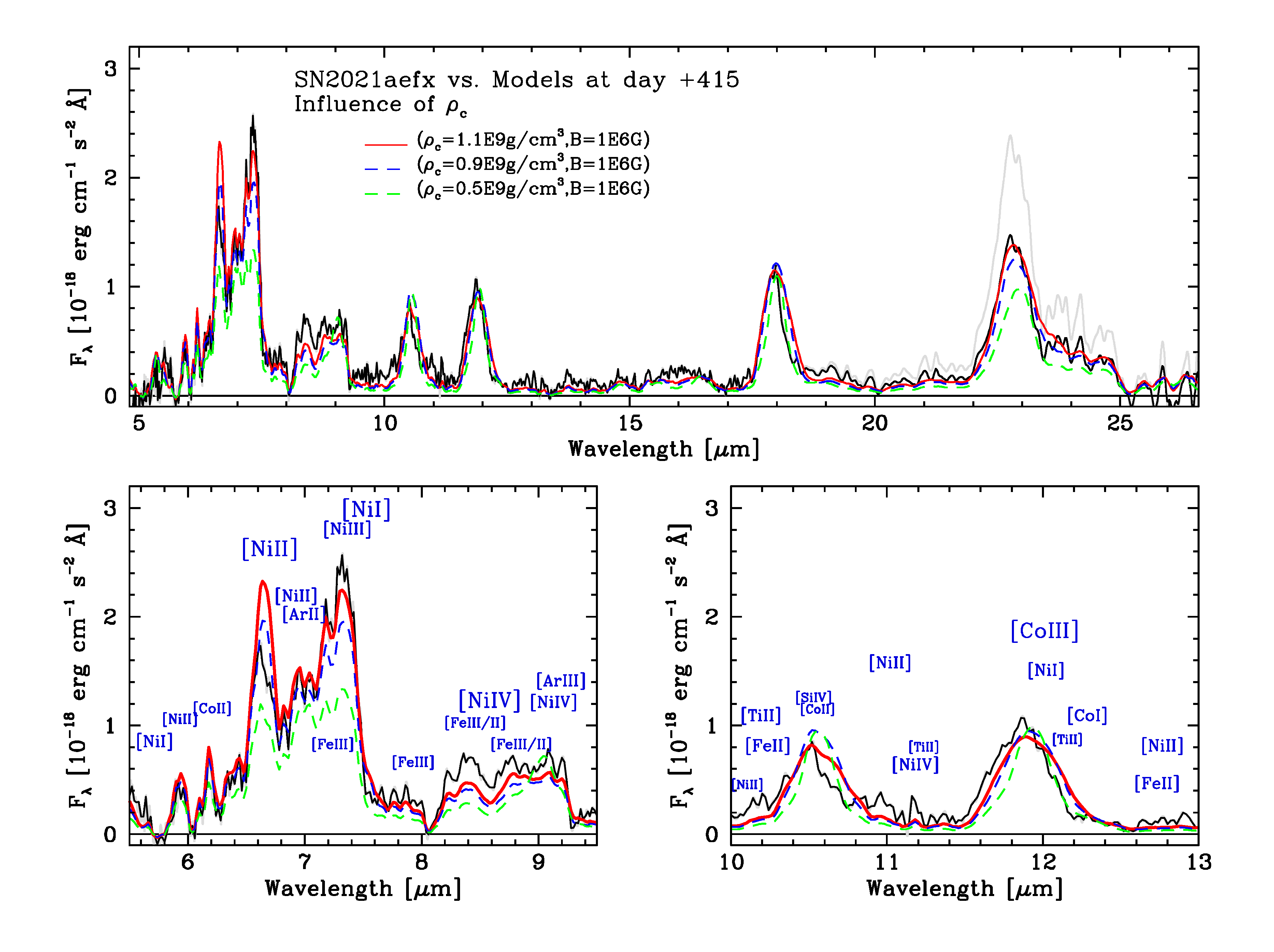}
  \caption{Same as Fig. \ref{fig:specmod} but the sensitivity of the  WD central density $\rho_c$ on the synthetic spectra is shown.
    The lower plots show the zoomed regions of 5.5-9.5 \mic (left) and 10-13 \mic (right), respectively. Note the sensitivity of $\rho_c$ on the overall 7 and 9 \mic features which, mostly, is a result of the decreasing $^{58}$Ni abundance and the change in the line ratio at e.g. 11.8 vs. 10.5 \mic.
    } 
  \label{fig:specmodRHO}
\end{figure*}
\begin{figure*}[ht]
  \includegraphics[width=\textwidth]{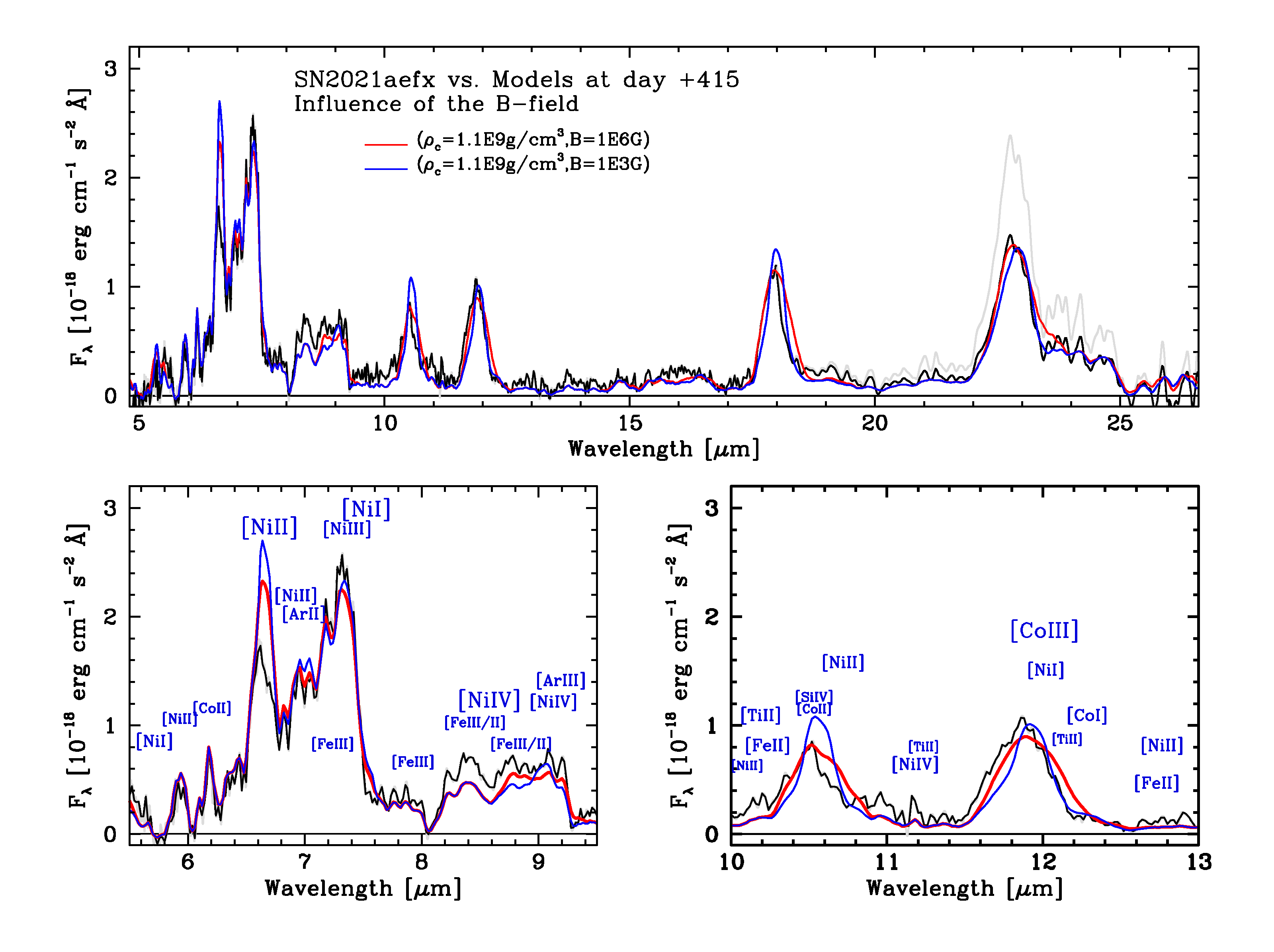}
  \caption{Same as Fig. \ref{fig:specmodRHO}, but the sensitivity of the initial magnetic
field of the WD, B,  on the synthetic spectra is shown.  The lines become narrower and more peaked with decreasing B. This is very
 similar to the low $\rho_c$ case. However, the effects can be separated by the overall spectral model. 
    } 
  \label{fig:specmodB}
\end{figure*}

\begin{figure*}[ht]
  \includegraphics[width=\textwidth]{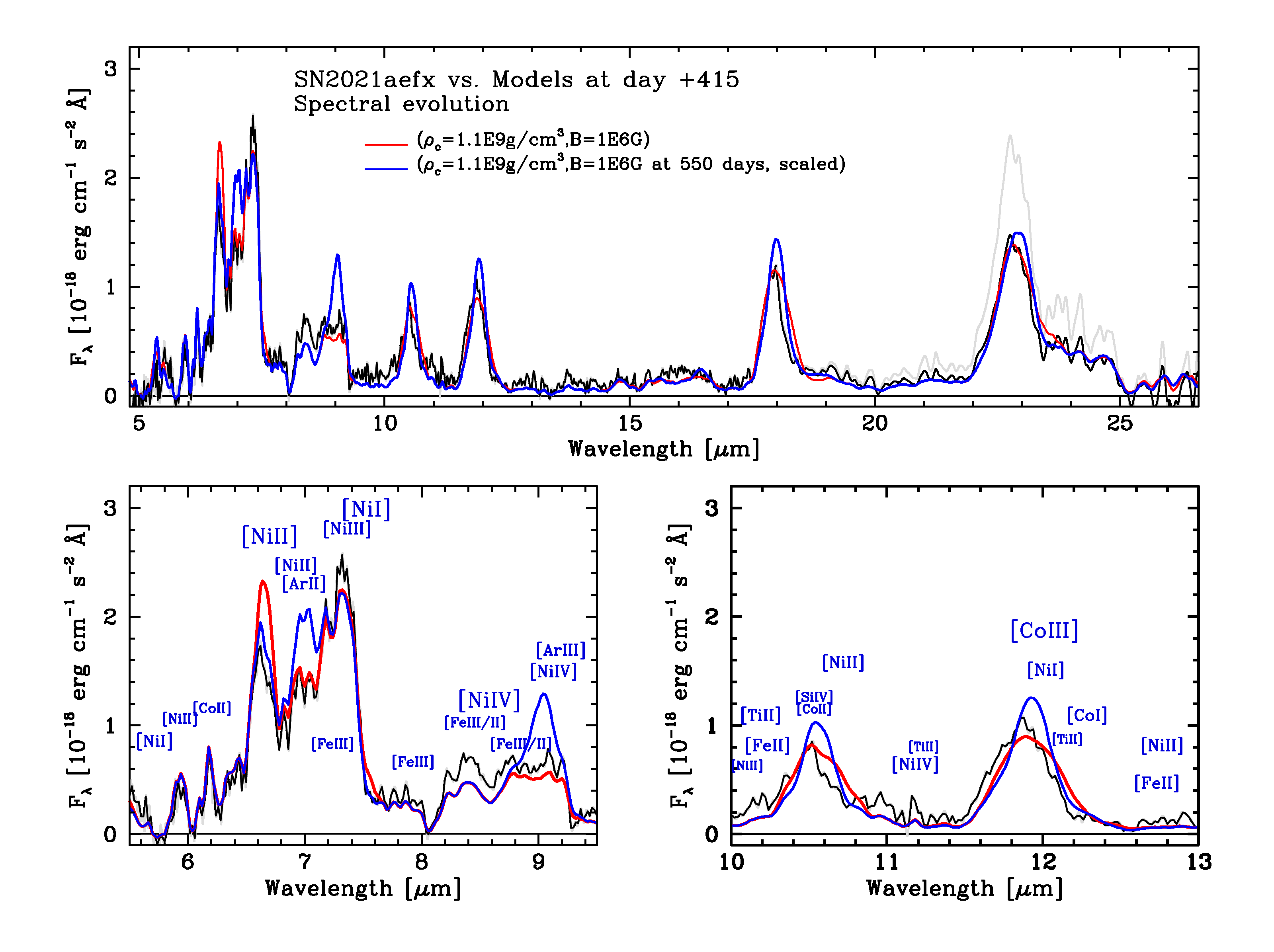}
  \caption{Same as Fig. \ref{fig:specmodRHO} but the evolution over one e-folding time for \Cofs\ decay is shown. The flux at +550~d is scaled up in luminosity to compensate for the $^{56}$Co decay. Thus, line profiles and ratios between the models can be compared.} 
  \label{fig:specmodC}
\end{figure*}

\subsubsection{Atomic data, line identifications, and strengths}\label{subsect:Atomic}

The atomic data used in our simulations come from \citet{Diamond_etal_2015,Diamond_etal_2018,vanHoof2018}, and references therein. In addition, the atomic data have been supplemented by fits of the lifetimes, $A_{ij}$, of unknown weak features between 5 and 27 \mic ~ based on the observed spectrum (see Appendix C, Tables \ref{tab:trans1} $\&$ \ref{tab:trans2}). Full lists with line strengths of transitions contributing to the flux spectrum of SN~2021aefx at +415~d from maximum are shown in Tables  \ref{tab:ir_lines} $\&$ \ref{tab:opt_lines}.  Line IDs are only shown for transitions with measured spontaneous lifetimes, \ie ~ Einstein $A_{ij}$ values. Our atomic models have been verified and used in previous papers \citep{DerKacy_etal_2023_21aefx,DerKacy23xkq},
and cross-checked with line identifications by \citet{Kwok2022}.

\subsubsection{Understanding the overall spectra and evolution} \label{subsec:Overall}

The observed and modeled spectra are presented with identifications of notable features in Fig. \ref{fig:specmod}. 
Here, we discuss the overall spectrum and underlying physics of our reference model before optimizing the parameters. The synthetic spectrum agrees reasonably well with the observations.

At first glance, the spectrum at +415~d resembles earlier epochs. 
However, this apparent similarity masks a physical regime change in the ejecta.  For example, at +415~d, most of the radioactive \Cofs\ has decayed to \Fefs\ (Fig. \ref{fig:models}) with only  $\sim$2\% of the initial \Cofs\ remaining. The observed and model spectra show broad features of [\FeII] and [\FeIII] at 18 and 23 \mic.
In the models the +415~d spectrum shows an equally strong, peaked feature at the location of  [\CoIII] 11.888~\micron\  rather than a  rounded profile seen a year earlier. This is also seen in several weaker features which  could be attributed to [\CoII] (e.g. 10.5 \mic) at earlier epochs.
Furthermore, the observed spectrum and model at +415~d also show strong features at 7 and 9~\micron. This is similar to earlier epochs, but both the profile shapes and dominant lines in the features have changed. 
At earlier times, the dominant ions have been identified as  
[\ArII] and [\ArIII] \citep{Gerardy2007,Telesco_etal_2015,DerKacy_etal_2023_21aefx,Kwok2022}.
In fact, at +415~d the 8.991~\micron\  feature is ``flat-topped'' rather than ``flat-tilted'', as was seen at earlier epochs. 
As we discuss below, the change in the profiles is not only caused by a spectral resolution effect (LRS vs. MRS) but hints towards a physical regime change at this epoch.

Our reference model can reproduce the 10.5 \mic feature reasonably well, and is dominated by forbidden,  singly-ionized Fe, Ti and Ni lines. Strong features at 7 and 9 \micron\ are still there but are now dominated by Ni I-III and Ni IV, respectively, with Ar being only a minor contributor.
There is a second, distinct spectral components consisting of many narrow features such as Fe, Co, and Ni between  5 to 7 \mic. These lines are well-known indicators of high-density burning ($\rho \geq 10^8$ \gcm; \eg\ \citealt{2017hsn..book.1955S}), and, being narrow, shows that there is little, or no mixing in the ejecta close to the center. 

The combination of line blending and the presence of two distinct spectral components, wide and narrow, is key to understanding the profiles and their evolution. Narrow lines of stable Ni are seen to dominate throughout because they produce sharply-peaked line profiles. 
Sharp profiles can also be produced by overlapping, broad lines if the
separation is small compared to the peak width, as demonstrated for the Fe/Co complex at 1.26 \mic \citep{Diamond_etal_2018,Hoeflich_2021_20qxp}. However, here, the small peaks are common in many lines with different blends.

\subsubsection{The [\CoIII] 11.888 $\micron$ evolution}
As shown in \autoref{sec:Cofsdecay}, to first order the peak emission of the [\CoIII] 11.888 $\micron $ feature follows the radioactive decay
of \Cofs\ as can be expected if the energy is deposited locally (e.g. by positrons) regardless of the distribution of elements (see Fig. \ref{fig:Covstime.pdf}). Because this line transitions to the ground state, the total emission is a direct measure of the ionization by hard radiation and non-thermal leptons independent of the temperature and other model details
\citep{Telesco_etal_2015}, very similar to the [\FeII] 1.644 \micron\ feature \citep{Hoeflich_etal_2004,Diamond_etal_2015,Kumar23}. 
In principle, the total \Nifs\
mass can be determined by the total line flux. 
 This  approach has been tried based on optical spectra by  but with mixed results due to the temperature-dependence of the optical transitions \citep{2015MNRAS.454.3816C}. This problem is not present when using the  [\FeII] 1.644 \micron\ and  [\CoIII] 11.888 $\micron $ features.

Analysis, of the absolute flux in this [\CoIII] is an important
test for the treatment of non-thermal leptons, and the production of \Nifs.
The overall quality of the spectral fits and fluxes may hint that our treatment is reasonably good. One limitation of the
direct, model-independent application to determine or test the
predicted \Nifs\ mass is that the number of ionizations per bound free absorptions by inner shell electrons will depend on details
of the atomic physics \citep{Berger1998}. Relative \Nifs\ masses between SNe~Ia should be reliable; however, the scaling factor between ionization and $^{56}$Ni mass needs further validation, as discussed by \citealt{Hoeflich_2021_20qxp}.

The presence of small deviations in the peak fluxes of [\CoIII] relative to the true \Cofs\ decay (see Fig.\ref{fig:Covstime.pdf}), hints at a more complex underlying physics. 
The relative contribution of $\gamma$-rays
to positrons decreases with  time,  resulting in the center becoming optically thin, and  a change of the central peak of the feature.
 At day +415, $\gamma$-rays still contribute significantly to the energy input in the low-velocity center with electron capture elements (Fig. \ref{fig:models}), namely $^{57}$Co, $^{58}$Ni and $^{54}$Fe. 
The isotopic shift between $^{56}$Co and $^{57}$Co is small, and the rest wavelengths of [\NiI] and [\CoI] are similar to [\CoIII]. This leads to narrower, additional components boosting the central peak of the 11.888 $\micron$~ feature.

\subsection{Sensitivity of the free parameters on the spectra and profiles}
\label{mod:spec}

 In this section, the formation of spectra and specific spectral profiles is considered in more detail, and discussed in the context of variations between models.

\subsubsection{Dependence on the WD central density}

Figure \ref{fig:specmodRHO} shows how the central density, $\rho_c$, of the WD (which determines the size of the inner \Nifs\ hole and the amount of electron  capture elements produced in the explosion) affects the spectral formation. 
With lower $\rho_c$, the intrinsic line profiles of \Fefs\ and \Cofs\ become narrower because an increase of emission at low velocities produces a rise in the peak \citep{Penney_etal_2014}. This can be seen in the broad [\FeII] and [\FeIII] dominated features at $\approx 18$ and $23 $\mic. Note that, though large $\rho_c$ increases the $^{58}$Ni production by a factor of 2, the specific energy input is halved, leading to similar electron capture line strengths, but different widths, in the features (Fig. \ref{fig:models}), resulting in similar strengths in $^{58}$Ni lines (Fig. \ref{fig:specmodRHO} vs. \ref{fig:specmodB}).  Furthermore, a high $\rho_c = 4\times 10^9$ \gcm can be ruled out
because it would produce profiles that are much broader than the data, by a factor of two  based on a spherical model series \citep{Diamond_etal_2015,2019A&A...630A..76G}. The MIR spectrum of such densities are not shown here because our simulation used do not have enough grid-points to resolve the complex radial structure of low $Y_e$ electron capture elements such as Mn and Cr that are important (see Figs. 25 and 26 in \citealt{2019A&A...630A..76G}).

Another spectral indicator of the central density, $\rho_c$,
is the trend that the features dominated by electron capture elements show a decreasing flux with decreasing $\rho_c$. This becomes obvious in the 6.7 and 7.4~\micron\ regions. The sensitivity to  $\rho_c$ is also seen by narrow mostly unblended weak [\NiII] and [\CoII] lines at 6 and 6.2 \mic, respectively.

Models with $\rho_c=1.1$ and $0.9\times  10^9$ \gcm do equally well at reproducing the observations.
The former shows better agreement with the flux level, whereas the latter produces slightly 
narrower forbidden [\FeII] and [\FeIII] features at 18 and 23~$\micron$.
Note, that for $\rho_c \ge 0.9 \times 10^9  $\gcm, the total emission of Ni features hardly depends on the $^{58}$Ni mass produced (in unmixed models) 
because the smaller electron capture core also increases the specific heating  which boosts the Ni emission (Fig. \ref{fig:models}).  
 Because heating is produced by $\gamma$-rays and positrons, the
 degeneracy can only be broken by time-series of spectra well beyond +500~d when positrons dominate everywhere. Moreover,  at very late times, the variations due to positron transport effects can separate macroscopic from microscopic mixing.

Both the [\FeII] and [\FeIII] features at 18 and 23~$\micron$ may favor the lowest $\rho_c\, ( 0.5 \times 10^9  $ \gcm).
However, this model is incompatible with the features and profiles at 7, 10.5 and 12~$\micron$  due to the low mass of electron capture elements. 

The 7~$\micron$ feature is complex with a structure dominated by narrow Ni components with a width of  
$\sim3000$~\kms, and a broad [\ArII] component (with a width of $\sim$ 7000–10000 \kms) which `fills the
emission gap' between the  [\NiI] and [\NiII] dominated peaks. Other lines such as  [\FeIII] contribute to the complexity. 

The emission complex between 8.1 and 9.3~$\micron$ shows two components. The blue part of this wavelength region ($\sim$8.4~$\micron$) is dominated by [\NiIV] 8.405~$\micron$.  Observations and models of SN~2021aefx at  +323~d show, that the red part of this region ($\sim$9~$\micron$)  was dominated by [\ArIII] \citep{DerKacy_etal_2023_21aefx}, which is similar to MIR observations of other SNe~Ia \citep{Gerardy2007,Telesco_etal_2015}. At these phases, SN~2021aefx showed a tilted profile, which was interpreted as evidence for off-center DDT \citep{DerKacy_etal_2023_21aefx}. 
Such a profile can be tilted
if the high-velocity region is asymmetric in the abundance distribution at the outer-edge, e.g. produced by an off-center DDT \citep{Hoeflich_2021_20qxp}.
However, in the latest spectrum of SN~2021aefx at +415~d, the 9~$\micron$ profile is  ``flat-topped'', not tilted, and is dominated by  [\NiIV] 8.945~$\micron$ and [\FeII]/[\FeIII]
lines with only a non-dominant contribution of [\ArIII]. 
This leads to a bump at the red end of the profile. 
Only about 50\% of the red end of the profile comes from Ar and is produced after the transition to the positron-dominated regime. The tilting vanishes because  
the outer, high-velocity Ar bulge is located outside the layers heated by positrons, resulting in a flat contribution to the profile at the latest epoch of observations.\footnote{At +323 days, heating was facilitated by $\gamma$-rays leading to  a ``flat-tilted'' Ar-dominated profile.}
About 50\% of the total emission can be attributed to [\NiIV] and multiple lines of [\FeII] and [\FeIII].

In the early nebular phase, blending by [\FeII] and [\FeIII] was suggested as a  main contributor to the [\ArIII] feature. However, this leads to an opposite tilt to the observed spectrum \citep{Blondin23}.\footnote{At +323d, our simulations  show blends at the same wavelength but weaker. If those blends dominate the profile,
they would cause a tilt opposite to the observation.}  Thus, the tilt and evolution of the feature supports our interpretation as a geometrical effect. 

It is well established that the 12 \mic feature at earlier times can be attributed
to [\CoIII] although some weaker additional components are present. 
 In the models by +415~d, [\NiI], [\TiII], and many weak Fe-lines, contribute about 1/3 of the flux. 
 The significant contribution beyond [\CoIII] also becomes evident from Fig. \ref{fig:specmodRHO} by
 showing a pointed peak compared to earlier observations. 

 Overall, at +415~d, the ionization shift towards lower ionization states with decreasing $\rho_c$.
 This is because for lower values of $\rho_c$
  the $^{56}$Ni region extends towards the more central layers \citep{Diamond_etal_2015}. 
  The can be seen in the effect of the  ratio of [\NiIII] 7.35~$\micron$/[\NiII] 6.92~$\micron$ features for various values of $\rho_c$. For $1.1 \times 10^9 $ \gcm this ratio is 1.7, for $0.9 \times 10^9 $ \gcm it is 1.5, and for $0.5 \times 10^9 $ \gcm the ratio is 1.1. 
  Therefore higher values of $\rho_c$ are consistent with the observations. 
  
  However, the 7.35~$\micron$ vs. the 6.64~$\micron$ peaks are close to 1 for our simulations, whereas the observed value is about 1.3, which may reflect the limitation in either atomic physics or the  underlying explosion models. Note, that mixing of electron capture
and $^{56}$Ni layers will produce the opposite trend \citep{DerKacy_etal_2023_21aefx} because with increasing $\rho_c$ the electron capture rich region increases in mass coordinate and into regions of Rayleigh–Taylor instabilities (see Sect. 6.6). Moreover, all the features in the 6 to 8 \micron ~region are heavily blended.
To derive $\rho_c$, the ratio between the peak of the [\CoIII] feature at 11.888 \micron~ and the peak of the feature at 7.35 \micron\ may be more appropriate. The corresponding value is 2.4 in both the observations and the model with  $\rho_c \approx 1.1 \times 10^9 $ \gcm. Where as the ratio is 1.2 for the low-density model.

As mentioned above, in both the models and data, most spectral features show a very narrow peak because they
are dominated by electron capture elements but still have a broad component. This characteristic is
valid for all strong Fe features in our model spectrum because the  \Fefs\ ~distribution traces \Cofs\ and \Nifs.

We note that microscopic mixing (i.e., mixing of species on atomic scales) over large scales would drastically change the evolution and was previously excluded for SN~2021aefx \citep{DerKacy_etal_2023_21aefx}.

\subsubsection{Dependence on the initial magnetic field of the WD }
Fig. \ref{fig:specmodB} shows the sensitivity of the initial magnetic field of the WD on the synthetic spectra. We use a turbulent morphology on scales of the Rayleigh-Taylor instabilities because even large-scale initial dipole fields would be tangled by passive flows during the deflagration phase of burning \citep{Hristov_etal_2021}. The changes of the spectra and profiles can be understood in the same way as the previous discussion. 
Prior to day $\approx $ 300, all
positrons locally annihilate, regardless of the strength of the initial magnetic field of the WD \citep{Penney_etal_2014}.
The changes in the specific energy distributions can be seen in Fig. \ref{fig:models}.

For weaker magnetic fields, positrons increasingly escape from the \Nifs\ layers to regions of higher velocities.
This leads to narrower mean half-width Fe and Co profiles, similar to changes in $\rho_c$, by reducing the emission at high Doppler shifts, namely \Cofs\ at high velocities (see Fig. \ref{fig:models}). The second effect is a boost of features where quasi-statistical equilibrium elements, such as Si and S, contribute significantly to the line
flux, e.g. see the feature at [\SiIV]  10.5 \mic, which is very similar to the potential appearance of the [\SiII] 1.3 \mic ~feature in the NIR \citep{Diamond_etal_2015}.  
The other effect is the shifting of the energy deposition towards the inner, higher-density layers, which results in a boost of the [\FeII] line at 18~\mic compared to the [\FeIII] at 23~\mic, because some positron escape leads to slightly lower temperatures and, thus, a shift in the flux from the optical to the MIR \citep{Penney_etal_2014,2022ApJ...930..107M}.

 Though still high, a B-field of $10^3$ G leads to strong leaking of positrons on both the inner edge and outer edge of the \Nifs\ layers, which leads to a boost of [\NiII], strong pumping of [\SiIV] and
a shift in the ionization balance towards [\CoII] (e.g. at 10.3 \micron), compared to [\CoIII]. 
Similarly, the profile of the 9 \micron ~ features changes from flat-topped with a B-field of $10^6$ G to a peaked profile with a B-field of $10^3$ G.
We require initial B-fields  $\approx 10^6 $G to keep the positron transport local, which is comparable with previous lower limits of $10^5$G  derived from other SNe~Ia \citep{Hoeflich_etal_2004,Diamond_etal_2015,Hristov_etal_2021,Kumar23}.  For the creation of high B fields, we possibly need a strong dynamo during the smoldering phase, as discussed in the above publications. 
 
\subsubsection{Future time evolution of the spectra}
 
In the discussion in \autoref{sec:specevolution}, we made use of past observations. 
From the models, we identified that the main physical driver of the evolution is the abundance change from  \Cofs\ to \Fefs, and the 
change of the mode of energy input, from a $\gamma$-ray to the positron-dominated regime.

Here, we consider the future evolution of SN~2021aefx using the  same model configuration and parameters presented above. This model has been evolved to one e-folding time of \Cofs\ later (+550~d, see Fig \ref{fig:specmodC}), hence the absolute flux is lower. 
The main change in the evolution of the spectral profiles is  a boost in features dominated by  intermediate mass elements, such as [\ArII]
at 7 \mic and [\ArIII] at 9 \mic. Furthermore, there is a narrowing of features due to a combination of positron transport effects and a decrease in the contribution from $\gamma$-rays by $\approx 60\% $ produced by the geometrical dilution.

 At +550~d the spectrum resembles an SN~Ia; however,  
 not unexpectedly, it does not match the data from +415 days.

 What would one learn if the prediction looks different from future data or if the spectral evolution is slow? 
One would have to restore positron trapping by increasing the magnetic field of the WD because B decreases with the square of time. In general, positron transport effects on the spectra
depend on the size and morphology of both the magnetic field and the 3D abundance distributions in the central region.

\subsubsection{Observable signatures of off-center DDTs prior to the nebular phase}

Prior to the nebular phase, the off-center DDT has two major effects: \textit{i}) It influences the rise, strength, and profile of the  [\CoIII] 11.888 \micron\ feature during the  transition from the photospheric to the nebular phase. This evolution has been identified as valuable diagnostics for the progenitor mass \citep{Telesco_etal_2015}. 
The rise is caused by the receding photosphere in combination with the rapidly dropping density \citep{Penney_etal_2014}  and, for asymmetric abundance distributions, will depend on the inclination; 
\textit{ii}) The DDT imposes a large-scale asymmetry in all abundances, including products of partial carbon and incomplete oxygen burning such as Mg/Si/S which results in significant line polarization. This is frequently seen in the \SiII\ 6355~\AA\  feature in many normal luminosity SNe~Ia during the photospheric phase 
\citep{Cikota_etal_2019}. It can be understood by selective depolarization by lines in scattering-dominated atmospheres when seen from positive $\Theta$ \citep{Yang2020,Hoeflich_etal_2023_19np}. Unfortunately, neither  polarization measurements nor earlier MIR spectra were obtained for SN~2021aefx.

\subsubsection{Progenitor signatures in SN~2021aefx}

The nebular spectrum presented provides a sensitive tool for studying the explosion mechanism and the thermonuclear runaway. As discussed in Sect. 5, the nebular spectra of SN~2021aefx closely resembles several normal SNe~Ia. However, nebular spectra provide limited information about the donor star and progenitor systems unless a significant amount of material is stripped from the companion \citep{2000ApJS..128..615M}. In our spectrum, all features have been identified without evidence for stripped material.

Early time spectra and light curves of SN~2021aefx show high spectral velocities and an early blue bump, suggesting an additional energy source \citep{Ashall22,Hosseinzadeh22,Ni23}. Observed early time variation between SNe~Ia suggests diversity in progenitor systems and in the path ways to the explosion.
At early times, only $10^{-3}$ to $ 10^{-4}~ \mathrm{M}_\odot$ of ejecta is visible (see Fig. 11 of \citealt[]{Hoeflich_etal_2023_19np}) \footnote{Note that from early time spectropolarimetry of SN~2019np, these corresponding layers have been found being very asymmetric \citep{Hoeflich_etal_2023_19np}.}
We refer to Sect. 6.1 of \citet{Hoeflich_etal_2023_19np} (and references therein) for a detailed discussion of possible imprints of the progenitor system and its environment on these early phases.
In short, these imprints may come from: \textit{i}) explosive surface burning of H/He of $\sim 10^{-3}$ to $ 10^{-4}~ \mathrm{M}_\mathrm{\odot}$ triggered by the outgoing detonation wave for some progenitor channels  within the delayed-detonation scenario (see Fig. 3 of \citealt[]{2019nuco.conf..187H}). This amount of He is very similar to progenitors with a acreeting He star, and is the minimum mass needed in He-triggered detonations in sub-$M_{Ch}$ explosions \citep{2015ApJ...805..150F,2022ApJ...932L..24R}. In both explosion scenarios, surface He-burning would produce high-velocity burning to  Si, S, Ca, Ti. \textit{ii}) Interaction of the outgoing shock wave with the circumstellar matter (CSM), namely a Roche-lobe or a strong wind from a companion star, and \textit{iii}) stellar rotation of the WD. Note that, for SN~2021aefx, Fig. \ref{fig:image} may suggest a dirty environment and, possibly, late-time interaction with the interstellar medium (ISM).

\subsection{Discussion and Implications for the Underlying Explosion Physics}\label{subsec:phys}

To prevent duplication of discussions on alternative scenarios, we direct readers to the analyses provided in earlier studies \citep{Hristov_etal_2021,Hoeflich_etal_2023_19np} and, for SN~2021aefx to \citet{DerKacy_etal_2023_21aefx}. Here,
we focus on our new findings and their implications.

SN~2021aefx can be understood in the framework of an off-center delayed-detonation model 
with central WD densities of $\rho_c \approx 0.9$ -- $1.1\times 10^9$ \gcm and with $M_{\mathrm{WD}}\approx 1.33-1.35~\mathrm{M}_\odot$. These parameters can be attributed to a near $M_{Ch}$ mass WD,  and place it into a regime of high-density burning with a proton to nucleon ratio Y$_\mathrm{e} \approx 0.49$. As discussed at the beginning of Sect. 6 in the context of the model construction,  the relatively low $\rho_c$ requires a large accretion rate that may be more compatible with He or C rather than H accretors.

For the +415~d spectra, the strength of the $^{58}$Ni hardly depends on the total mass of 
 $^{58}$Ni which is governed by $\rho_c$.
 As is obvious from Fig. \ref{fig:models}, the amount of 
 $^{58}${Ni} may vary by a factor of  2, but this can be compensated by the specific energy input, i.e. the energy input per gram. Time series of data and models are needed that cover the $\gamma$-ray dominated nebular regime ($\sim$200~d) to the positron-dominated regime, which starts at $\sim$500 days after the explosion.

The result is that the main isotopes in the core
are stable electron capture elements e.g. $^{58}$Ni, $^{57}$Co, and $^{54}$Fe. This places SN~2021aefx into a similar physical regime as  other normal-luminosity SNe~Ia
such as SN~2014J \citep{Telesco_etal_2015}. 

 However, in our models not all SNe~Ia have the same $\rho_c$. In fact, they seem to span a wide range between $5-50\times 10^8$ \gcm, where there is evidence for high $\rho_c$ in sub-luminous SNe~Ia such as SN~2016hnk, SN~2020qxp, and SN~2022xkq  \citep{Hoeflich_etal_2004,Penney_etal_2014,Diamond_etal_2015,Diamond_etal_2018,2019A&A...630A..76G,DerKacy23xkq}\footnote{High central densities, close to an Accretion Induced Collapse, found in many SNe Ia do not imply one specific unique explosion/progenitor scenario for all SNe~Ia.}.
 
 Probing the transition between the $\gamma$-ray  and positron dominated regimes is  important for establishing the Urca-process, its inner working, its impact on the energy balance during the smoldering phase, and the nature of weak interactions (see \citealt{Diamond_etal_2018,2022ApJ...926L..25T}, and references therein).  Moreover, nuclear cross-sections played a role (e.g. \citealt{2000ApJ...536..934B,2006NewAR..50..470H}) and will continue to play a central role \citep{2019nuco.conf..125T} in our understanding of the explosion mechanism and thermonuclear runaway.
 For example, improvements in the electron capture rates \citep{2000NuPhA.673..481L} resulted in a drastic change of the [\FeII] at 1.644 \mic  nebular line profile \citep{Hoeflich2006} from flat-topped to rounded \citep{Penney_etal_2014}, which demonstrated the need for high $\rho_c$ WDs close to where an accretion induced collapse would occur (e.g. \citealt{2021A&A...656A..94G}).

\subsection{Implications for the physics of the thermonuclear runaway}\label{runaway}

The comparison of synthetic and observed spectral profiles places strong constraints on the thermonuclear runaway in the delayed-detonation scenario, specifically regarding the initiation of thermonuclear explosive burning.

Detailed multi-dimensional simulations of the central single-point runaway show only mixing of the inner electron capture layers during the deflagration phase (see e.g. Figs. 1, 2, and 15  in  \citealt{2000astro.ph..8463K}).  Mixing occurs only during the subsonic deflagration phase and not during the subsequent detonation phase of DDT models, which 
 leads to overall structures very similar to spherical DDT models  (see Fig. 3  in \citealt{Gamezo_etal_2005}).  In contrast, multiple-spot, off-center ignitions mix the electron capture rich material to high velocities \citep{2014MNRAS.438.1762F,2024arXiv240211010P}. 
The differences between these hydro-simulations can be simply understood  by the 
lack of gravitation in the center, which results in  close-to-laminar burning for about 
1 second of high-density burning before Rayleigh-Taylor  instabilities develop.
Whereas, in multi-spot far off-center ignitions, the  Rayleigh-Taylor-dominated burning phase starts right away. As shown by 
\citet{2000astro.ph..8463K}, the long delay between ignition happens regardless of
pre-existing turbulence produced during the smoldering phase \citep{Hoeflich_Stein_2002} or
low C/O-ratios (e.g. \citealt{2001ApJ...557..279D}).  Pre-existing turbulence will introduce some mixing by dragging material, but this effect is limited
to the innermost slowly expanding layers.
As found in the analysis of previous SNe~Ia  \citep{2019A&A...630A..76G,Hoeflich_2021_20qxp}, again the spectra here are consistent with 
a near-to-central ignition and inconsistent with multi-spot strong-off-center ignition.

Moreover, the narrow width of the lines produced by electron capture elements puts tight limits on the macroscopic mixing, and confines it to the inner $\approx 3000 $ \kms which may be expected from the passive drag of electron capture material by pre-existing turbulent fields produced during the smoldering phase (\eg ~\citealt{Hoeflich_Stein_2002}). Note that
high WD magnetic fields  ($B \ge 10^6$G) are required based on studies of light curves\footnote{To measure B fields of the order of $10^6$ G, an accuracy of $0.025$~mag is required for late-time bolometric light curves \citep{Hristov_etal_2021}} and spectra.
Such fields may decrease the mixing of electron capture element even further.
For detailed discussions,  see \citet{Hoeflich_etal_2004,Diamond_etal_2015,Hristov_etal_2021,Hoeflich_etal_2023_19np,DerKacy23xkq}.  A turbulence field can be expected to develop during the smoldering phase \citep{Hoeflich_Stein_2002}, which drags the freshly formed 
products of high-density burning by passive flow. The result is an imhomogenous mixture of non-radioactive electron capture elements and $^{56}$Ni material which is heated over days by radioactive decay. Pressure equilibrium will compress the electron capture elements and non-radioactive layers into thin sheets \citep{Hoeflich_2017}.
Pre-existing turbulence may produce caustic
distributions with a wall thickness of $\approx 1000-1500$~\kms\ but limited to the inner region \citep{2015ApJ...804..140F,Hoeflich_2017}.  
The models used here do not take into account these structures, but their existence has been recently indicated by late-time polarization \citep{2022ApJ...939...18Y}.

\section{Conclusions}\label{sec:conclusions}
We present nebular phase \textit{JWST} MIRI/MRS observations of  SN~2021aefx  at +415~d past maximum.
Our work  demonstrates how combining  MIR medium-resolution data with detailed spectral models allows for the physics of SNe~Ia to be understood in a way that was previously not possible. \textit{JWST} promises to transform this area of research over the coming years.
The main results of our study can be summarized as follows:

\begin{itemize}

\item 
These new data and models covering 5-27~\mic\ have allowed us to produce an extended list of line identifications during the iron-dominated nebular phase (see, \autoref{sec:Data} \& Table \ref{tab:lines_model}). The higher resolution of the spectrum relative to previous epochs has allowed for many spectral features to be resolved. The spectrum is dominated by iron-group elements, with a strong contribution of stable Ni.  

\item 
The \textit{JWST} MIRI/MRS spectrum at +415~d has been analyzed in conjunction with previous MIRI/LRS observations of SN~2021aefx, allowing for a time series analysis from +255 to +415~d to be performed. 
We find that the peak evolution of the resonance [\CoIII] 11.888~\mic\ feature is consistent with the half-life of radioactive decay of  \Cofs\ (see Sect. \autoref{sec:Cofsdecay} \& Fig. \ref{fig:Covstime.pdf}).

\item
The spectrum has been analyzed using a new series of off-center delayed-detonation models. All models had a central point-of-ignition, and a point of the DDT at 0.5 $M_\odot$, which results in asymmetric abundance distributions (see \autoref{sec:models}). The viewing angle $\Theta $ is measured between the line-of-sight and the line defined  by  the point of DDT and the kinematic center. 
The spectrum at +415~d can be reproduced using the same $\Theta \approx -30^o$ that has been found in our previous study of SN~2021aefx at +323~d \citep{DerKacy_etal_2023_21aefx}. 
For other SNe~Ia, a wide range of values for $\Theta $ from $-90^o$ to $+90^o$ have been found between SNe.  This supports our interpretation that the profiles of SN~2021aefx are produced by a geometric effect.

\item Variations in the central density and initial magnetic field of the WD in the models were studied. We find that the progenitor of SN~2021aefx had a WD mass of $\approx 1.33$ -- $1.35$~\Msun, and central density of $\rho_c=0.9-1.1\times 10^9$~\gcm, and an initial magnetic field of $\ge10^6$G.

\item 
Comparison between these new data and models has revealed a profound change in the physics of the spectral formation compared to previous epochs. At +415~d the SN made the transition from a \Cofs-dominated to a mostly \Fefs-dominated regime. 
For the energy input, this is the transition phase from a globally $\ \gamma$-ray-dominated regime towards a positron-dominated regime. However, at +415~d $\gamma$-rays  still dominate the energy input in the central region. The evolution of the line profiles can be understood in terms of the above physics (see \autoref{mod:spec}).

\item 
The fact that we observe a narrow region of electron capture elements means there is very limited mixing in the inner regions of the ejecta. Thus, the point of ignition in SN~2021aefx is consistent with being near-to-central, and is inconsistent with being strongly off center if burning started as a deflagration front (see \autoref{runaway}).

\item 
At +415~d, the strength of the Ni features is dominated by the size of the B-field in the WD and is rather insensitive to the amount of Ni produced through high central density burning $\ge 10^9 $ \gcm.

\item 
At earlier epochs, the spectral features at $7$ and $9~\micron$ were mostly attributed to [\ArII] and [\ArIII]. The profiles were interpreted as being caused by asymmetric abundance distribution produced by the off-center nature of the DDT. 
This  resulted in  `flat-tilted' line profile because the entire Ar region was excited by gamma-rays. 
At +415~d these features have significantly changed.
The Ar-region is no longer excited by $\gamma$-rays but by positrons, which deposit their energy locally.
Hence,   the emission is governed by the overlapping region between \Nifs\ and Ar. 
The Ar contribution to the $7$ and $9~\micron$ features is small and flat topped, because the contribution of the asymmetric component from the outer layers is suppressed relative to earlier epochs. 
Overall, these features are dominated by iron-group elements at this phase (\autoref{mod:spec}).

\item 
Spectra and line profiles are sensitive to the initial $B$ field of the WD because positron transport alters the energy input relative to the abundances \citep{Penney_etal_2014,Diamond_etal_2015}. 
Here, the importance of positron transport and the need for high B fields has been demonstrated. 
Where we find the need for B fields which are larger than fields typically found in WDs. These values of B may be produced during the smoldering phase prior to the explosion. Later time-spectra are needed to push the limits to the regime where they heavily modify the properties of nuclear-burning fronts
\citep{2014ApJ...794...87R,Hristov_etal_2018}.

\end{itemize}

Finally, we want to emphasize the future prospects for our analysis of SN~2021aefx.  After about 550 days, the spectra enter the full positron-dominated regime. The positron transport effects have been demonstrated  and the lower limit of B has been derived for SN~2021aefx, but  spectra  taken after day 700 \citep{2023jwst.prop.3726D,2024jwst.prop.6582D} will be more sensitive to the morphology of the magnetic field of the WD and will provide the measurements of B beyond the current limit of $10^6$ G.  Current \textit{JWST} observations indicate a low density and high mass WD which had a  high magnetic field. This hints both towards a lack of full understanding of the accretion and smoldering phase, and the Urca cooling, which may be affected  by large B-fields and WD rotation. Future \textit{JWST} observations may address these problems. 

Overall, SN~2021aefx has demonstrated how intensive studies with an extended time-series of data consisting of low, and medium-resolution spectra can revolutionize our understanding  of the physical processes governing SNe~Ia. However, SN~2021aefx is just one object and future low to medium resolution \textit{JWST} observations of SNe~Ia are required to examine the true diversity both within  the physical processes in the ejecta, and the explosion scenarios within the universe.

\begin{acknowledgements}

C.A, P.H., E.B., J.D., and K.M. acknowledge support by NASA grants JWST-GO-02114, JWST-GO-02122, JWST-GO-03726, JWST-GO-04436, and JWST-GO-04522.  M.S. acknowledges support by NASA grants JWST-GO-03726, JWST-GO-04436, and JWST-GO-04522.  Support for programs \#2114, \#2122, \#3726, \#4436, and \#4522 were  provided by NASA through a grant from the Space Telescope Science  Institute, which is operated by the Association of Universities for  Research in Astronomy, Inc., under NASA contract NAS 5-03127.  P.H. and E.B. acknowledge support by NASA grant 80NSSC20K0538.  P.H. acknowledges support by NSF grant AST-2306395. 

L.G. acknowledges financial support from the Spanish Ministerio de Ciencia e Innovaci`on (MCIN) and the Agencia Estatal de Investigaci'on (AEI) 10.13039$/$501100011033 under the PID2020-115253GA-I00 HOSTFLOWS project, from Centro Superior de Investigaciones Cient'ificas (CSIC) under the PIE project 20215AT016 and the program Unidad de Excelencia Mar'ia de Maeztu CEX2020-001058-M, and from the Departament de Recerca i Universitats de la Generalitat de Catalunya through the 2021-SGR-01270 grant. I.D. is supported by the project PID2021-123110NB-I00 financed by MCIN$/$AEI
$/$10.13039$/$501100011033$/$ $\&$ FEDER A way to make Europe, UE.
\end{acknowledgements}

\facilities{\textit{JWST} (MIRI/MRS), MAST (\textit{JWST})}

\software{jwst (\citealp[ver. 1.9.4,][]{Bushouse2023_JWSTpipeline}),
          jdaviz (\citealp[ver. 3.2.1,][]{jdaviz}),
          HYDRA \citep{Hoeflich2003,Hoeflich2009,Hoeflich_etal_2017},
          Astropy \citep{astropy:2013, astropy:2018, astropy:2022},
          NumPy \citep{numpy2020}, SciPy \citep{SciPy2020}, 
          Matplotlib \citep{matplotlib}.}

\bibliographystyle{aasjournal}

\bibliography{ms}

\appendix
\vspace{-0.01cm}
\renewcommand\thefigure{\thesection.\arabic{figure}}   
\setcounter{figure}{0} 

\section{Data reduction}\label{sec:Datared}

\begin{figure*}
    \centering
    \includegraphics[width=0.99\textwidth]{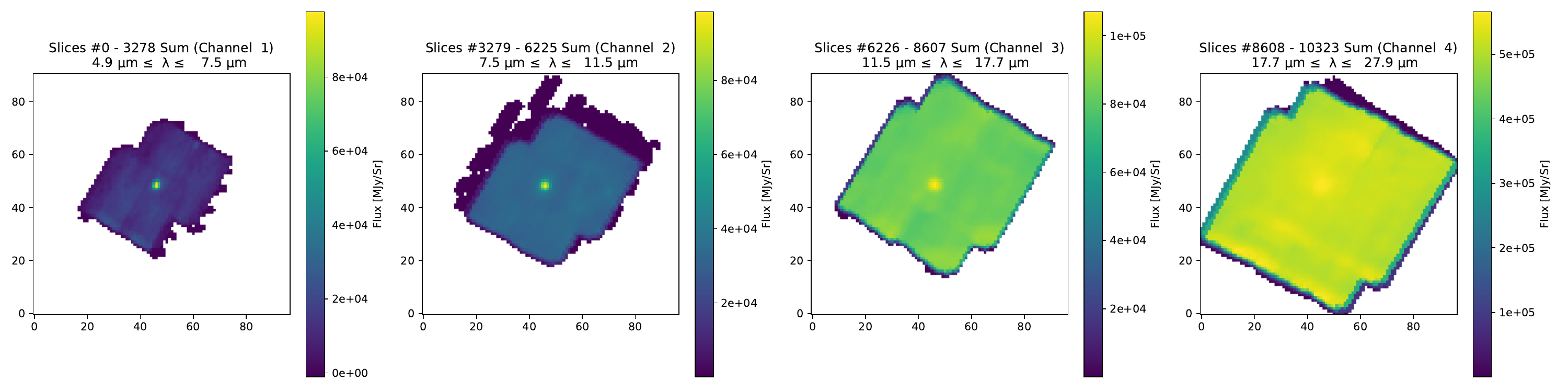}
     \includegraphics[width=0.99\textwidth]{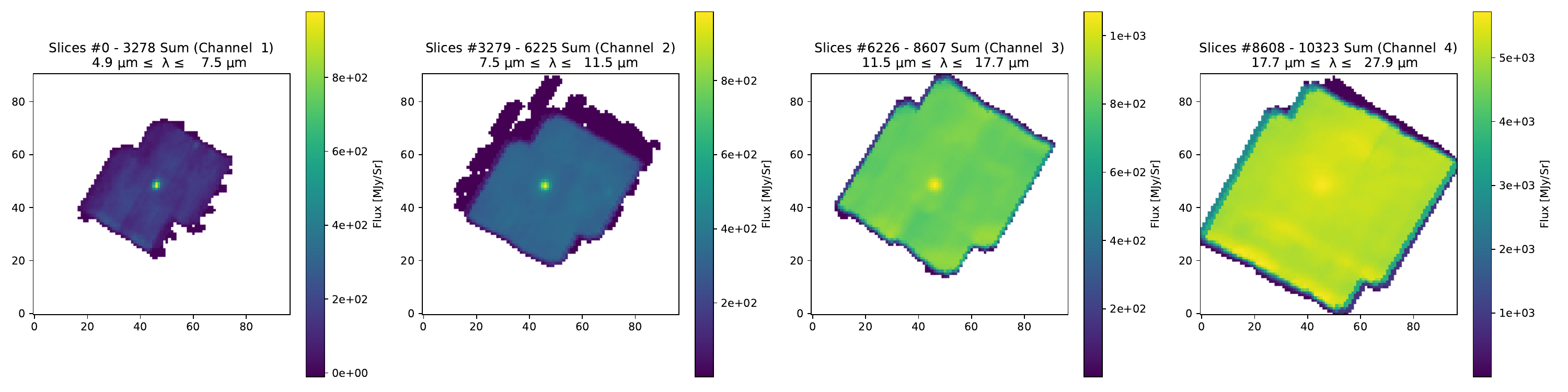}
    \caption{\textit{Top panel:} MIRI/MRS cube of SN~2021aefx before background subtraction divided into 4 channels. Each channel is collapsed sum of all its slices.
    \textit{Bottom panel:} MIRI/MRS cube of SN~2021aefx after background subtraction.}
    \label{fig:cubeimage}
\end{figure*}

The data were reduced using a custom-built pipeline\footnote{\url{https://github.com/shahbandeh/MIRI_MRS}} designed to extract observations of faint point sources that have complex backgrounds in MIRI/MRS data cubes (see \citealt{Shahbandeh24}). 
In short, the pipeline creates a
master background based on 20 different positions away from the source within the field of view.
This master background is subtracted from
the whole data cube. 
The SN flux is then extracted along the data cube, using the {\it Extract1dStep} in stage 3 of the {\it JWST}  reduction pipeline. 
The resulting data cube subtraction is shown in Fig. ~\ref{fig:cubeimage}.
All the previously unpublished data used in this paper can be found in this Digital Object Identifier (DOI)\footnote{\dataset[doi: 10.17909/f37y-gn67]{https://doi.org/10.17909/f37y-gn67}
}.

With this custom-made pipeline the flux level in Channel 4 (which covers wavelengths longer than 17.7\micron) is uncertain due to the dominant instrumental background flux which increases with wavelength. 
Therefore, for the spectral analysis, 
we choose to extract the data from Channel 4 separately using a manually-selected background region, to ensure that the continuum level is close to flat, as would be expected from a normal SN~Ia in the nebular phase (see \autoref{sec:Ch4} for more details). 
We emphasize that this may cause large uncertainties in terms of line strengths and ratios in Channel 4, and the reduction of Channel 4 data may change  as the \textit{JWST} MIRI/MRS pipeline improves. 
The reduction shown here utilized version 1.9.4 of the \textit{JWST} Calibration pipeline
\citep{Bushouse2023_JWSTpipeline} and Calibration Reference Data System files version 11.16.20.

\section{Channel 4 reduction}\label{sec:Ch4}
\begin{figure}[h!]
    \centering
    \includegraphics[width=0.9\textwidth]{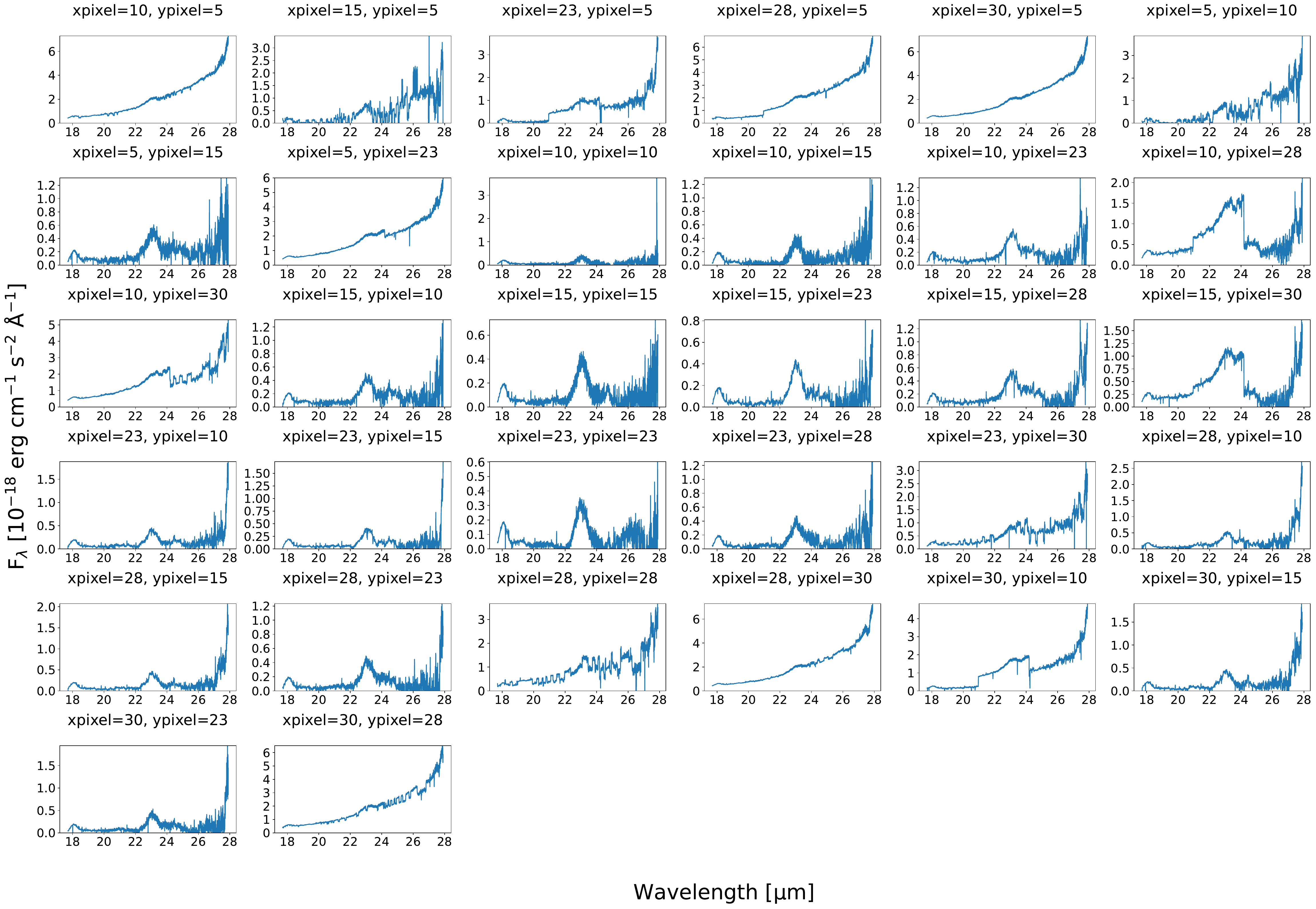}
    \caption{32 spectra extracted using different background positions within the IFU data cube.   }
    \label{fig:backgroundpositions}
\end{figure}

Due to the increased noise from the background in Channel 4, we chose to manually extract the SN flux in this region. 
To do this we use an STScI notebook 
\footnote{\url{https://spacetelescope.github.io/jdat_notebooks/notebooks/ifu_optimal/ifu_optimal.html}}
designed to perform extraction on a point source in JWST NIRSpec IFU data. We adapt this for use with MIRI/MRS data, using the final s3d data cube obtained from MAST. 
We extract the spectrum using a linearly expanding circular aperture (cone) because the PSF size increases with wavelength.
The spectrum at the position of the SN is obtained by extracting the raw flux using
various-size apertures at the location of the SN. 
After visual inspection, we chose to proceed with an aperture radius of 0.273~arcsec (this is 1 pixel in radius). 
Although this aperture size is likely to be smaller than the true PSF of the SN, using a larger size dramatically increases the noise in the data.
To remove the background contribution, individual background spectra are constructed at 32 locations away from the SN  across the field of view. 
For each  spectrum, the background is subtracted from the raw SN flux, see Fig. \ref{fig:backgroundpositions}. There is a large variation in the final flux depending on the location of the background selected.
We opt to use a background where the continuum is roughly flat.
Our chosen optimum background position is centered around pixels x=15, y=23. 

\begin{figure}[h!]
    \centering
    \includegraphics[width=0.5\textwidth]{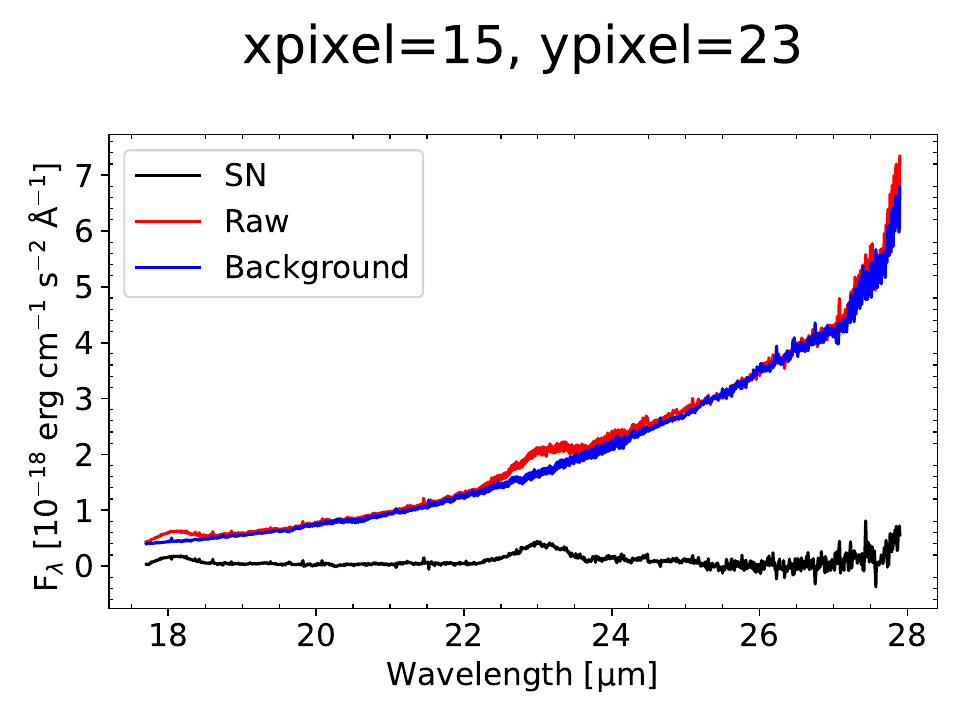}
    \caption{ The raw flux at the position of the SN (red), with the flux at the position of the optimal background region centered on pixel x=15, y=23. The final Channel 4, produced from subtracting the extractions  is plotted in black.}
    \label{fig:finalbackgroundpositions}
\end{figure}

Figure \ref{fig:finalbackgroundpositions} shows the final SN spectrum in Channel 4. Due to the small aperture this spectrum has to be scaled by a factor of 6 to match the flux in Channel 3. Overall, we highlight that the flux in Channel 4 is uncertain, but we are confident in the width and location of the features identified. 
Finally, we note that combining all 32 background spectra before subtracting 
from the raw SN spectrum does not successfully remove the background flux and leaves a large excess at longer wavelengths.

\section{Velocities} \label{sec:vel}
In this section, we fit spectral line profiles of the dominant ions in selected wavelength ranges to determine the region in the ejecta in which they are formed. As the spectral resolution of MIRI/MRS is much higher than the previous MIRI/LRS observations, this analysis has the potential to allow us to determine more accurately the location of the emitting regions within the ejecta.  However, as discussed in the main text, many of the atomic line transitions are not known in this region and detailed spectral modeling is required to make significant progress. Despite this, we fit the spectra below. 

For the fitting process, the \textit{scipy.optimize} package is used, and the best fits are determined via using the Levenberg-Marquardt algorithm and the least squares statistic option.  
  In each wavelength region, the emission features within the regions are assumed to be composed of Gaussian profiles. 
  Although treating emission line profiles as Gaussians makes assumptions about the distribution of the material in the emitting region, it does provide us with a quantitative way to analyze the data. 
 Priors and bounds are provided to each fit to ensure that they are consistent with the line IDs provided from the models in  \autoref{sec:models}.  If there is a region in which a line ID is not provided from the models, but there is a clear feature we also add an emission profile in this region. The spectral fits can be found in Fig. \ref{fig:specfits}, and the corresponding values, plotted in Fig. \ref{fig:peakvsFWHM}, are presented in table \ref{tab:fits}.
Finally, we choose not to fit the feature between 8-10~$\micron$ due to the complex blend around the ``flat-topped'' Ar region, and we do not fit longwards of 18.6~$\micron$ due to the uncertainty in the 
 Channel 4 data, and the fit around the [\CoIII]~11.888~$\micron$ feature is shown in the main body of the text (see \autoref{sec:Cofsdecay}). 

\renewcommand\thetable{\thesection.\arabic{table}}   
\setcounter{table}{0}

\begin{deluxetable}{c c c c c c c c}[h]
  \tablecaption{The best fit parameters from the ions used in  Fig. \ref{fig:peakvsFWHM}. \label{tab:fits} }
  \tablehead{\colhead{Ion} & \colhead{Wavelength} & \colhead{v$_{\mathrm{peak}}$} 
    & \colhead{Error v$_{\mathrm{peak}}$} & \colhead{Amplitude} & \colhead{Error Amplitude}
    & \colhead{$\sigma$} & \colhead{Error $\sigma$}}
    \startdata
    \hline
    &\micron &\micron &F$_{\mathrm{\lambda}}$&F$_{\mathrm{\lambda}}$&\micron &\micron \\
    \hline
{[\ion{Ni}{2}]} 	&	6.636	&	6.644	&	0.004	&	1.699	&	0.052	&	0.092	&	0.008\\
{[\ion{Fe}{2}]} 	&	6.721	&	6.771	&	0.005	&	1.040	&	0.109	&	0.064	&	0.008\\
{[\ion{Ar}{2}]} 	&	6.985	&	6.966	&	0.008	&	1.501	&	0.051	&	0.086	&	0.012\\
{[\ion{Ni}{3}]} 	&	7.349	&	7.360	&	0.005	&	2.564	&	0.024	&	0.139	&	0.013\\
{[\ion{Co}{1}]} 	&	7.202	&	7.202	&	0.004	&	1.288	&	0.167	&	0.054	&	0.005\\
{[\ion{Co}{3}]} 	&	7.103	&	7.093	&	0.006	&	1.203	&	0.152	&	0.066	&	0.012\\
{[\ion{Co}{1}]} 	&	7.507	&	7.472	&	0.003	&	0.489	&	0.195	&	0.048	&	0.011\\
{[\ion{Fe}{2}]} 	&	17.936	&	17.926	&	0.020	&	1.011	&	0.245	&	0.100	&	0.042\\
{[\ion{Co}{1}]} 	&	18.265	&	18.282	&	0.023	&	0.587	&	0.079	&	0.100	&	0.036\\
{[\ion{Fe}{2}]} 	&	10.189	&	10.212	&	0.011	&	0.337	&	0.019	&	0.100	&	0.016\\
{[\ion{Co}{3}]} 	&	10.523	&	10.555	&	0.010	&	0.784	&	0.162	&	0.085	&	0.023\\
{[\ion{Co}{2}]} 	&	10.682	&	10.712	&	0.027	&	0.439	&	0.059	&	0.100	&	0.040\\
{[\ion{Co}{3}]} 	&	11.888	&	11.917	&	0.002	&	1.013	&	0.008	&	0.196	&	0.002\\
\hline
\enddata
\end{deluxetable}

\renewcommand\thefigure{\thesection.\arabic{figure}}   
\setcounter{figure}{0} 

\begin{figure*}
    \centering
    \includegraphics[width=0.9\textwidth]{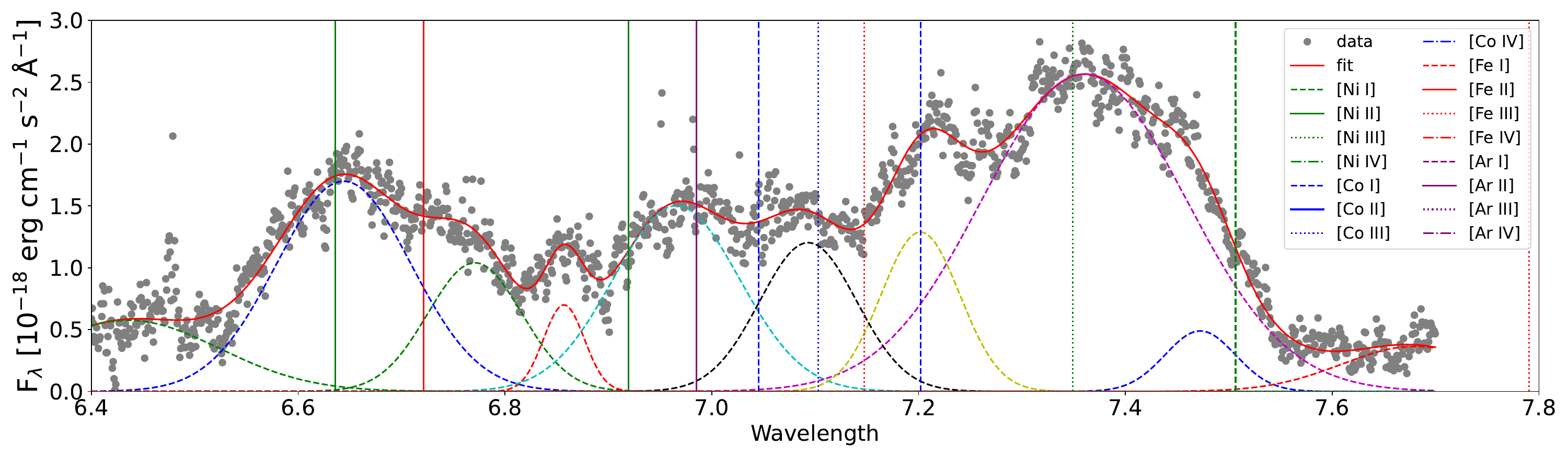}
    \includegraphics[width=0.9\textwidth]{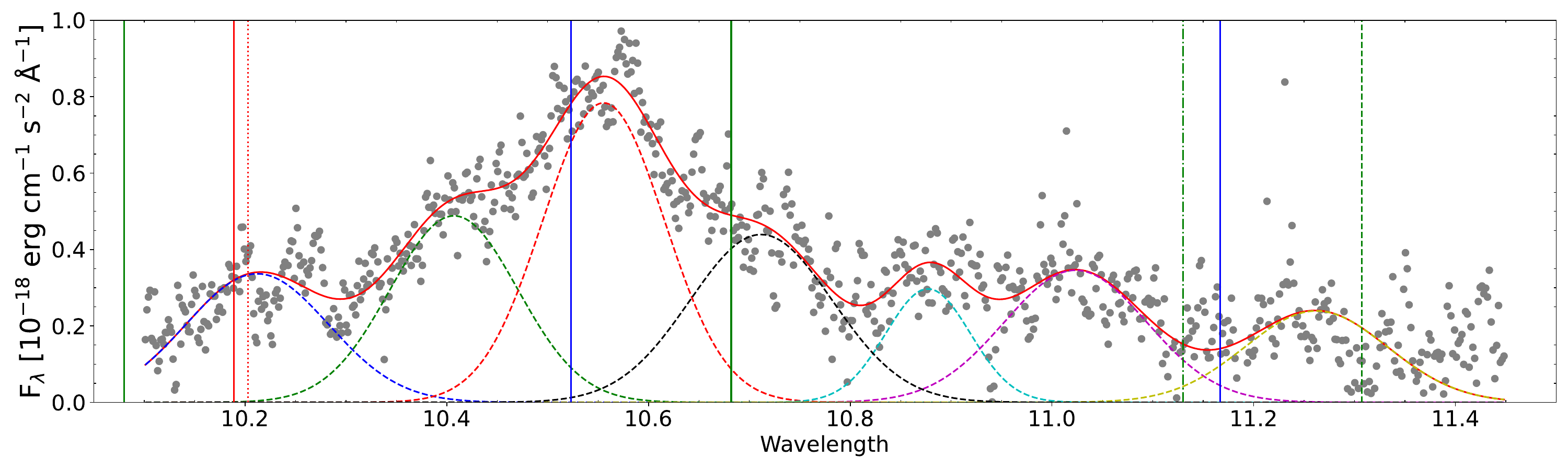}
    \includegraphics[width=0.9\textwidth]{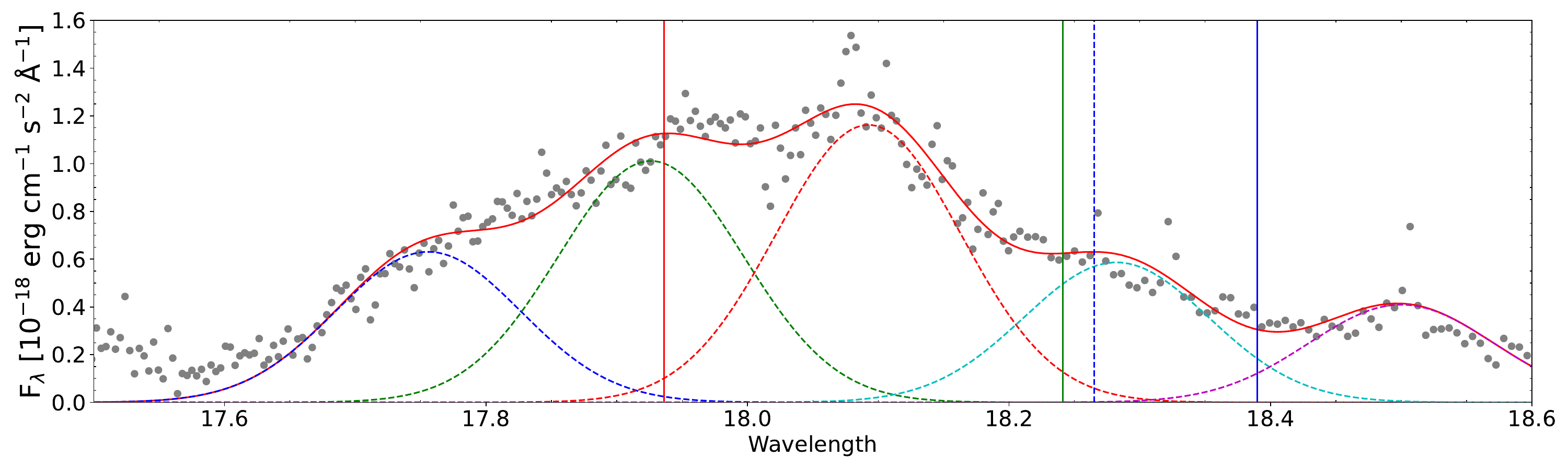}
    \caption{Spectral fits of three main regions in the data.}
    \label{fig:specfits}
\end{figure*}

\section{Optical to 4.7~$\micron$ line list from Models} 
\renewcommand\thetable{\thesection.\arabic{table}}   
\setcounter{table}{0} 

\label{sec:opt_line_List}
For completeness, in Table \ref{tab:opt_lines} the line list from our models in the wavelength regime shorter than our observations is provided. \\

\begin{deluxetable}{rcl|rcl|rcl|rcl|rcl}[h!]
  \tablecaption{Line contributions to the spectra from the optical to 4.7 \micron ~ at  +415~d from the Reference Model with $\rho_c=1.1\times 10^9 $ \gcm and $B=10^6$G. \label{tab:opt_lines}}
  \tablehead{\colhead{\bf S} & \colhead{$\lambda$~[\mic]} & \colhead{Ion} 
    & \colhead{\bf S} & \colhead{$\lambda$~[\mic]} & \colhead{Ion}
    & \colhead{\bf S} & \colhead{$\lambda$~[\mic]} & \colhead{Ion}
    & \colhead{\bf S} & \colhead{$\lambda$~[\mic]} & \colhead{Ion}
    & \colhead{\bf S} & \colhead{$\lambda$~[\mic]} & \colhead{Ion}}
    \startdata
       	&	0.3689	&	 {[\ion{Co}{2}]} 	&
      	&	0.4104	&	 {[\ion{Co}{2}]} 	&
      	&	0.4116	&	 {[\ion{Fe}{2}]} 	&
      	&	0.4178	&	 {[\ion{Fe}{2}]} 	&
      {$\ast \ast\ \ast\ $} 	&	0.4245	&	 {[\ion{Fe}{2}]} 	\\
       {$\ast\ $}	&	0.4246	&	 {[\ion{Fe}{2}]} 	&
       {$\ast\ \ast\ $}	&	0.4278	&	 {[\ion{Fe}{2}]} 	&
      {$\ \ \ast\ $} 	&	0.4289	&	 {[\ion{Fe}{2}]} 	&
      	&	0.4307	&	 {[\ion{Fe}{2}]} 	&
      {$\ast\ $}	&	0.4321	&	 {[\ion{Fe}{2}]} 	\\
      	&	0.4327	&	 {[\ion{Ni}{2}]} 	&
      {$\ast\ $}	&	0.4348	&	 {[\ion{Fe}{2}]} 	&
      {$\ast\ $}	&	0.4354	&	 {[\ion{Fe}{2}]} 	&
      	&	0.436	&	 {[\ion{Fe}{2}]} 	&
      	&	0.4361	&	 {[\ion{Fe}{2}]} 	\\
     {$\ast\ $} 	&	0.4374	&	 {[\ion{Fe}{2}]} 	&
      	&	0.4415	&	 {[\ion{Fe}{2}]} 	&
   {$\ast\ $}   	&	0.4418	&	 {[\ion{Fe}{2}]} 	&
      	&	0.4453	&	 {[\ion{Fe}{2}]} 	&
     {$\ast\ $} 	&	0.4459	&	 {[\ion{Fe}{2}]} 	\\
      	&	0.4476	&	 {[\ion{Fe}{2}]} 	&
      	&	0.449	&	 {[\ion{Fe}{2}]} 	&
      	&	0.4494	&	 {[\ion{Fe}{2}]} 	&
      	&	0.4501	&	 {[\ion{Co}{3}]} 	&
      	&	0.4608	&	 {[\ion{Fe}{3}]} 	\\
      	&	0.4624	&	 {[\ion{Co}{2}]} 	&
 	&	0.4641	&	 {[\ion{Fe}{2}]} 	&
 {$\ast\ \ast\ \ast $} 	&	0.4659	&	 {[\ion{Fe}{3}]} 	&
  {$\ast\ $}	&	0.4703	&	 {[\ion{Fe}{3}]} 	&
  {$\ast\ $}	&	0.4729	&	 {[\ion{Fe}{2}]} 	\\
   {$\ast\ $}	&	0.4735	&	 {[\ion{Fe}{3}]} 	&
 	&	0.4749	&	 {[\ion{Co}{2}]} 	&
{$\ast\ $} 	&	0.4756	&	 {[\ion{Fe}{3}]} 	&
 {$\ast\ $} 	&	0.4771	&	 {[\ion{Fe}{3}]} 	&
 	&	0.4804	&	 {[\ion{Co}{2}]} 	\\
 {$\ast\ $}	&	0.4882	&	 {[\ion{Fe}{3}]} 	&
{$\ast\ $} 	&	0.4891	&	 {[\ion{Fe}{2}]} 	&
 	&	0.4932	&	 {[\ion{Fe}{3}]} 	&
 {$\ast\ \ast\ $} 	&	0.5013	&	 {[\ion{Fe}{3}]} 	&
 	&	0.5086	&	 {[\ion{Fe}{3}]} 	\\
 {$\ast\ $}	&	0.5113	&	 {[\ion{Fe}{2}]} 	&
 {$ \ast\ \ast\ $} 	&	0.516	&	 {[\ion{Fe}{2}]} 	&
 	&	0.5222	&	 {[\ion{Fe}{2}]} 	&
 {$\ast\ $} 	&	0.5263	&	 {[\ion{Fe}{2}]} 	&
 	&	0.527	&	 {[\ion{Co}{2}]} 	\\
 {$\ast\ \ast\ $} 	&	0.5272	&	 {[\ion{Fe}{3}]} 	&
 	&	0.5298	&	 {[\ion{Fe}{2}]} 	&
 {$\ast\ $}	&	0.5335	&	 {[\ion{Fe}{2}]} 	&
{$\ast\ $} 	&	0.5378	&	 {[\ion{Fe}{2}]} 	&
 	&	0.5414	&	 {[\ion{Fe}{3}]} 	\\
 	&	0.5472	&	 {[\ion{Co}{2}]} 	&
 	&	0.5548	&	 {[\ion{Co}{2}]} 	&
 	&	0.5562	&	 {[\ion{Co}{2}]} 	&
 {$ $} 	&	0.589	&	 {[\ion{Co}{3}]} 	&
 	&	0.5976	&	 {[\ion{Fe}{1}]} 	\\
 	&	0.6016	&	 {[\ion{Fe}{2}]} 	&
 	&	0.6129	&	 {[\ion{Co}{3}]} 	&
 	&	0.6197	&	 {[\ion{Co}{3}]} 	&
 	&	0.6578	&	 {[\ion{Co}{3}]} 	&
 	&	0.6586	&	 {[\ion{Co}{1}]} 	\\
 	&	0.6669	&	 {[\ion{Ni}{2}]} 	&
 	&	0.6855	&	 {[\ion{Fe}{2}]} 	&
 	&	0.6934	&	 {[\ion{Co}{2}]} 	&
 	&	0.7138	&	 {[\ion{Ar}{3}]} 	&
 	&	0.7155	&	 {[\ion{Co}{3}]} 	\\
 {$\ \ast\ \ast\ \ast\ $} 	&	0.7157	&	 {[\ion{Fe}{2}]} 	&
 {$\ \ \ast\ $} 	&	0.7174	&	 {[\ion{Fe}{2}]} 	&
 	&	0.7249	&	 {[\ion{Co}{1}]} 	&
 {$\ \ast\ \ast\ \ast\ $} 	&	0.7293	&	 {[\ion{Ca}{2}]} 	&
 {$\ \ast\ \ast\  $} 	&	0.7326	&	 {[\ion{Ca}{2}]} 	\\
 {$\ \ast\  $} 	&	0.738	&	 {[\ion{Ni}{2}]} 	&
 {$\ast\ $} 	&	0.739	&	 {[\ion{Fe}{2}]} 	&
 {$\ \ \ast\ $} 	&	0.7414	&	 {[\ion{Ni}{2}]} 	&
 {$\ast\ \ast\ $} 	&	0.7455	&	 {[\ion{Fe}{2}]} 	&
 	&	0.7541	&	 {[\ion{Co}{2}]} 	\\
 {$\ast\ $}	&	0.764	&	 {[\ion{Fe}{2}]} 	&
{$\ast\ $} 	&	0.7689	&	 {[\ion{Fe}{2}]} 	&
 	&	0.7892	&	 {[\ion{Ni}{3}]} 	&
 	&	0.803	&	 {[\ion{Co}{2}]} 	&
 {$\ \ast\ \ast\ $} 	&	0.8123	&	 {[\ion{Co}{2}]} 	\\
 	&	0.8123	&	 {[\ion{Co}{2}]} 	&
 	&	0.8303	&	 {[\ion{Ni}{2}]} 	&
 	&	0.8336	&	 {[\ion{Co}{2}]} 	&
 	&	0.8466	&	 {[\ion{Co}{2}]} 	&
 	&	0.8469	&	 {[\ion{Co}{2}]} 	\\
 	&	0.8502	&	 {[\ion{Ni}{3}]} 	&
 	&	0.8546	&	 {[\ion{Co}{1}]} 	&
 {$\ \ \ast\ $} 	&	0.8574	&	 {[\ion{Co}{2}]} 	&
 	&	0.8583	&	 {[\ion{Co}{2}]} 	&
 	&	0.8597	&	 {[\ion{Co}{1}]} 	\\
 {$\ \ast\ \ast\ $} 	&	0.8619	&	 {[\ion{Fe}{2}]} 	&
  {$\ast\ \ast\ \ast\ $} 	&	0.8894	&	 {[\ion{Fe}{2}]} 	&
 {$\ast\ $}	&	0.9036	&	 {[\ion{Fe}{2}]} 	&
 {$\ast\ \ast\ $}	&	0.9054	&	 {[\ion{Fe}{2}]} 	&
 	&	0.9071	&	 {[\ion{S}{3}]} 	\\
{$\ast\ \ast\ $} 	&	0.9229	&	 {[\ion{Fe}{2}]} 	&
 {$\ast\ $} 	&	0.927	&	 {[\ion{Fe}{2}]} 	&
 {$ \ast\ $} 	&	0.9339	&	 {[\ion{Co}{2}]} 	&
 {$\ast\ $} 	&	0.9345	&	 {[\ion{Co}{2}]} 	&
 	&	0.9447	&	 {[\ion{Fe}{3}]} 	\\
 	&	0.9474	&	 {[\ion{Fe}{2}]} 	&
 	&	0.9533	&	 {[\ion{S}{3}]} 	&
 	&	0.9642	&	 {[\ion{Co}{2}]} 	&
 {$\ \ \ast\ $} 	&	0.9642	&	 {[\ion{Co}{2}]} 	&
 	&	0.9697	&	 {[\ion{Co}{1}]}  	\\
 	&	0.9705	&	 {[\ion{Fe}{3}]}  	&
 	&	0.9983	&	 {[\ion{Fe}{2}]}  	&
 {$\ast\  $} 	&	0.9483	&	 {[\ion{Fe}{2}]}  	&
 {$\ast\ \ast\ \ast\ $} 	&	1.0191	&	 {[\ion{Co}{2}]}  	&
 {$\ \ast\ $} 	&	1.0248	&	 {[\ion{Co}{2}]}  	\\
 	&	1.0283	&	 {[\ion{Co}{2}]}  	&
 {$ $} 	&	1.0283	&	 {[\ion{Co}{2}]}  	&
 	&	1.0611	&	 {[\ion{Fe}{3}]}  	&
 	&	1.0718	&	 {[\ion{Ni}{2}]}  	&
 	&	1.0718	&	 {[\ion{Ni}{2}]}  	\\
 {$\ \ast\ \ast\ $} 	&	1.0824	&	 {[\ion{S}{1}]}  	&
 	&	1.0885	&	 {[\ion{Fe}{3}]}  	&
 {$\ \ \ast\ $} 	&	1.0976	&	 {[\ion{Co}{2}]}  	&
 	&	1.0994	&	 {[\ion{Si}{1}]}  	&
 {$\ \ \ast\ $} 	&	1.1283	&	 {[\ion{Co}{2}]}  	\\
 {$\ast\ $} 	&	1.1309	&	 {[\ion{S}{1}]}  	&
 	&	1.1616	&	 {[\ion{Ni}{2}]}  	&
 	&	1.2122	&	 {[\ion{Fe}{2}]}  	&
 	&	1.2525	&	 {[\ion{Fe}{2}]}  	&
 {$\ast\ \ast\ \ast\ $} 	&	1.257	&	 {[\ion{Fe}{2}]}  	\\
 {$\ast\ \ast\ $} 	&	1.2707	&	 {[\ion{Fe}{2}]}  	&
 {$\ \ \ast\ $} 	&	1.2791	&	 {[\ion{Fe}{2}]}  	&
 {$\ \ \ast\ $} 	&	1.2946	&	 {[\ion{Fe}{2}]}  	&
 {$\ast\ $}	&	1.2981	&	 {[\ion{Fe}{2}]}  	&
    {$\ast\ \ast\ $} 	&	1.3209	&	 [\ion{Fe}{2}] 	\\
   {$\ast\ $}  	&	1.321	&	 [\ion{Fe}{1}] 	&
    {$\ast\ \ast\ $} 	&	1.3281	&	 [\ion{Fe}{2}] 	&
    	&	1.3422	&	 [\ion{Fe}{1}] 	&
    	&	1.3556	&	 [\ion{Fe}{1}] 	&
    	&	1.3676	&	 [\ion{Fe}{1}] 	\\
    {$\ast\ $} 	&	1.3722	&	 [\ion{Fe}{2}] 	&
    {$\ast\ \ast\ $}	&	1.3733	&	 [\ion{Fe}{1}] 	&
    	&	1.3762	&	 [\ion{Fe}{1}] 	&
    	&	1.4055	&	 [\ion{Co}{2}] 	&
     { $\ast\ \ast\ $ } 	&	1.4434	&	 [\ion{Fe}{1}] 	\\
     	&	1.4972	&	 [\ion{Co}{2}] 	&
    {$\ast\ \ast\ \ast $} 	&	1.5339	&	 [\ion{Fe}{2}] 	&
    {$ $} 	&	1.5474	&	 [\ion{Co}{2}] 	&
    	&	1.5488	&	 [\ion{Co}{3}] 	&
    	&	1.5694	&	 [\ion{Co}{2}] 	\\
    {$\ast\ \ast\ \ast\ $} 	&	1.5999	&	 [\ion{Fe}{2}] 	&
    	&	1.6073	&	 [\ion{Si}{1}] 	&
    {$ $} 	&	1.6267	&	 [\ion{Co}{2}] 	&
    {$\ast\ $} 	&	1.6347	&	 [\ion{Co}{2}] 	&
    {$\ast\ \ast\ \ast\ $} 	&	1.644	&	 [\ion{Fe}{2}] 	\\
    {$\ast\ $} 	&	1.6459	&	 [\ion{Si}{1}] 	&
    {$\ast\ \ast\ \ast\ $} 	&	1.6642	&	 [\ion{Fe}{2}] 	&
    {$\ast\ \ast\ $} 	&	1.6773	&	 [\ion{Fe}{2}] 	&
    {$\ast\ $} 	&	1.7116	&	 [\ion{Fe}{2}] 	&
     	&	1.7289	&	 [\ion{Co}{2}] 	\\
    {$ $} 	&	1.7366	&	 [\ion{Co}{2}] 	&
    	&	1.7413	&	 [\ion{Co}{3}] 	&
    {$\ast\ \ast\ $} 	&	1.7454	&	 [\ion{Fe}{2}] 	&
    {$\ast\ \ast\ $} 	&	1.7976	&	 [\ion{Fe}{2}] 	&
    {$\ast\ \ast\ $} 	&	1.8005	&	 [\ion{Fe}{2}] 	\\
    {$\ast\ $} 	&	1.8099	&	 [\ion{Fe}{2}] 	&
    	&	1.8119	&	 [\ion{Fe}{2}] 	&
    {$\ast\ $} 	&	1.904	&	 [\ion{Co}{2}] 	&
    {$\ast\ \ast\ $} 	&	1.9393	&	 [\ion{Ni}{2}] 	&
    	&	1.9581	&	 [\ion{Co}{3}] 	\\
    	&	2.0028	&	 [\ion{Co}{3}] 	&
    	&	2.0073	&	 [\ion{Fe}{2}] 	&
    	&	2.0418	&	 [\ion{Ti}{2}] 	&
    {$\ast\ \ast\ $} 	&	2.0466	&	 [\ion{Fe}{2}] 	&
    	&	2.0492	&	 [\ion{Ni}{2}] 	\\
    	&	2.0979	&	 [\ion{Co}{3}] 	&
    	&	2.1334	&	 [\ion{Fe}{2}] 	&
   {$\ast\  $} 	&	2.1457	&	 [\ion{Fe}{3}] 	&
    	&	2.1605	&	 [\ion{Ti}{2}] 	&
 {$\ast\ \ast\ $} 	&	2.2187	&	 [\ion{Fe}{3}] 	\\
 {$\ast\ \ast\ $}	&	2.2425	&	 [\ion{Fe}{3}] 	&
 {$\ast\ \ast\ $} 	&	2.2443	&	 [\ion{Fe}{2}] 	&
 	&	2.3086	&	 [\ion{Ni}{2}] 	&
 {$\ast\ \ast\ \ast\ $} 	&	2.3486	&	 [\ion{Fe}{3}] 	&
 	&	2.3695	&	 [\ion{Ni}{2}] 	\\
  {$\ast\ \ast\ \ast\ $}	&	2.4781	&	 [\ion{Fe}{2}] 	&
 {$\ast\ \ast\ $} 	&	2.5255	&	 [\ion{Co}{1}] 	&
  {$\ast\  $}	&	2.6521	&	 [\ion{Co}{1}] 	&
 {$\ast\ \ast\ $} 	&	2.7173	&	 [\ion{Fe}{3}] 	&
 {$\ast\ $} 	&	2.8713	&	 [\ion{Co}{1}] 	\\
{$\ast\ \ast\ $}  	&	2.8742	&	 [\ion{Fe}{3}] 	&
 {$\ast\ \ast\ $} 	&	2.9048	&	 [\ion{Fe}{3}] 	&
 {$\ast\ \ast\ $} 	&	2.9114	&	 [\ion{Ni}{2}] 	&
 {$\ast\ $} 	&	2.9542	&	 [\ion{Co}{1}] 	&
 {$\ast\ \ast\  $} 	&	 2,9610 	&	 [\ion{Fe}{2}] 	\\
 	&	3.0305	&	 [\ion{Co}{1}] 	&
 {$\ast\ \ast\ $} 	&	3.0439	&	 [\ion{Fe}{3}] 	&
 	&	3.0457	&	 [\ion{Co}{1}] 	&
 {$\ast\ \ast\ \ast\ $} 	&	3.12	&	 [\ion{Ni}{1}] 	&
 	&	3.2294	&	 [\ion{Fe}{3}] 	\\
 	&	3.3942	&	 [\ion{Ni}{3}] 	&
 	&	3.4917	&	 [\ion{Co}{3}] 	&
 {$\ast\ $} 	&	3.6334	&	 [\ion{Co}{1}] 	&
  	&	3.7498	&	 [\ion{Co}{1}] 	&
 	&	3.8023	&	 [\ion{Ni}{3}] 	\\
 {$\ast\ \ast\ $} 	&	3.9524	&	 [\ion{Ni}{1}] 	&
 {$\ast\ \ast\ \ast\ $} 	&	4.0763	&	 [\ion{Fe}{2}] 	&
 	&	4.082	&	 [\ion{Fe}{2}] 	&
 {$\ast\ \ast\ $} 	&	4.115	&	 [\ion{Fe}{2}] 	&
 	&	4.3071	&	 [\ion{Co}{2}] 	\\
 	&	4.5196	&	 [\ion{Ni}{1}] 	&
 {$\ast\ \ast\ $} 	&	4.6077	&	 [\ion{Fe}{2}] 	&
    \enddata
      \tablecomments{The relative strengths are indicated by the number of $*$. For transitions without known 
    lifetimes (marked by $^{\dagger}$), $A_{i,j}$ are assumed from the equivalent iron levels. }
\end{deluxetable}

\vfill\eject
\section{MIR Transitions without cross-sections} \label{sec:opt_lines}
\renewcommand\thetable{\thesection.\arabic{table}}   
\setcounter{table}{0} 

Weak MIRI/MRS features are used to calibrate the atomic models.
One of the main uncertainties in the modeling is the lack of
atomic data for many weak transitions  \citep{Hoeflich_2021_20qxp} or, here, the
lifetimes or Einstein coefficients $A_{l,u}$ 
for the spontaneous decay.

  In principle, we follow the method employed by \citet{1993PhST...47..110K,1995HiA....10..579K}, who calibrated allowed cross-sections by comparing synthetic spectra with the observed solar spectrum. Similarly,  we 
make use of the MIRI spectrum of SN~2021aefx to 
estimate Einstein A-values. We do not change the atomic models
and only adjust atomic cross-sections between known energy levels but without previously measured values. Because of the low densities
and non-thermal excitation, the full rate equations are simulated
for the given background.

However, the applicability of the method 
is limited by the S/N $\approx 5-10$ in
the continuum and the uncertainty in the absolute calibration between the channels (Sect. \ref{subsec:Overall}). Moreover, the features are smeared out over several 1000 \kms. Many weak features are present in the quasi-continuum in both the observations and synthetic spectra (see Fig. \ref{fig:specmod}), e.g. between 13 and 23 \micron, ,but not always: Two features are predicted at 15.9 \micron\ and 16.8 \micron, ~ whereas a single, broad feature has been observed without a corresponding transition in Table \ref{tab:trans1} to fill the gap. 
The method can be made more complete by future SNe~Ia observations with MIRI/MRS  during the early-time nebular phase. For SN~2021aefx at day +415, only a few features are well above the noise level and can be used individually. However, the underlying flux is produced by many weak lines 
of Fe, Co, and Ni I-IV and, at +415~d, a quasi-continuum of free-free and bound-free and some allowed lines produced in inner layers \citep{Diamond_etal_2015}. The $A_{lu}$ values have been found by a Monte-Carlo scheme for some individual transitions by attributing the `missing' continuum flux by a global {\sl default} $A_{lu}$ value, with the latter determined by the residual flux in the continuum around 9.5$\pm$0.4~\micron\ and $15\pm 2$~$\mathrm{\mu}$m. The values adopted are given in Tabs. \ref{tab:trans1} \& \ref{tab:trans2}. The corrections to the flux are on the percent level.

\begin{deluxetable*}{llllll|llllll}[ht]
\label{tab:trans1}
\tabletypesize{\scriptsize}
  \tablecaption{Life-times of transitions without experimental values have been calculated based
on a combination of observations and the base model.\label{tab:trans}}
  \tablehead{\colhead{$\lambda$~[\mic]} &
   \colhead{Ion} & 
   \colhead{Term} &
    \colhead{$J_{l,u}$} &
     \colhead{$A_{lu}$} &
      \colhead{ $E_{l,u} [cm^{-1}]$} &
   \colhead{$\lambda$~[\micron]} & 
   \colhead{Ion} 
   & \colhead{Term} &
    \colhead{$J_{l,u}$} &
     \colhead{$A_{lu}$} & 
     \colhead{ $E_{l,u} [cm^{-1}]$} 
   }
\startdata
 5.022021&[Fe II]&a2G-a2P&7/2,3/2&1E-3& 16369.41-18360.64 &
 5.051034     &    [Fe II]     &       b4P-b4F    &         5/2,7/2 &  1E-3    &      20830.55-22810.35 \\
5.022021&{[\ion{Fe}{2}]}&a2G-a2P&7/2,3/2&1E-3&16369.41-18360.64&
5.051034&{[\ion{Fe}{2}]}&b4P-b4F&5/2,7/2&1E-3&20830.55-22810.35\\
5.0623456&{[\ion{Fe}{2}]}&a6D-a4F&3/2,5/2&1E-3&862.61-2837.98&
5.105920&{[\ion{Co}{2}]}&b1G-a3H&4,6&1E-3&25147.23-27105.74\\
5.144614&{[\ion{Fe}{2}]}&a2P-b4P&1/2,5/2&1E-3&18886.77-20830.55&
5.149301&{[\ion{Co}{2}]}&a1D-a3P&2,0&1E-3&11651.28-13593.29\\
5.157755&{[\ion{Co}{2}]}&b3F-a3P&2,2&1E-3&11321.86-13260.69&
5.282999&{[\ion{Fe}{2}]}&a2D2-b4P&5/2,1/2&1E-3&20516.95-22409.82\\
5.301864&{[\ion{Fe}{2}]}&a4G-a2F&11/2,7/2&1E-3&25428.79-27314.92&
5.3401693&{[\ion{Fe}{2}]}&a6D-a4F&9/2,9/2&1E-2&0.00-1872.60\\
5.3736547&{[\ion{Fe}{2}]}&a6D-a4F&1/2,5/2&1E-2&977.05-2837.98&
5.396797&{[\ion{Co}{2}]}&a3G-b3P&3,1&1E-3&22414.43-24267.38\\
5.439462&{[\ion{Co}{2}]}&b3F-a1D&4,2&2E-3&9812.86-11651.28&
5.456068&{[\ion{Fe}{2}]}&b2P-a2F&3/2,5/2&1E-3&25787.58-27620.40\\
5.460209&{[\ion{Fe}{2}]}&a2H-b4F&9/2,9/2&1E-3&20805.76-22637.19&
5.477981&{[\ion{Fe}{1}]}&a5P-a3P&2,1&1E-3&17726.99-19552.48\\ 
5.509410&{[\ion{Fe}{2}]}&a4G-a2F&9/2,5/2&1E-3&25805.33-27620.40&
5.535132&{[\ion{Fe}{2}]}&b4P-b4F&5/2,9/2&1E-3&20830.55-22637.19\\
5.596278&{[\ion{Co}{2}]}&b3P-c3P&2,0&1E-3&24074.42-25861.32&
5.604112&{[\ion{Co}{2}]}&a3D-a3D&3,1&1E-3&27484.37-29268.78\\
5.6316&{[\ion{Co}{3}]}&a4P-a2G&5/2,9/2&1E-3&15201.90-16977.60&
5.6739070&{[\ion{Fe}{2}]}&a6D-a4F&5/2,7/2&1E-2&667.68-2430.14\\
5.801083&{[\ion{Co}{2}]}&c3P-a1P&0,1&1E-3&25861.32-27585.14&
5.802874&{[\ion{Fe}{2}]}&a2D2-b4F&3/2,3/2&1E-3&21308.00-23031.28\\
5.826&{[\ion{Co}{4}]}&3H-3F2&6,4&1E-3&23679.50-25396.00&
5.845&{[\ion{Ni}{4}]}&4P-2G&5/2,9/2&1E-3&18118.60-19829.60\\
5.868&{[\ion{Co}{4}]}&3H-3F2&5,3&1E-3&24031.80-25735.90&
5.893&{[\ion{Co}{4}]}&3H-3F2&4,2&1E-3&24272.00-25969.00\\
6.024136&{[\ion{Co}{2}]}&a3G-b3P&3,2&5E-3&22414.43-24074.42&
6.102182&{[\ion{Fe}{2}]}&a4G-a2F&7/2,5/2&5E-3&25981.65-27620.40\\
6.129884&{[\ion{Fe}{2}]}&a2D2-b4F&3/2,5/2&5E-3&21308.00-22939.35&
6.134290&{[\ion{Fe}{2}]}&a2P-a2D2&1/2,5/2&5E-3&18886.77-20516.95\\
6.153483&{[\ion{Fe}{1}]}&a5P-a3P&1,1&5E-3&17927.38-19552.48&
6.227663&{[\ion{Fe}{2}]}&a4H-a6S&7/2,5/2&5E-3&21711.90-23317.64\\
6.260968&{[\ion{Co}{2}]}&a3F-a3F&4,2&3E-2&0.00,1597.20&
6.332062&{[\ion{Fe}{2}]}&b4P-b4P&5/2,1/2&5E-3&20830.55-22409.82\\
6.3794832&{[\ion{Fe}{2}]}&a6D-a4F&3/2,7/2&2E-2&862.61-2430.14&
6.389812&{[\ion{Fe}{2}]}&a4G-a2F&5/2,5/2&1E-3&26055.41-27620.40\\
6.547345&{[\ion{Fe}{2}]}&b2P-a2F&3/2,7/2&1E-3&25787.58-27314.92&
6.624310&{[\ion{Fe}{2}]}&a4G-a2F&9/2,7/2&1E-3&25805.33-27314.92\\
6.626905&{[\ion{Co}{2}]}&b3F-b3F&4,2&2E-3&9812.86-11321.86&
6.62998&{[\ion{Ni}{1}]}&3D-3D&3,1&1E-2&204.79-1713.09\\
6.641916&{[\ion{Fe}{2}]}&b4P-a6S&3/2,5/2&1E-3&21812.05-23317.64&
6.656252&{[\ion{Fe}{2}]}&a2D2-b4F&3/2,7/2&1E-3&21308.00-22810.35\\
6.721277&{[\ion{Fe}{2}]}&a6D-a4F&7/2,9/2&5E-2&384.79-1872.60&
6.831&{[\ion{Co}{4}]}&3H-3F2&4,3&1E-3&24272.00-25735.90\\
6.890&{[\ion{Co}{3}]}&a4F-a4F&9/2,5/2&1E-2&0.00-1451.30&
6.986&{[\ion{Ni}{4}]}&4F-4F&7/2,3/2&5E-3&1189.70-2621.10\\
7.201&{[\ion{Co}{4}]}&3P2-3H&2,4&1E-3&22883.30-24272.00&
7.217029&{[\ion{Fe}{2}]}&a4H-b4F&13/2,9/2&1E-3&21251.58-22637.19\\
7.246432&{[\ion{Fe}{2}]}&a4H-b4F&11/2,7/2&1E-3&21430.36-22810.35&
7.290488&{[\ion{Fe}{2}]}&a2H-a4H&11/2,7/2&1E-3&20340.25-21711.90\\
7.330&{[\ion{Co}{4}]}&3H-3F2&5,4&1E-3&24031.80-25396.00&
7.48092&{[\ion{Ni}{1}]}&3D-3F&2,2&2E-2&879.82-2216.55\\
7.500338&{[\ion{Fe}{2}]}&a4G-a2F&7/2,7/2&1E-3&25981.65-27314.92&
7.579280&{[\ion{Fe}{2}]}&a4H-b4F&7/2,3/2&1E-3&21711.90-23031.28\\
7.721459&{[\ion{Fe}{2}]}&a2D2-b4P&5/2,3/2&1E-3&20516.95-21812.05&
7.888699&{[\ion{Fe}{2}]}&b2H-a2F&9/2,5/2&1E-3&26352.77-27620.40\\
7.939619&{[\ion{Fe}{2}]}&a4G-a2F&5/2,7/2&1E-3&26055.41-27314.92&
8.055622&{[\ion{Fe}{2}]}&a2H-a4H&11/2,9/2&1E-3&20340.25-21581.62\\
8.066&{[\ion{Co}{4}]}&3P2-3F2&1,2&1E-3&24729.20-25969.00&
8.138479&{[\ion{Fe}{2}]}&a4H-b4F&9/2,7/2&1E-3&21581.62-22810.35\\
8.146939&{[\ion{Fe}{2}]}&a4H-b4F&7/2,5/2&1E-3&21711.90-22939.35&
8.201847&{[\ion{Fe}{2}]}&b4P-b4F&3/2,3/2&1E-3&21812.05-23031.28\\
8.262448&{[\ion{Co}{2}]}&a5F-a5F&5,3&2E-3&3350.49-4560.79&
8.286112&{[\ion{Fe}{2}]}&a4H-b4F&11/2,9/2&1E-3&21430.36-22637.19\\
8.299328&{[\ion{Fe}{2}]}&a6D-a4F&5/2,9/2&1E-3&667.68-1872.60&
8.317945&{[\ion{Co}{2}]}&a5P-a1G&3,4&1E-3&17771.51-18973.73\\
8.368601&{[\ion{Fe}{2}]}&a2D2-a4H&5/2,7/2&1E-3&20516.95-21711.90&
8.37951&{[\ion{Ni}{1}]}&3F-1D&2,2&1E-3&2216.55-3409.94\\
8.735629&{[\ion{Fe}{2}]}&b2H-a2F&11/2,7/2&1E-3&26170.18-27314.92&
8.87015&{[\ion{Ni}{1}]}&3D-3F&3,3&1E-2&204.79-1332.16\\
8.870707&{[\ion{Fe}{2}]}&b4P-b4F&3/2,5/2&1E-3&21812.05-22939.35&
8.897&{[\ion{Co}{4}]}&3H-3F2&4,4&1E-3&24272.00-25396.00\\
9.075902&{[\ion{Fe}{2}]}&a2D2-b4P&3/2,1/2&1E-3&21308.00-22409.82&
9.103740&{[\ion{Fe}{2}]}&a4H-b4F&7/2,7/2&1E-3&21711.90-22810.35\\
9.173384&{[\ion{Fe}{2}]}&a2H-a4H&11/2,11/2&1E-3&20340.25-21430.36&
9.279&{[\ion{Co}{4}]}&5D-5D&4,2&1E-2&0.00-1077.70\\
9.321289&{[\ion{Co}{2}]}&b3P-b1G&2,4&1E-3&24074.42-25147.23&
9.392655&{[\ion{Fe}{2}]}&a2D2-a4H&5/2,9/2&1E-3&20516.95-21581.62\\
9.473466&{[\ion{Fe}{2}]}&a4H-b4F&9/2,9/2&1E-3&21581.62-22637.19&
9.522585&{[\ion{Co}{2}]}&b3P-c3P&1,1&1E-3&24267.38-25317.52\\
9.750&{[\ion{Co}{3}]}&a4F-a4F&7/2,3/2&1E-2&841.20-1866.80&
9.933&{[\ion{Co}{4}]}&3P2-3F2&1,3&1E-3&24729.20-25735.90\\
10.017024&{[\ion{Fe}{2}]}&b4P-b4F&3/2,7/2&1E-3&21812.05-22810.35&
10.077445&{[\ion{Fe}{1}]}&a3F-a3F&4,2&1E-3&11976.24-12968.55\\
10.257231&{[\ion{Co}{2}]}&c3P-c3P&2,0&1E-3&24886.40-25861.32&
10.358606&{[\ion{Fe}{2}]}&a4F-a4F&9/2,5/2&1E-3&1872.60-2837.98\\
10.393377&{[\ion{Fe}{2}]}&b2H-a2F&9/2,7/2&1E-3&26352.77-27314.92&
10.605061&{[\ion{Co}{2}]}&b3F-a1D&3,2&1E-3&10708.33-11651.28\\
10.612332&{[\ion{Co}{2}]}&a5P-a1G&2,4&1E-3&18031.43-18973.73&
10.807321&{[\ion{Fe}{2}]}&a4H-b4F&7/2,9/2&1E-3&21711.90-22637.19\\
10.822771&{[\ion{Fe}{2}]}&a4G-b2H&11/2,9/2&1E-3&25428.79-26352.77&
10.856891&{[\ion{Co}{2}]}&a5F-a5F&4,2&1E-3&4028.99-4950.06\\
10.972887&{[\ion{Fe}{2}]}&a2H-a4H&11/2,13/2&1E-3&20340.25-21251.58
\enddata
    \tablecomments{The relative strengths of individual features are fitted. For many weak lines,
    a standard value has been assumed based on the overall spectra (see text).}
\end{deluxetable*}

\begin{deluxetable*}{llllll|llllll}[ht]
\label{trans2}
\tabletypesize{\scriptsize}
  \tablecaption{Same as Table 4 but for 11-30 \mic . \label{tab:trans2}}
  \tablehead{\colhead{$\lambda$~[\mic]} & 
  \colhead{Ion}  & 
  \colhead{Term} & 
  \colhead{$J_{l,u}$} &
   \colhead{$A_{lu}$} &
    \colhead{ $E_{l,u} [cm^{-1}]$} &
   \colhead{$\lambda$~[\mic]} & 
   \colhead{Ion} &
    \colhead{Term} &
     \colhead{$J_{l,u}$} & 
     \colhead{$A_{lu}$} &
      \colhead{ $E_{l,u} [cm^{-1}]$} 
   }
\startdata
11.015432&{[\ion{Fe}{2}]}&b4P-a6S&1/2,5/2&1E-3&22409.82-23317.64&
11.035906&{[\ion{Fe}{2}]}&a2H-a4H&9/2,7/2&1E-3&20805.76-21711.90\\
11.037065&{[\ion{Co}{2}]}&b3P-c3P&0,1&1E-3&24411.48-25317.52&
11.36601&{[\ion{Ni}{1}]}&3F-3D&4,2&2E-2&0.00-879.82\\
11.398313&{[\ion{Fe}{2}]}&a4G-b2P&5/2,1/2&1E-3&26055.41-26932.73&
11.825&{[\ion{Co}{4}]}&3G-3G&5,3&1E-3&29021.80-29867.50\\
11.908&{[\ion{Ni}{4}]}&4P-4P&5/2,1/2&1E-3&18118.60-18958.40&
12.077222&{[\ion{Fe}{1}]}&a5P-a3P&3,2&1E-3&17550.18-18378.19\\
12.315574&{[\ion{Co}{2}]}&b3P-c3P&2,2&1E-3&24074.42-24886.40&
12.503266&{[\ion{Fe}{1}]}&a5F-a5F&5,3&2E-3&6928.27-7728.06\\
12.55624&{[\ion{Co}{2}]}&a3H-a3H&6,4&1E-3&27105.74-27902.16&
12.659782&{[\ion{Co}{2}]}&a3G-a3G&5,3&1E-3&21624.53-22414.43\\
12.889058&{[\ion{Fe}{2}]}&a2H-a4H&9/2,9/2&1E-3&20805.76-21581.62&
13.314485&{[\ion{Fe}{2}]}&b4P-a4H&5/2,9/2&1E-3&20830.55-21581.62\\
13.48815&{[\ion{Fe}{2}]}&a4G-b2H&11/2,11/2&1E-3&25428.79-26170.18&
13.790212&{[\ion{Fe}{2}]}&a4D-a4D&7/2,3/2&1E-3&7955.32-8680.47\\
13.924&{[\ion{Co}{4}]}&5D-5D&3,1&1E-2&639.10-1357.30&
14.204404&{[\ion{Fe}{1}]}&a5D-a5D&4,2&1E-2&0.00-704.01\\
14.54189&{[\ion{Fe}{2}]}&b2P-a2F&1/2,5/2&1E-3&26932.73-27620.40&
14.548612&{[\ion{Fe}{2}]}&a4F-a4F&7/2,3/2&5E-3&2430.14-3117.49\\
14.69637&{[\ion{Fe}{2}]}&b4F-a6S&9/2,5/2&1E-3&22637.19-23317.64&
14.977170&{[\ion{Fe}{2}]}&a6D-a6D&9/2,5/2&1E-2&0.00-667.68\\
15.35631&{[\ion{Fe}{1}]}&a5P-a3P&2,2&1E-3&17726.99-18378.19&
15.53014&{[\ion{Co}{2}]}&a5F-a5F&3,1&1E-3&4560.79-5204.70\\
16.01042&{[\ion{Fe}{2}]}&a2H-a4H&9/2,11/2&1E-3&20805.76-21430.36&
16.09101&{[\ion{Fe}{2}]}&b4P-b4F&1/2,3/2&1E-3&22409.82-23031.28\\
16.407&{[\ion{Co}{3}]}&a4P-a4P&5/2,1/2&1E-3&15201.90-15811.40&
16.41982&{[\ion{Fe}{1}]}&a5F-a5F&4,2&5E-3&7376.76-7985.78\\
16.878&{[\ion{Co}{4}]}&3H-3H&6,4&1E-3&23679.50-24272.00&
17.572&{[\ion{Ni}{3}]}&3P-3P&2,0&1E-3&16661.60,17230.70\\
17.63255&{[\ion{Co}{2}]}&a5P-a5P&3,1&1E-3&17771.51-18338.64&
18.08790&{[\ion{Fe}{2}]}&a4G-a4G&11/2,7/2&1E-3&25428.79-25981.65\\
18.26684&{[\ion{Fe}{2}]}&a4G-b2H&9/2,9/2&1E-3&25805.33-26352.77&
18.88455&{[\ion{Fe}{2}]}&b4P-b4F&1/2,5/2&1E-3&22409.82-22939.35\\
19.71262&{[\ion{Fe}{2}]}&b4F-a6S&7/2,5/2&1E-3&22810.35-23317.64&
19.83944&{[\ion{Fe}{2}]}&a2D2-b4P&3/2,3/2&1E-3&21308.00-21812.05\\
19.8624&{[\ion{Ni}{1}]}&3D-3F&1,2&2E-3&1713.09,2216.55&
20.928182&{[\ion{Fe}{2}]}&a6D-a6D&7/2,3/2&3E-3&384.79-862.61\\
20.94479&{[\ion{Fe}{2}]}&b4P-a2D2&5/2,3/2&3E-3&20830.55-21308.00&
21.05613&{[\ion{Co}{2}]}&b3P-c3P&0,2&2E-3&24411.48-24886.40\\
21.17751&{[\ion{Fe}{1}]}&a5D-a5D&3,1&2E-3&415.93-888.13&
22.1069&{[\ion{Ni}{1}]}&3D-3F&2,3&1E-2&879.82-1332.16\\
22.18259&{[\ion{Fe}{1}]}&a5P-a3P&1,2&1E-3&17927.38-18378.19&
22.43057&{[\ion{Fe}{2}]}&a2H-a4H&9/2,13/2&2E-3&20805.76-21251.58\\
23.23350&{[\ion{Fe}{2}]}&a4P-a4P&5/2,1/2&2E-3&13474.45-13904.86&
23.43820&{[\ion{Fe}{1}]}&a5F-a5F&3,1&2E-3&7728.06-8154.71\\
23.93558&{[\ion{Co}{2}]}&a3D-a3H&3,4&2E-3&27484.37-27902.16&
24.04&{[\ion{Co}{4}]}&5D-5D&2,0&1E-2&1077.70-1493.60\\
24.5422&{[\ion{Ni}{1}]}&3P-3P&2,0&1E-3&15609.84-16017.31&
24.75877&{[\ion{Fe}{2}]}&a2D2-a4H&3/2,7/2&1E-3&21308.00-21711.90\\
26.2520&{[\ion{Ni}{1}]}&3F-3D&3,1&2E-3&1332.16-1713.09&
26.43517&{[\ion{Fe}{2}]}&b4F-a6S&5/2,5/2&1E-3&22939.35-23317.64\\
26.51106&{[\ion{Fe}{1}]}&a5P-a5P&3,1&1E-3&17550.18-17927.38&
26.63&{[\ion{Co}{3}]}&a2D2-a2H&5/2,9/2&1E-3&23058.80-23434.30\\
26.94532&{[\ion{Fe}{2}]}&a4G-b2H&7/2,9/2&1E-3&25981.65-26352.77&
27.40822&{[\ion{Fe}{2}]}&a4G-b2H&9/2,11/2&1E-3&25805.33-26170.18 
\enddata
    \tablecomments{The relative strengths of individual features are fitted. For many weak lines,
    a standard value has been assumed based on the overall spectra (see text).}
\end{deluxetable*}

\end{document}

%% file: authors.tex
\author[0000-0002-5221-7557]{ C.~Ashall}
\affiliation{Department of Physics, Virginia Tech, Blacksburg, VA 24061, USA}

\author[0000-0002-4338-6586]{P.~Hoeflich}
\affiliation{Department of Physics, Florida State University, 77 Chieftan Way, Tallahassee, FL 32306, USA}
\author[0000-0001-5393-1608]{E.~Baron}
\affiliation{Planetary Science Institute, 1700 East Fort Lowell Road, Suite 106,
 Tucson, AZ 85719-2395 USA}
\affiliation{Hamburger Sternwarte, Gojenbergsweg 112, D-21029 Hamburg, Germany}

\author[0000-0002-7566-6080]{M.~Shahbandeh}
\affiliation{Space Telescope Science Institute, 3700 San Martin Drive, Baltimore, MD 21218-2410, USA}

\author[0000-0002-7566-6080]{J.~M.~DerKacy}
\affiliation{Department of Physics, Virginia Tech, Blacksburg, VA 24061, USA}

\author[0000-0001-7186-105X]{K.~Medler}
\affiliation{Department of Physics, Virginia Tech, Blacksburg, VA 24061, USA}

\author[0000-0003-4631-1149]{B.~J.~Shappee}
\affiliation{Institute for Astronomy, University of Hawai'i at Manoa, 2680 Woodlawn Dr., Hawai'i, HI 96822, USA }

\author[0000-0002-2471-8442]{M.~A.~Tucker}
\altaffiliation{CCAPP Fellow}
\affiliation{Center for Cosmology and AstroParticle Physics, The Ohio State University, 191 W. Woodruff Ave., Columbus, OH 43210, USA}

\author[0009-0001-9148-8421]{E.~Fereidouni}
\affiliation{Department of Physics, Florida State University, 77 Chieftan Way, Tallahassee, FL 32306, USA}

\author[0000-0001-5888-2542]{T.~Mera}
\affiliation{Department of Physics, Florida State University, 77 Chieftan Way, Tallahassee, FL 32306, USA}

\author[0000-0003-0123-0062]{J.~Andrews}
\affiliation{Gemini Observatory/NSF’s NOIRLab, 670 North A`ohoku Place, Hilo, HI 96720-2700, USA}

\author[0000-0003-1637-9679]{D.~Baade}
\affiliation{European Organization for Astronomical Research in the Southern Hemisphere (ESO), Karl-Schwarzschild-Str. 2, 85748 Garching b. M\"unchen, Germany}

\author[0000-0002-4924-444X]{K.~A.~Bostroem}
\altaffiliation{LSSTC Catalyst Fellow}
\affiliation{Steward Observatory, University of Arizona, 933 North Cherry Avenue, Tucson, AZ 85721-0065, USA}

\author[0000-0001-6272-5507]{P.~J.~Brown}
\affiliation{George P. and Cynthia Woods Mitchell Institute for Fundamental Physics and Astronomy, Texas A\&M University, Department of Physics and Astronomy, College Station, TX 77843, USA}

\author[0000-0003-4625-6629]{C.~R.~Burns}
\affiliation{Observatories of the Carnegie Institution for Science, 813 Santa Barbara Street, Pasadena, CA 91101, USA}

\author[0000-0002-5380-0816]{A.~Burrow}
\affiliation{Homer L. Dodge Department of Physics and Astronomy, University of Oklahoma, 440 W. Brooks, Rm 100, Norman, OK 73019-2061, USA}

\author[0000-0001-7101-9831]{A.~Cikota}
\affiliation{Gemini Observatory/NSF's NOIRLab, Casilla 603, La Serena, Chile}

\author[0000-0001-6069-1139]{T.~de~Jaeger}
\affiliation{Sorbonne Université, CNRS/IN2P3, LPNHE, F-75005, Paris, France}

\author[0000-0003-3429-7845]{A.~Do}
\affiliation{Institute of Astronomy and Kavli Institute for Cosmology, Madingley Road, Cambridge, CB3 0HA, UK}

\author[0000-0002-7937-6371]{Y.~Dong}
\affiliation{Department of Physics, University of California, 1 Shields Avenue, Davis, CA 95616-5270, USA}

\author[0000-0002-3827-4731]{I. Dominguez}
\affiliation{Universidad de Granada, 18071, Granada, Spain}

\author[0000-0003-2238-1572]{O. Fox}
\affiliation{Space Telescope Science Institute, 3700 San Martin Drive, Baltimore, MD 21218-2410, USA}

\author[0000-0002-1296-6887]{L.~Galbany}
\affiliation{Institute of Space Sciences (ICE, CSIC), Campus UAB, Carrer de Can Magrans, s/n, E-08193 Barcelona, Spain}
\affiliation{Institut d’Estudis Espacials de Catalunya (IEEC), E-08034 Barcelona, Spain}

\author[0000-0003-1039-2928]{E.~Y.~Hsiao}
\affiliation{Department of Physics, Florida State University, 77 Chieftan Way, Tallahassee, FL 32306, USA}

\author[0000-0002-6650-694X]{K.~Krisciunas}
\affiliation{George P. and Cynthia Woods Mitchell Institute for Fundamental Physics and Astronomy, Texas A\&M University, Department of Physics and Astronomy, College Station, TX 77843, USA}

\author[0009-0005-0311-0058]{B.~Khaghani}
\affiliation{Department of Physics, Virginia Tech, Blacksburg, VA 24061, USA}

\author[0000-0001-8367-7591]{S.~Kumar}
\affiliation{Department of Astronomy, University of Virginia, 530 McCormick Rd, Charlottesville, VA 22904, USA}

\author[0000-0002-3900-1452]{J.~Lu}
\affil{Department of Physics and Astronomy, Michigan State University, East Lansing, MI 48824, USA}

\author[0000-0003-0733-7215]{J.~R.~Maund}
\affiliation{Department of Physics and Astronomy, University of Sheffield, Hicks Building, Hounsfield Road, Sheffield S3 7RH, U.K.}

\author[0000-0001-6876-8284]{P.~Mazzali}
\affiliation{Astrophysics Research Institute, Liverpool John Moores University, UK}
\affiliation{Max-Planck Institute for Astrophysics, Garching, Germany}

\author[0000-0003-2535-3091]{N.~Morrell}
\affiliation{Las Campanas Observatory, Carnegie Observatories, Casilla 601, La Serena, Chile}
	
\author[0000-0002-0537-3573]{F.~Patat}
\affiliation{European Organization for Astronomical Research in the Southern Hemisphere (ESO), Karl-Schwarzschild-Str. 2, 85748 Garching b. M\"unchen, Germany}

\author[0000-0002-7305-8321]{C.~Pfeffer}
\affiliation{Department of Physics, Virginia Tech, Blacksburg, VA 24061, USA}

\author[0000-0003-2734-0796]{M.~M.~Phillips}
\affiliation{Las Campanas Observatory, Carnegie Observatories, Casilla 601, La Serena, Chile}

\author{J. Schmidt}
\affiliation{Citizen scientist}

\author[0000-0001-5570-6666]{S.~Stangl}
\affiliation{Homer L. Dodge Department of Physics and Astronomy, University of Oklahoma, 440 W. Brooks, Rm 100, Norman, OK 73019-2061, USA}

\author[0000-0003-0763-6004]{C.~P.~Stevens}
\affiliation{Department of Physics, Virginia Tech, Blacksburg, VA 24061, USA}

\author[0000-0002-5571-1833]{M.~D.~Stritzinger}
\affiliation{Department of Physics and Astronomy, Aarhus University, Ny Munkegade 120, DK-8000 Aarhus C, Denmark}

\author[0000-0002-8102-181X]{N.~B.~Suntzeff}
\affiliation{George P. and Cynthia Woods Mitchell Institute for Fundamental Physics and Astronomy, Texas A\&M University, Department of Physics and Astronomy, College Station, TX 77843, USA}

\author[0000-0002-0036-9292]{C.~M.~Telesco}
\affiliation{Department of Astronomy, University of Florida, Gainesville, FL 32611 USA}

\author[0000-0001-7092-9374]{L.~Wang}
\affiliation{Department of Physics and Astronomy, Texas A\&M University, College Station, TX 77843, USA}

\author[0000-0002-6535-8500]{Y.~Yang}
\altaffiliation{Bengier-Winslow-Robertson Postdoctoral Fellow}
\affiliation{Department of Astronomy, University of California, Berkeley, CA 94720-3411, USA}